\documentclass[prodmode]{acmsmall}

\usepackage[ruled]{algorithm2e}
\usepackage{color}
\usepackage{amssymb}
\usepackage{subfigure}
\usepackage{url}
\usepackage{rotating}
\usepackage{multirow}
\usepackage{hyperref}

\usepackage{listings}

\newtheorem{thm}{Theorem}[section]

\newtheorem{lem}[thm]{Lemma}

\renewcommand{\algorithmcfname}{ALGORITHM}
\SetAlFnt{\small}
\SetAlCapFnt{\small}
\SetAlCapNameFnt{\small}
\SetAlCapHSkip{0pt}
\IncMargin{-\parindent}

\newcommand{\comment}[1]{{\textcolor{magenta}{\,[\emph {#1}\,]}}}

\newcommand{\oss}{ \ensuremath{{^{O}\hspace{-0.5mm}S}^2_{n+1} }}
\newcommand{\nss}{ \ensuremath{{^{N}\hspace{-0.5mm}S}^2_{n+1} }}
\newcommand{\ossthree}{ \ensuremath{{^{O}\hspace{-0.5mm}S}^2_{3} }}
\newcommand{\nssthree}{ \ensuremath{{^{N}\hspace{-0.5mm}S}^2_{3} }}
\newcommand{\zone} { \ensuremath{{^{_1}\hspace{-0.5mm}z}}}
\newcommand{\ztwo} { \ensuremath{{^{_2}\hspace{-0.5mm}z}}}
\newcommand{\zi} { \ensuremath{{^{_i}\hspace{-0.5mm}z}}}
\newcommand{\zii} { \ensuremath{{^{_{i+1}}\hspace{-0.5mm}z}}}
\newcommand{\znn} { \ensuremath{{^{_{n+1}}\hspace{-0.5mm}z}}}

\def\var{\mathop{var}\nolimits}
\def\cov{\mathop{cov}\nolimits}

\begin{document}

% Page heads
\markboth{T. Kalibera, R.E. Jones}{Quantifying Performance Changes with Effect Size Confidence Intervals}

% Title portion
\title{Quantifying Performance Changes with Effect Size Confidence Intervals}
\author{TOMAS KALIBERA, RICHARD JONES
\affil{University of Kent}
}

\begin{abstract}

Measuring performance and quantifying a performance change are core
evaluation techniques in programming language and systems research.  Out of
122 recent scientific papers published at PLDI, ASPLOS, ISMM, TOPLAS, and
TACO, as many as 65 included experimental evaluation that quantified a
performance change using a ratio of execution times.
Unfortunately, few of these papers evaluated their results with the level of
rigour that has come to be expected in other experimental sciences.  The
uncertainty of measured results was largely ignored.  Scarcely any of the
papers mentioned uncertainty in the ratio of the mean execution times, and
most did not even mention uncertainty in the two means themselves. 
Furthermore, most of the papers failed to address the non-deterministic
execution of computer programs (caused by factors such as memory placement,
for example), and none addressed non-deterministic compilation (when a
compiler creates different binaries from the same sources, which differ
in performance, for example again because of impact on memory placement).
It turns out that the statistical methods presented in the computer systems
performance evaluation literature for the design and summary of experiments
do not readily allow this either.  This poses a hazard to the repeatability,
reproducibility and even validity of quantitative results.

Inspired by statistical methods used in other fields of science, and
building on results in statistics that did not make it to introductory
textbooks, we present a statistical model that allows us both to quantify
uncertainty in the ratio of (execution time) means and to design experiments
with a rigorous treatment of those multiple sources of non-determinism that
might impact measured performance.  Better still, under our framework
summaries can be as simple as ``system A is faster than system B by 5.5\%
$\pm$ 2.5\%, with 95\% confidence'', a more natural statement than those
derived from typical current practice, which are often misinterpreted.

\end{abstract}




%\acmformat{Kalibera, T., Jones, R.E. 201X. Quantifying Performance Changes with
%Effect Size Confidence Intervals.}

\begin{bottomstuff}
%This work is supported by -- add grants here.

Author's addresses: T. Kalibera (R.E. Jones), School of Computing, University
of Kent, Canterbury, CT2 7NF, UK.

\emph{A preliminary version of a portion of this work was presented at
the Third European Performance Engineering Workshop.}
\end{bottomstuff}

\maketitle

\section{Introduction}

Quantification of performance change is a common task in experimental
computer science.  Out of all 122~papers published in ASPLOS, ISMM, PLDI,
TACO and TOPLAS in 2011 up to 2nd August, 65~included empirical evaluation
that quantified performance change by giving the ratio of execution times
(for example, speed-ups of optimisations, or overheads of new techniques). 
Quantification of performance change is also part of the software
development process (for example, detecting performance regressions), both
open-source and commercial.

Execution times for quantification are obtained by running benchmark
applications using the two systems to compare.  Computer systems are
becoming more complex and increasingly resemble the direct products of
nature observed by physicists and natural scientists --- an overwhelming
number of factors influence performance.  Some are unknown, some are
out of the experimenter's control, and some are non-deterministic.  Repeated
executions of the same benchmark on the same system, even when the
experimenter does everything possible to enforce the same conditions, always
report slightly different execution times.  This is even more pronounced if
the same experiment is repeated by an independent experimenter using
different equipment.  Thus, for credible quantification of performance
change, one needs to design experiments and summarize their results in a
repeatable and reproducible manner.  Experiment design and statistical
inference are the fields of statistics that address these issues, and are
widely applied in physics, natural sciences and social sciences.

Unfortunately, the practice in computer science lags behind. Of the 122
papers at ASPLOS, ISMM, PLDI, TACO, and TOPLAS we looked at, 90 evaluated
execution time based on experiments.  This includes ratio of execution times
for two systems and also the execution times for a single system.  71 of
these 90 papers completely ignored the question of uncertainty in the
measured times.  This lack of rigour makes repeatability difficult and
undermines the validity of the results.  We hardly dare mention that simple
rules for summarizing and reporting uncertainties are taught to every
student of physics and natural sciences~\cite{kirkup}, are part of
engineering practice~\cite{nistunc}, and are based on statistics taught in
introductory courses.  Moreover, advice on elementary experiment design and
statistical inference has been readily available to computer scientists in
the form of textbooks aimed and used for teaching~\cite{jain,lilja}.  A
subset of the statistics of the latter textbook has been advocated in the
context of Java runtime performance evaluation~\cite{georges07} and compared
with the practice in scientific papers, with results not particularly
flattering to our field.

Nevertheless, even though the best quantification method recommended so far
in our field (\cite{georges07} based largely on~\cite{lilja}) is ahead of
current practice, it has a number of flaws.  It exacts a high price for
statistical rigour in summarizing results: \emph{we do not get a reliable
estimate of the metric that we are ultimately interested in, which is the
ratio of execution times}.  Maybe this flaw on its own is a significant
detractor from wider adoption of the method in practice.  Even worse, the
rigour that the method does provide is highly questionable, as the method is
based on statistical significance, a concept with a number of shortcomings,
some of which have been known for seven decades~\cite{cohen}.  While
statistical significance still dominates elementary statistical textbooks
and is still used by some researchers even in other fields, its limits have
been well described and arguments for its deprecation have been made in
statistics~\cite{statnhst}, psychology~\cite{cohen}, education~\cite{coe},
ecology, medicine, bio-medicine, and biology~\cite{nagawa}.  Critical views
of statistical significance have also been published in the context of
research in sociology, criminology, economics, marketing, chemistry, and
nursing (see~\citeN{fidlerps} for references and details).  Some journals
explicitly require alternative methods~\cite{eshill}.  And such methods are
available.  As we demonstrate in this paper, the specialized statistical
literature ~\cite{fieller,davison} provides methods to estimate the ratio of
execution times we are interested in --- inference of the ratio of means is
possible, and a so-called \emph{effect size} confidence interval can be
constructed for it.

Apart from the problems with the summary of results, \citeANP{georges07}'s
quantification method, while ahead of the prevailing practice in the field,
lacks rigour in experimental design.  It is based on a
two-level~\footnote{Please note that we use the term `level' informally. In
our text, it is \emph{not} what`factor level' means in statistics.}
hierarchical experiment that repeats executions of a benchmark, each
consisting of repeated measurements.  For example, to evaluate an
optimisation in a Java virtual machine, the method requires us to run many
invocations of the JVM and have each invocation run and measure many
iterations of a test application.  Two levels are not enough.  It is well
known that a large number of factors that influence performance
are inherently non-deterministic or need to be randomized by
experimenter to avoid measurement bias~\cite{envsize}.  The first group
includes for example context switches, hardware interrupts, memory placement
due to virtual-to-physical memory mapping~\cite{mascots05}, randomized
algorithms in compilation~\cite{epew06}, or decisions of a just-in-time
compiler on which methods to compile.  Non-deterministic compilation by
itself means that compilation needs to be repeated.  The second group
includes symbol names of methods and variables in source code which impacts
binary code layout~\cite{verbrugge}, the size of the UNIX process
environment or the linking order~\cite{envsize}.  

To efficiently run
experiments with such as number of sources of non-determinism, we need more
than two levels.  \citeANP{georges07}'s quantification method, however, does
not allow this.  It is, in fact, based on one-level only summarization,
which looks at the means from the lowest level (repeated measurements)
instead of the measurement themselves.  
%
% Georges probably does not propose this - put it back if there is evidence
%
%Or, alternatively, all measurements
%from different executions are treated as a single level, assuming that they
%are independent identically distributed, which they are not.  
%
There is a
better way.  While not included in elementary statistical textbooks, there
are statistical methods which model so-called random
effects~\cite{mcculloch}, which can represent non-deterministic factors in
our experiments.  To our knowledge, however, none of the existing models can
be used directly.  Thus, we extend one such model to be more robust and to
support an arbitrary number of experiment levels.  We also derive the
optimum number of repetitions at each level (with the best recommended
method so far, there is no systematic guidance to select the number of
repetitions).

\medskip
\noindent
Our contributions in this work are:

\begin{enumerate}
\item {\bfseries Statistical Inference for Quantification.} A statistical
model of a hierarchy of random effects that contribute to fluctuations in
performance.  This allows us to build a confidence interval for the mean
performance metric (such as execution time) within such a hierarchy.  This
is our first original contribution to statistics.
\item {\bfseries Experiment Planning for Quantification.} After some initial
measurements, the model can also be used to guide the experiment.  Repeating
all experiments blindly is often very expensive.  However, the model will
give the optimum number of repetitions of experiments at different levels of
the hierarchy that are necessary to provide the narrowest confidence
interval in available experimentation time.  This is our second original
contribution to statistics.
\item {\bfseries Asymptotic Parametric Quantification Method.} A parametric
approach to constructing a confidence interval for the ratio of mean metric
values, based on Fieller's theorem~\cite{fieller}, asymptotic normality of
the mean, and our statistical model.  Hence, for example, the result of a
study might be that the change is 5.5\% $\pm$ 2.5\%, with 95\% confidence. 
Such a statement provides a clear but rigorous account of the magnitude of
the performance change and the uncertainty involved.  
%This is an application
%of an existing statistical method, though not known in our field.
%
\item {\bfseries Bootstrap Quantification Method.} An alternative
non-parametric approach to constructing the same confidence interval using
hierarchical random re-sampling (a bootstrap method).  This is an
application of an existing statistical method, used in our field much less
than it deserves.
\item {\bfseries Evaluation.} Thorough evaluation of the two approaches on a
set of benchmarks.  We estimate the true coverage of the intervals that can
be obtained.  We also estimate a false alarm rate (the case where a change
is detected although there is none), given a threshold of how large changes
need to be for us to care.  A particularly important benefit of our method
is that it supports a rigorous way to use such a threshold.  This is our
third original contribution.
\end{enumerate}
Contributions 1, 2, 4, and partially 3 and 5 also apply to evaluations of
a single system. 
For contributions 1 and 2, we extend and correct a preliminary version
published in~\cite{epew06}, which in turn extends~\cite{mascots05}.  For
item 5, we use our benchmark data from~\cite{epew06} as inputs for (new)
statistical simulations.

\section{BACKGROUND: Current Quantification Methods}

In this section, we provide an overview of quantification methods currently
used, we describe their limitations, and we summarize the best-so-far
quantification methods recommended for overcoming these
limitations.

\subsection{Currently Used Methods}

To map the current quantification methods used in computer science, we
analysed papers published at selected conferences (PLDI, ASPLOS, ISMM) and
journals (TOPLAS, TACO).  We restricted our survey to papers published
between 1st January and 2nd August, 2011.  We thus cover all papers from the
conferences, issues 1--4 of TOPLAS, and issues 1--2 of TACO --- 122 papers
in total.

As summarized in Table~\ref{tPapers}, nearly three quarters of the papers
studied included empirical performance evaluation that measured and reported
execution time.  The majority of the execution time evaluations (over~70\%)
reported the ratio of execution times as a measure of performance change
(the remaining 30\% reported execution time for a single system).  In total,
over half of all papers reported the ratio of execution times.  We also
noted papers where execution time was important for the evaluation (counts
in parentheses in the table).  More than half of the papers had empirical
evaluations with execution time as an important metric.  Out of these papers
where execution time was an important metric, in nearly 90\% it was the
ratio of execution times that was reported (only in 10\% of cases it was the
absolute execution time for a single system).  Nearly half of all papers had
empirical evaluation with ratio of execution times as an important metric. 
The ratio of execution times thus seems to be a metric that people care
about and report.  Anecdotal evidence confirms that this metric is also
important to industrial developers.

In practice, where decisions have to be made based on the results of
performance quantification, there are certain minimum thresholds for
relative performance change above/below which it is considered of no
practical interest.  These thresholds may depend on the evaluation context
(a particular system, regression test, optimisation, new feature).  These
thresholds are also assumed and used in~\citeN{georges07}.  Of the papers in
our survey, thresholds on ratio of execution times are explicitly part of an
auto-tuning algorithm by~\citeN{mapreduce}, where they have to be set
(manually, based on experience) large enough to cater for variabilities in
the data.  However, most papers that quantify performance change report only
the ratio of (mean) execution times, and thus imply that this ratio alone,
maybe along with an understanding of the context of the study, is sufficient
to assess importance of a performance change.  The use of `thresholds' to
assess importance is implied.  We also have confirmation from a senior
member of a well-known industrial research laboratory of the use of such
thresholds in industry.

\begin{table}
\tbl{Current Practice of Performance Quantification\label{tPapers}}{
\begin{tabular}{|r|c|c|c|c|c||c|}
\hline
& PLDI & ASPLOS & ISMM & TOPLAS & TACO & Total\\
\hline
Number of Papers & 55 & 32 & 13 & 13 & 9 & 122\\
Evaluated Execution Time & 42 (28) & 25 (20) & 12 (11) & 5 (2)  & 6 (6) & 90 (67)\\
Ignored Uncertainty In Measurement & 39 (24) & 18 (12) & 5 (5) & 4 (1) & 5 (5) & 71 (47)\\
Evaluated Ratio of Execution Times & 27 (25) & 22 (19) & 9 (8) & 1 (1) & 6 (6) & 65 (59)\\
\hline
\end{tabular}
}
\begin{tabnote}
\Note{Source:}{Scientific papers at selected venues published in 2011, up to
2nd August. Counts are given in form `A (M)', where A is the number of
all papers that reported the metric, and M the number of papers where the
metric was important or the main result in the evaluation.
}
\end{tabnote}
\end{table}

Nearly 80\% of the papers studied that evaluated execution time (both
absolute and relative) failed to mention anything about uncertainty in the
figures they reported.  The numbers are not that much better if we only
focus on papers where execution time was an important metric: 70\% of these
papers failed to mention uncertainty.  Note that we were very generous in
the classification --- if a paper reported that uncertainty was low, without
any further details, we gave it the benefit of the doubt and counted it as
having taken uncertainty into account.  Ignoring uncertainty represents a
severe lack of rigour in face of the fact that execution time on today's
computer systems is always subject to variation in performance.  While
majority of the papers studied ran experiments on real hardware, it should
be mentioned that some (particularly from ASPLOS) were based on execution
time measurements in a simulator.  It is certainly possible to have a
simulator in which execution time is deterministic, and we did not attempt
to check the particular simulators used in individual papers we analysed. 
However, if a simulator is deterministic, it is not a realistic
one~\cite{alameldeen}.  Thus, we have not attempted to extract simulations
from the summary --- using a deterministic simulator in our opinion falls
into the category of ignoring uncertainty.

We used to hear arguments that, for very large speed-ups (say 2$\times$),
summarizing uncertainty is a waste of effort.  This implies that
experimenter experience is that, in the given system, uncertainty would be
much smaller.  However,  in systems research much
smaller performance changes are usually reported.  \citeN{envsize} found that
out of 88 papers from ASPLOS 2008, PACT 2007, PLDI 2007, CGO 2007 with
experimental evaluation in a dedicated section, the median of reported
speed-ups was only 10\%.  In their work they show that the measurement bias
with SPEC CPU 2006 benchmarks can easily obfuscate this speed-up when
evaluating compiler optimisations.  This means proper randomization to avoid
this bias would have lead to higher uncertainty than 10\%, and hence proper
handling of this uncertainty would be necessary (such a randomization tool
was later provided by~\citeN{stabilizer}).  A senior member of a well-known
industrial research lab told us that he would care only about differences of 10\%
for new work, but look for differences of as little as 2-3\% or even 1\% when
looking for regressions (where he would look at uncertainty too).  Since
such small performance changes are often of interest, ignoring uncertainty
in systems research represents a serious threat to validity.  When speed-ups are
large compared to expected uncertainty, validity would not be threatened,
but ignoring it would still be a serious lack of rigour and in our view not
a property of good research.

Out of the 19 papers that touched on uncertainty in any way, most reported
uncertainty of single systems in the form of a standard deviation or of 95\%
confidence intervals.  It was exceptional for a paper to specify how the
confidence intervals were constructed; most were presumably based on the
normal distribution.  While most papers reported the ratio of execution
times, they mostly showed uncertainty only for the means in isolation of
each of the two systems compared.  Only three of the papers reported the
uncertainty of the ratios.  However, it seems from the text of two of these
that they did not take uncertainty of both systems into account, but only
normalized the uncertainty of the new system against the mean of the old
system.  The third paper was ours, and it took the uncertainty of both systems
into account, but we have to admit that our text did not say that.
The remaining papers which address uncertainty do so by saying that it was
low (without giving numbers), and one says that a proposed optimisation
reduced it (without giving numbers).

The papers we studied gave extremely little information on the experimental
design used.  From some papers, it is not even obvious if benchmarks were
executed more than once, and hence if the reported execution time is not
just a single measurement.  Only very few papers mention that their
experiments repeat at two levels (i.e.\ execution of a benchmark and measurement
within that benchmark).  However, all of those that do also seem to treat
the data as if it came from a one-level experiment.  If executions were
repeated, the numbers of repetitions seemed to be arbitrary.  Some
experiments used an adaptive number of iterations based on a heuristic 
--- the benchmark code itself calculates the standard deviation of the last few
measurements taken and terminates the experiment once the standard deviation
is sufficiently small.

Some papers claim to use certain  evaluation methodologies, such as those of
\citeN{envsize}, \citeN{georges07} or \citeN{georges08}.  However, these
papers rarely made it clear how the concrete experiments were performed. 
One paper claims to use the method proposed by~\citeANP{envsize} to reduce
variance, without any further details.  However, the cited paper argues for
dealing with measurement bias through randomization, which in fact
increases the variance, but for very good reasons.  Some papers mention
uncertainty due to the decisions of a just-in-time compiler of which methods
to compile.
Confidence intervals or standard deviation error bars are usually shown for
a single system with one particular configuration, even when multiple
configurations were used.  Uncertainty quantified in this way is prone to
bias.

\paragraph{Summary}
\noindent
The ratio of (mean) execution times is the major metric for quantification
of performance change.  A threshold for the minimum/maximum ratio is used as
a measure of the practical importance of a difference.  Uncertainty in the
ratio of execution times is almost never quantified, and when it is, only in
an unclear way.  Sometimes separate uncertainties for the two systems
(means) are shown, but more commonly even these are ignored.  Benchmarks are
mostly repeated only at one level, or exceptionally at two levels
(executions and measurements).  Two-level experiments, if done at all, are
evaluated as if they were one-level only (we believe by merging all the
measurements from the different executions, but the texts are not explicit
here).

\subsection{What Is Wrong With Currently Used Methods}

Any paper that reports execution time, yet ignores the uncertainty of the
final metric, be it absolute execution time of one system or the ratio of
execution times for two systems, is methodologically flawed.  Moreover,
failing to address uncertainty makes the results unrepeatable.  Repeating
any experiment will lead to slightly different results due to variability in
measurement.  Without quantified uncertainty, we cannot know if we got a
`slightly' different result due to variability in measurement, or a very
different result due to an error.  Note that a failure to address
uncertainty is not the only way to get unrepeatable results.  A common
problem is measurement bias, when the experimenter fails to identify and
then describe or vary a factor that largely influences the
results~\cite{envsize}.  In such a case, statistics even properly used for
factors that were varied cannot help and the summarized uncertainty would be
misleadingnly small.  The responsibility for identifying these factors lies
fully on the experimenter and no statistics can help. But even when
important factors are fixed and described or varied properly, failing to address
uncertainty is a potential threat to validity when quantifying performance
change, and this threat is particularly high when the observed change is
relatively small (e.g. 10\%), which it often is.  When such quantification is used to show that an optimisation
provides a good speed-up, there is a risk that the `improvement' lies within
the range of uncertainty.  A similar risk applies to some extent when
quantification is used to show that the overhead of a certain feature is
small --- if the uncertainty of the new system is much larger than the old
one, the price may in practice be high even if the mean overhead is small.

%Note that particularly ISMM and also ASPLOS papers out of our set were
%better at addressing uncertainties than other venues.
%REJ: that's why I say that ISMM is such a good conference :-)
%TK: point taken; absurdly, pldi is particularly bad in statistics

Failure to report how many repetitions of a benchmark were executed makes
repeatability harder.  Often, the reader is only left with suspicion that
insufficient repetitions were made and in particular that the researchers
did not repeat whole executions of benchmarks (that in turn repeat
measurements within a single execution).  Sometimes, the reported number of
measurements used for averaging is suspiciously small (e.g.\ three or five). 
Failure to repeat enough times, or indeed to repeat at all, increases the
risk of obtaining unrepeatable results and reaching incorrect conclusions.

In summary, our survey of published papers suggests that the current
practice of quantification of performance change in scientific papers is
weak.  We are not the first to find this, but we were surprised at how
prevalent this practice is even in the most recent papers.  The lack of
reporting of uncertainties while quantifying a performance change in Java
runtimes research was documented by~\citeN{georges07}.  \citeN{envsize}
discovered that measurement bias was not addressed in experimental computer systems
performance evaluation; this is a more general problem, but it also applies
to the quantification of performance change.  These shortcomings have arisen
despite well-known textbooks on computer systems performance
evaluation~\cite{jain,lilja} having described a number of errors common in
performance evaluation.

%\begin{figure} \centering
%\includegraphics[angle=-90,width=0.45\linewidth,trim=0cm 0.5cm 1cm
%2cm,clip=true]{figures/fftrs.pdf}
%
%\caption{FFT}
%
%\label{fFFTrs} \end{figure}

\subsection{Currently Proposed Methods}

Experimental design and statistical inference are mature fields of
mathematical statistics with many results and described in a number of
texts.  Basic recommendations on conducting experiments in physics and
natural sciences, measuring uncertainties, calculating propagation of
uncertainties, and reporting the results are available in textbooks such
as~\cite{kirkup}.  More advanced rules of propagation of uncertainties can
be found in a recommendation by NIST~\cite{nistunc}.  High-level overview
and pointers to works on experiment design and the scientific method in
general can be found in~\cite{wilson}.  While physics does not deal
with exactly the same problems as experimental computer science, many ideas are
general.  Moreover, basic methods of statistical inference and experimental
design, in the context of computer systems performance evaluation, have then
been presented in books~\cite{jain,lilja}.  These books have been and are
being used as textbooks for teaching performance evaluation courses. 
\citeN{georges07} argued for wider adoption of a performance change
quantification method based on Lilja's book, in the context of performance
evaluation of Java runtimes.  They also added some practical recommendations
for the number of repetitions of benchmark measurements.

In this section, we summarize the current recommended best practice for
quantification of performance change, based on~\citeN{jain}, \citeN{lilja},
and~\citeN{georges07}.

\subsubsection{Experiment Design}

\citeN{georges07} recommends repeating measurements within a single
benchmark, as well as repeating executions of the whole benchmark.  The
number of measurements within a benchmark is controlled by a heuristic in
the benchmark --- as soon as the standard deviation of the last few
measurements is small enough, the benchmark's execution is finished.  The
number of benchmark executions is an arbitrary choice of the experimenter,
or, alternatively, additional executions are adaptively added until
confidence interval for the mean is sufficiently narrow (its width is within
1\% or 2\% of the sample mean).

\citeN{lilja} and particularly~\citeN{jain} also recommend experimental
designs that cope with various combinations of factors that can be fixed by
the experimenter.  These designs apply to factors that can take only a
finite, usually small, number of values (i.e.  a particular processor or benchmark).  The
effect of choosing a particular value (say the processor) on performance is
of interest to the experimenter.  Such effects are called \emph{fixed
effects} in statistics (see~\cite{mcculloch} and~\cite{varcomp} for more detailed discussion of
fixed and random effects).  The experimental designs then help in planning
which combinations of values of which factors to run to get reasonable
information given reasonable experimentation time.  The methods recommended
by~\citeN{jain} and ~\citeN{lilja}, however, do not address effects of
factors that can take a large number of values, usually at random and often
not under our control, for which the performance effect of
a particular choice is not interesting.  Such a factor can be, say, a particular mapping of
virtual to physical pages chosen by the operating system.  In statistics,
these are called \emph{random effects}.  In this work, we only address
random effects.  Our work could be extended to include fixed effects,
as there indeed are statistical models that include both types of
effects~\cite{mcculloch}, but we do not attempt it here.

\citeN{georges08} proposes a particular experiment design for performance
evaluations within managed environments with just-in-time compilation.  The
design assumes generation of several compilation plans and then running both
systems using these plans.  The motivation for this design, in contrast to
just keeping the compilation plans random (unpaired) for the two systems, is
smaller uncertainty within given experimentation time.

\subsubsection{Summary of Experiment Results}

A performance change is quantified using statistical
significance~\cite{georges07,lilja,jain}.  The output is a binary decision:
``the systems (may) have the same performance'' or ``(it is likely that) the
systems do not have the same performance''.  This decision is based on the
probability that the actually observed difference or larger in (sample) means of the
two systems would occur if the (true) means were the same.  This probability
is called the \emph{p-value} in statistical tests.  The p-value is compared
against a pre-defined threshold called \emph{significance level} (i.e. 
5\%).  If the p-value is smaller, then the second decision, that the systems
do not perform the same, is chosen.  Otherwise, the first decision, the
default that they may have the same performance, is chosen.

There are two recommended alternative methods that provide this kind of
quantification.  The preferred one is a visual test using confidence
intervals.  The other one is a statistical test.  
The visual test is as follows. We construct two confidence intervals for the
means of the two systems.  We check if they overlap.  If they do not, the
decision is that ``it is likely that the systems do not have the same
performance''.  If they do overlap, the method recommended by~\citeN{lilja}
and~\citeN{georges07} finishes, concluding that ``the systems may have the
same performance''.  The method by~\citeN{jain} adds another step --- it
falls back to a statistical test if the intervals overlap only slightly,
that is if centre of neither interval lies within the other.
The visual test is preferred, because in addition to the binary decision, it
is easy to give a visual measure of how large the difference in the means
actually is, compared to their uncertainty.  While this is a clear aid for
an analyst, it lacks a rigorous semantics (and in his dissertation,
~\citeN{georgesphd} recommends using the statistical test instead).

The confidence interval used for the visual test has the same underlying
statistical backing as the statistical test.  It is assumed that
measurements are independent identically distributed and follow the normal
distribution (the normality assumption is incorrectly omitted in many texts as we
discuss in greater detail in Section~\ref{sBadSummarization}). \citeN{jain} and~\citeN{lilja} suggest using Student's t-test
for unequal variances (or the corresponding interval).  \citeN{georges07}
additionally recommend (an interval based on) the z-test for sample sizes
over~30.  \citeANP{georges07} also suggest working with sample means of
measurements from individual executions, instead of with the measurements
themselves.  The corresponding test/interval is hence only for these sample
means.  An alternative solution that is sometimes used, but not part of
their recommendation, is to join all the measurements from the different
executions and treat them as coming from a one-level experiment.

\citeN{lilja} and particularly~\citeN{jain} then describe analysis of
variance (ANOVA) methods to summarize experiments where multiple fixed
factors are varied (i.e. processor, memory size, etc.). \citeN{georges07}
summarizes also some of the ANOVA methods from~\cite{lilja}, but does not
recommend it for use in (Java) performance evaluations as the outputs are too
hard to understand and doing all the measurements to provide the inputs is
too time consuming. We do not address these types of factors (`fixed
effects') in this work, but our method should be extensible to do so.

The summarisation method that~\citeN{georges08} recommends for the replay
compilation design leads to a test/interval based on paired t-test.  The
inputs are differences of means obtained from the old and new systems,
averaging over multiple executions with a given compilation plan.  We do not
address paired comparisons in this work, but most of our criticism of the
recommended quantification using the test/interval we mention later applies
to it as well.

\section{What Is Wrong With The Currently Proposed Methods}

Even though the current recommended best practice for quantification of
performance change by~\cite{jain}, ~\cite{lilja}, and~\cite{georges07} is
ahead of the pitiful practice in the field, it has a number of problems.

\subsection{Experiment Design}

Repetition of executions (and measurements) is not enough. It is necessary
to include all inherently random factors in the experiment, and hence to
repeat experiments at level(s) even higher than execution of the benchmark. 
For example, randomized compilation necessitates both compilation and
running the experiments for different binaries.  We have found this to be
essential in our earlier work with the GNU C++ Compiler when benchmarking
CORBA middleware and when benchmarking applications within the Mono platform
~\cite{epew06}.  The empirical evaluation in this paper confirms this need
on the Mono platform.
Furthermore, we need to randomize and include additional (non-random)
factors that it does not make sense to fix.  These factors can arise from
the operating system or the language environment.  \citeN{envsize} provide
evidence that unexpected factors such as the size of UNIX process
environment or the linking order can impact performance.  \citeN{verbrugge}
observed a performance impact of the names of identifiers in the source
code. It is obvious that many additional factors of this kind have not yet
been discovered to have significant performance impact, so they cannot be
really fixed in experiments. Sufficient randomization, called for
by~\citeN{envsize}, is hence necessary. \citeN{stabilizer} provide a
randomization tool applicable to applications studied in~\cite{envsize}.

Note in particular that randomization of identifier names or linking order
leads in turn again to randomized compilation/building.  Repeating
compilation or repeating at an even higher level can be very expensive,
because large systems take long to compile.  Repeating too many times is a
waste of resources, but repeating too few times undermines repeatability and
validity.  Hence, a good experiment design method should also allow the
researcher to derive an optimum number of repetitions at different levels,
so as to make efficient use of the time available for experimentation in
order to reduce uncertainty in the final result.  The present recommended
best practice does not offer this.

\subsection{Summarization of Experiment Results}
\label{sBadSummarization}

The single most significant problem of the best recommended method is that
\emph{it does not tell us what we want to know}.  It does not give us the
metric we are ultimately interested in, a reliable estimate of the ratio of
execution times.  Of course, we can use the recommended method and, in the
case that the (binary) conclusion is that the systems are likely to differ
in performance, we can in addition report the ratio of the sample means. 
But the recommended method would not give us the uncertainty of this ratio,
so we do not know how much of the ratio is due to uncertainty.  Since we do
not report uncertainty of the ratio, the results we provide are unrepeatable
and may not be valid.  This problem also means that the recommended method
does not allow us to compare the ratio against a threshold (the
minimal/maximal change that is important to us).

The problem of not supporting comparison against a threshold is more
significant than it may seem at first --- the recommended method looks for a
performance difference however small as long as it is unlikely to be by pure
chance.  However, the larger the sample size is (the more measurements we
have), the more unlikely even a very small difference becomes.  Hence, the
decision of the method is influenced by the number of measurements.  In
practice this means that with a large sample size (and in our field it is
easy to generate very large samples), the decision will nearly always be
``it is likely that the systems do not have the same performance'', no
matter how small or large the difference actually is.  The method then
becomes of very little use --- it just adds an illusion of rigour to the
results and, worse, only to the results we are not interested in.

The fact that this does not tell us what we want, and that it often gives
the same answer only because the sample size is very large, is a fundamental drawback
to the statistical significance.  This drawback has been known for the last
seven decades~\cite{cohen}.  It has been reiterated by researchers in
psychology~\cite{cohen}, education~\cite{coe}, and more recently in
bio-medicine and biology~\cite{nagawa}.  This and other drawbacks of the
method have been brought up in criticism of statistical significance also in
sociology, criminology, economics, marketing, chemistry, and nursing
(see~\citeN{fidlerps} for references and details).  The drawbacks, however,
seem to be quite unknown to experimental computer scientists.
Even worse, methods based on statistical significance are notoriously hard
to interpret.  This may be partially because they do not give us the answer
to what we want to know, but they also offer temptations, such as the belief
that the p-value is actually the probability that the systems have the same
performance.  The interpretation of the results of statistical tests is so
difficult that even statistical textbooks sometimes get it wrong (examples
are given by~\citeN{cohen}).  In a study by~\citeN{oakes} cited
by~\citeN{cohen}, 68 of 70 psychologists made an error in the interpretation
of a statistical test.  Interpretation of the visual test using confidence
intervals seems simpler, but it is not much so.  This is well reflected by
the wording used and stressed by~\citeN{lilja}, and repeated
by~\citeN{georges07} and~\citeN{georges08}, to describe the positive outcome
(the systems differ in performance) of the visual test
--- ``there is no evidence to suggest that there is not a statistically
significant difference''.

The wording is rather cryptic as it tries to be precise, which we do not
find it to be.  In frequentist statistics which the texts rely on, the
wording is incorrect.  The goal of the comparison is to learn something
about the difference in the \emph{true} but unknown means of two random
variables.  The true means are either equal or different, and the
statistical significance does \emph{not} speak about this difference (it
speaks about the difference of the \emph{sample} means instead).  Hence, the
second part of the wording, ``statistically significant difference'' is
incorrect.  The first part of the wording that speaks about non-existence of
evidence is incorrect as well.  The test is about verifying if the data we
have measured are unlikely provided the true unknown means are the same. 
Hence, we are checking if the data we have form an evidence \emph{against}
the assumed zero performance change.  We can make no claims about
non-existence of evidence. And we are not looking for evidence \emph{for}
the true means being equal, but rather \emph{against} it.  

\citeN{lilja} (also cited by~\citeN{georges07}) explain the wording by
stating that there is always a certain probability, `$\alpha$', that a large
observed difference was due to random fluctuations.  This is correct, but
does not justify the wording in our opinion.  Even more, the context in the
texts suggests that in the visual test of confidence interval overlap, this
probability ($\alpha$) is the significance of the confidence intervals used
(i.e.  $\alpha=0.05$ for 95\% confidence interval), which it is not. In
fact, with the visual test this probability is not known. If we use 95\%
confidence intervals, the probability of such an error is not 5\%, but under
the normality assumption is below 1\%~\cite{overlapp}.  This makes the
visual test far more conservative than it may seem, and hence its results
are even harder to interpret, and more likely to mislead.  Also, one would
need far more experimentation time to show a performance change.  Further
information on the error of the visual test can be found
in~\citeN{overlapp,schenker}.

With the (non-visual) statistical test, the theory tells us the probability
of erroneously concluding that compared systems do not have the same
performance when actually they do.  The probability of this error is the
threshold we are comparing the p-value against (often 5\%).  For this
reason,~\citeN{schenker} (cited by~\citeN{georgesphd}) prefer statistical test
over the visual test.  Our understanding is that the visual test is
generally preferred, though, in our field, and that both method share their
key problems.

Yet another issue of visual and non-visual test is the use of parametric
methods on data that violate their assumptions.  Computer performance
measurements cannot be assumed to be normally distributed.  Often they are
multi-modal, with long-tails to the right.  Deviations from normality may
not be fatal for the t-test/confidence interval though, and some
practitioners in other fields ignore them as well.  The sample mean is
asymptotically normal due to the Central Limit Theorem.  Many texts in our
field omit the normality assumption of the t-test or incorrectly state that
the Central Limit Theorem is enough to overcome it
(\cite{lilja,jain,georges07}).  The Central Limit Theorem is \emph{not}
enough, but some more involved studies show that, under certain deviations
from normality, the parametric methods work well, even for reasonably small
sample sizes (see for example~\citeN{cfirobust} or \citeN{robustsim}) given
a reasonably high confidence (95\% confidence intervals and wider).  Still,
there is no general agreement that ignoring the normality assumption is an
acceptable practice, one should certainly make it clear if such assumptions
are ignored.  Robust statistical methods do exist (see~\citeN{robustas} for
summary and references), and there has been no study of how the
t-test/confidence interval is affected by violations from normality common
in computer performance data.

\section{How to do things better}

A good quantification method needs to provide an estimate of the ratio of
execution time means and an estimate of its uncertainty.  These estimates
should be based on experiments that include all random factors to which the
measured system is subject, both factors naturally random and factors that
we randomize to avoid measurement bias.  A statistical model of such an experiment can
then offer some way of planning the experiment: how many repetitions are
needed at each level.

\subsection{Experiment Design with Random Effects}

There are factors that influence performance, which are inherently random
and we cannot control them.  To get valid results (avoid bias), we need to
repeat experiments at a high enough level to include all random factors.  For example, if
our system is prone to randomized compilation which has an effect on
performance, we need to repeat compilations and measure multiple binaries. 
With managed runtimes, the `binary' may be the binary of the virtual machine
or the byte-code of the application.  If we took an extreme position, it
would be all compiled code that influences performance observed by the
benchmark.  In theory, we could start from the top level for each
measurement, thus compiling the binary, executing it only once and running
only one iteration (say after dropping the initial measurements that were
prone to warm-up noise).  This approach would allow use of a one-level
model.  However, it would be an extravagant waste of time if the variation
in binaries had a far smaller effect on performance than the variation in,
say, executions or measurements.  In this case, intuitively it should be
possible to do better by executing each binary multiple times and reporting
more than one measurement in an execution.

This situation can be modeled mathematically by a random effect models with
$n$-way classification.  Such random effects model is based on a hierarchy
of $n$ random ways,\footnote{In statistical texts, one can say also `$n$
factors', but we refrain from doing so here to avoid confusion with the
informal meaning of a `factor' used throughout the text. In our text,
a `factor' refers to a technical/real cause that may impact performance and
a `way' refers to a statistical model of (some of) such causes.} which have effect
on the random distribution of the actual measurements.  Hence, a model with
$n$-way classification corresponds to an $n+1$ level experiment.  In
particular, a three-level experiment with repeated compilations, executions,
and measurements corresponds to a model in 2-way classification
(measurements influenced by execution and compilation).  Random effects
models are sometimes also referred to as random effects ($n$-factor) ANOVA. 
References to specialized literature on related models can be found
in~\citeN{mcculloch}, though we have not found a model that would apply
directly.  Derivation of such a model is one of the contributions of this
work.

\subsection{Quantification with Effect Size Confidence Interval}

We propose to construct a confidence interval for the ratio of mean
execution times.  This confidence interval will be a measure of uncertainty
of the metric that is of ultimate interest.  Reported results can thus be
for instance that system A is 4\%$\pm$1.5\% faster than system B, with 95\%
confidence.  Multiple intervals would also lend themselves easily to
graphical visualisation.
The method gives us a way to rigorously compare against a threshold. Say
that we only care about differences larger than 3\%.  Here, we would
conclude that performance of two systems is different if the upper bound of the
confidence interval is less than 0.97 or its lower bound greater than 1.03,

Note also that the method still allows a significance-based binary decision. 
We conclude that the systems ``may have the same performance'' if the
interval includes 1.  Otherwise, we would conclude that the systems are
``likely not to have the same performance''.  The chance of erroneously
concluding that the systems differ in performance here is the confidence of
the interval (5\% for a 95\% confidence interval).  Hence, even for the
significance based method only, we have the advantage over the recommended
practice that we know this error.  In summary, the confidence interval for
the ratio of means subsumes the currently best recommended practice.

Constructing a confidence interval for a metric that measures the difference
in two systems is a known concept in statistics.  Such a metric is called
the \emph{effect size} and hence the \emph{effect size confidence interval}. 
Effect size confidence intervals have been proposed for quantification as a
replacement for significance methods in psychology~\cite{cohen},
education~\cite{coe}, medicine, bio-medicine, and biology~\cite{nagawa}. 
The change in statistical methods is not smooth, but evidence can be found
that effect size is already being used, at least to some extent, in
medicine~\cite{oddsratio} and psychology~\cite{espsych}.  To support the
change, the use of effect size has sometimes been made an official
requirement (\citeN{eshill} list 23 journals mostly in psychology and
education that require reporting effect size, and the \citeN{apa} requires
it in its Publication Manual).

There are different metrics used for measure the effect size in different
fields and one of them is the ratio of means.  There are statistical methods
for interval construction, although they did not make it to introductory
textbooks (more information and references can be found
in~\cite{ratiocficmp,reportfieller,mratios}).  From several available
methods, we have a closer look at two.  One is based on statistical
simulation (bootstrap)~\cite{davison}.  The other is based on Fieller's
theorem~\cite{fieller}, which gives a confidence interval for the ratio of
means of normally distributed variables.  We show how both methods can be
applied in the experimental design we propose, and empirically evaluate
them.

\subsection{Related Efforts}

Some of the things we propose here have been proposed earlier in computer
science, but have not been widely adopted.

A bootstrap-based method for construction of confidence intervals is being
used in the Haskell community, supported by the Criterion benchmarking
library~\cite{criterion}.  The library is for measuring performance of one
system only in single-level experiments.  Apart from using robust methods it
has also the advantage of detecting outliers and auto-corellation of the
data.  The tool is based on~\cite{robustj}.

Confidence interval for speed-up has been proposed by~\citeN{luo04} in the
context of processor simulation.  The uncertainty does not come from random
effects, but rather from random sub-sampling -- only randomly chosen
execution intervals from the whole application execution are simulated, the
same intervals in both of the systems to compare.  The confidence interval
proposed for this problem comes from~\cite{cochran}, it is a parametric
interval based on asymptotic normality of not only the means, but also their
ratio, which is only possible for large samples.  The method
by~\citeN{fieller} that we use in this work is more general.

\section{Outline of the New Method}

The rest of our paper includes a detailed description of our quantification
method and its empirical evaluation based on real benchmark results.
We offer two alternative descriptions of our method. Section~\ref{sDirty}
provides a guide for practitioners, while Section~\ref{sStatistics} is a
variant for scientists with some statistical background. 
Section~\ref{sDirty} describes the method in practical terms and with
recommendations on how it can be used.  Section~\ref{sStatistics} formulates
the method in statistical terms, gives the assumptions and necessary proofs,
and discusses further alternatives.  We use the terminology common in the
field.

In statistical terms, the core of the method is our statistical model of
random effects in $n$-way classification, which models performance in one
system.  We apply it independently to both systems we aim to compare.  For
clarity, the model is described for 2-way classification first (3 levels in
the experiment, Section~\ref{sTwoWay}), but later in the general form of
$n$-way classification (an arbitrary number of levels, Section~\ref{sNWay}). 
While we keep referring to `execution time' in the statistics to make
presentation concrete, the model is sufficiently general for any response
variable, given the stated assumptions.
We then show how to construct the confidence interval for the mean within
this model, still for the mean of each system separately
(Section~\ref{sOneSystem}).  We give two alternative methods, a parametric
method based on asymptotic normality and a non-parametric method based on
bootstrap.  Later we show how to construct the confidence interval for the
ratio of the means of two systems, again using a parametric and
non-parametric method (Section~\ref{sTwoSystems}).  The non-parametric
bootstrap method is a natural extension of the method for a single system. 
The parametric method uses Fieller's theorem~\cite{fieller}.

\section{For Practitioners: The New Method without Statistics}
\label{sDirty}

The technique we propose here comprises of the actual statistical method
(proven and precisely defined in the following section) and of practical
advice how to use it.  We feel that it is important to stress that while the
statistical method is general, our advice may not be.  The practicalities
depend on the context (the goal of the study, the criticality of the results
and the consequences of possible errors, the kind of computer system
evaluated, the runtime environment, infrastructure, and benchmarks).  We
understand that having a single best technique that could be mechanically
followed in every study might be appealing, but we strongly believe that
such practice is an illusion, and we definitely do not provide one here.

\subsection{Designing the Experiments}
\label{sDesignPractitioner}

We describe our technique sequentially, but in practice it would be an
iterative process.  
At the end of this section, we provide a small worked example.

\paragraph{Initial number of levels} 
When planning experiments, we first decide what the highest level of the
experiment should be.  All factors influencing performance above that level
need to be controlled (and documented).  The highest level might be
compilation (of the benchmark, runtime environment, or all).  If compilation
is found to be fully deterministic, this may not be needed.  Though note
that even symbol names have impact on performance~\cite{verbrugge}, so we
may choose to randomize these or otherwise randomize the building process to
get rid of other sources of measurement bias.

On the other hand, in practice, if repeating say compilation is very
expensive, and we know that it has relatively little impact on performance
or we find it out during the following process, we may choose not to repeat
it.  The threshold we then use for comparisons, however, needs to be set
accordingly.

After deciding on the highest level, we continue by identifying lower
levels.  We need to add an additional lower level to the experiment if the
variability (say in executions) is expected or known to be higher than in
the level above (say compilations), while the cost of repeating at the lower
level is smaller than at the higher level.  If we create a level where it is
in fact not needed, we may detect this later in the following process. 
Also, we can measure the costs of repetition in the following process as a
side-effect (for example, such cost is the time to build a new binary).  In
the formulae below, the number of levels is $n+1$ (where $n$ is the number
of `ways' in the statistical model, which is one less than the number of
levels in the experiment).

\paragraph{Numbers of repetitions} Having decided on initial levels of the
experiment, we run the experiment a few times to estimate variability at
different levels.  With our method, we always use the same repetition counts
at a particular level (thus all executions have the same number of
measurements, all binaries same number of executions, etc.).  We denote the
repetition counts as $n_i$, where $1 \le i \le n+1$ is the level of the
experiment ($n_{n+1}$ is the number at highest level, say compilation).

For the initial experiment, we can set all these counts to the same value,
say 30.  Note that these counts do not include warming-up various components
of the experiment.  Usually, we want to evaluate only measurements from a
steady-state of every execution.  Say that we do so by discarding the first
$c_1$ measurements of each execution before collecting the $n_1$ we need. 
$c_1$ must be found by the experimenter.  Repetitions at higher levels,
$c_i$, $1 < i \le n$, also may have a non-zero cost.  For instance,
compilation has notable cost as it takes time to compile.  All costs are
represented as counts of measurements --- the cost of compilation is the
compilation time divided by the (average) time for a measurement.  Precision
here is not crucial, as the costs are only needed for optimisation.

To run the initial experiment, we need to have $c_1$ already and we use
$n_i=30$.  During the experiment we collect measurements and also measure
the costs for repetition $c_i$ for $1 < i \le n$.  Based on the costs and
the measurements, we decide on optimum counts of repetitions at all but the
highest level (the derivation is provided in Sections~\ref{sPlanning}
and~\ref{sPlanningDerivation}) :
$$
n_1 = \left\lceil\,\sqrt{c_1 \frac{T_{1}^2}{T_{2}^2}}\,\right\rceil, \qquad
  \forall i, 1 < i \le n \,\, n_i = \left\lceil\,\sqrt{ \frac{c_i}{c_{i-1}} \frac{T_{i}^2}{T_{i+1}^2}}\,\right\rceil.
$$

We obtain $T_i$, the unbiased estimator of the variance at level $i$, through an
iterative process. First, we calculate $S_i$, the biased estimator of the variance at level $i$
(formulae below). Then, we start calculating $T_i$ as (the derivation is
provided in Section~\ref{sVarEst})
\begin{eqnarray*}
T_1^2 &=& S_1^2, \\
%T_2^2 &=& S_2^2 - \frac{T_1^2}{n_1}, \\
\forall i, 1 < i \le n+1 \,\, T_i^2 &=& S_i^2 - \frac{S_{i-1}^2}{n_{i-1}}.
\end{eqnarray*}
If we should get $T_i^2 \le 0$ (or at least very small), then this level of the experiment induces  
little variation so we can remove level $i$ from the
experiment. This is semantically equivalent of running the experiment all
again with fewer levels, but we obtain the same effect by dropping data from
the repetitions. We calculate $S_i$s as follows (the used notation is
detailed in Sections~\ref{sNWay} and~\ref{sModelProperties})
$$
S_1^2 = \frac{1}{\prod_{k=2}^{n+1} n_k} \, \frac{1}{n_1-1} \sum_{j_{n+1}=1}^{n_{n+1}} \dots \sum_{j_{1}=1}^{n_1} 
\left( Y_{j_{n+1} \dots j_1} - \overline{Y}_{j_{n+1} \dots j_2 \bullet} \right)^2
$$
for $i$, $2 \le i \le n$
$$
S_i^2 = \frac{1}{\prod_{k=i+1}^{n+1} n_k} \, \frac{1}{n_i-1} \sum_{j_{n+1}=1}^{n_{n+1}} \dots \sum_{j_{i}=1}^{n_i} 
\left( \overline{Y}_{j_{n+1} \dots j_i \underbrace{\bullet \dots \bullet}_{i-1}} - \overline{Y}_{j_{n+1} \dots j_{i+1} \underbrace{\bullet \dots \bullet}_{i}} \right)^2
$$
and finally
$$
S_{n+1}^2 = \frac{1}{n_{n+1}-1} \sum_{j_{n+1}=1}^{n_{n+1}} 
\left( \overline{Y}_{j_{n+1} \underbrace{\bullet \dots \bullet}_{n}} - \overline{Y}_{\underbrace{\bullet \dots \bullet}_{n+1}} \right)^2
$$
$Y$ denotes the measurements, indexed by experiment levels (highest to
lowest).  The bar over $Y$ denotes an arithmetic mean.  The mean is always
calculated over all the indexes that are denoted by a bullet.  While the
formulas for $S_i$s may seem involved in full generality, it is easy to see
how they are constructed.  Suppose we have functions $M$ and $V$ that
calculate sample mean and variance of a vector,
$$
M(x_1 \dots x_k) = \frac{1}{k} \sum_{i=1}^{k} x_i \qquad 
V(x_1 \dots x_k) = \frac{1}{k-1} \sum_{i=1}^{k} \left(x_i - M(x_1 \dots x_k)\right)^2
$$
Any statistical software would have these functions.  For $n=2$, each of
$S^2_1$, $S^2_2$, and $S^2_3$ is created by three applications of these
functions to the data, where $V$ is applied once and $M$ two times, but
always in different order (see Table~\ref{tSinterp}).  Each application of
$M$ or $V$ reduces the dimension of the data.  For example, to calculate
$S_2^2$ we first apply $M$ on measurements from each execution, hence
getting a two-dimensional matrix (binaries $\times$ execution means) from a
three-dimensional matrix (binaries $\times$ executions $\times$
measurements).  Second, we apply $V$ on the execution means, getting a
vector (one element per binary) of variances of the execution means.  Third,
we apply $M$ on this vector and get $S_2^2$.

\begin{table}
\tbl{Interpretation of $S_i^2$ in Three-level Experiment\label{tSinterp}}{
\begin{tabular}{lcccl}
 & Binaries & Executions & Measurements & Interpretation \\
\hline
$S_1^2$  & $M$ & $M$ & $\mathbf{V}$ & Mean {\bfseries variance} at level 1 (due
to {\bfseries measurements}) \\[0.8mm]
$S_2^2$  & $M$ & $\mathbf{V}$ & $M$ & Mean {\bfseries variance} at level 2 (of
{\bfseries execution} means) \\[0.8mm]
$S_3^2$  & $\mathbf{V}$ & $M$ & $M$ & {\bfseries Variance} at level 3 (of
{\bfseries binary} means) \\[0.8mm]
\end{tabular}
}
\end{table}

If we obtain
a small number of repetitions at some level, we may decide to remove that
level to simplify the experiment.  Indeed for practical purposes we can also
increase the number of repetitions at the lowest level (i.e.\ measurements),
if a measurement is very short.  As there is no further overhead for
additional measurements at the lowest level (since we do not have to repeat
higher levels and nor perform further warm-up), it is cheap to get more data
useful for other things, such as spotting unusual behaviour.  This method
would not work well for very small repetition numbers (say, below 5, but
this depends on experimenter's judgment) --- if at all possible, we should
not use such small counts. Note that in practice, to make the process
simpler at the cost of more expensive experiments in the end, we could
assume $T_i=S_i$. In that case we would not get problems with the negative
numbers, but the resulting numbers of repetitions at higher levels may be
higher than needed.

Once we have decided on the final number of levels and repetition counts, we
re-run the experiments and continue as follows.  In the following, $n_i$ and
$n$ refer to the new setting.  In practice, the established setting would be
re-used for very similar experiments (same platform, benchmarks, etc).

\paragraph{Example}

Let us assume that we are to dimension an experiment based on three-level
data ($n=2$) shown in Table~\ref{tWEData} (Raw Data).  We have measurements
from 3 binaries, each executed 2 times, and reporting 2 measurements per
each execution.  In a real experiment we will have more, but this is to
demonstrate the method.  To calculate the $T^2_i$ we need the $S^2_i$, which we
will calculate iteratively as demonstrated in the table.
First, we calculate the (matrices of) execution means and variances ($\bullet\bullet M$
and $\bullet\bullet V$), 
%e.g.  the second execution of third binary provided
%measurements 2 and 4, which leads to mean $\frac{2+4}{2}=3$ and variance
%$\frac{1}{1} ((2-3)^2+(4-3)^2) = 2$,
as recorded in the second and third table.  Next, we calculate the column means and
variances of $\bullet\bullet M$, getting vectors $\bullet MM$ and $\bullet
VM$.  Also we calculate $\bullet M V$.  Finally, we get $S_3^2\doteq3.6$
($VMM$), $S_2^2\doteq2.6$ ($MVM$), and $S_1^2\doteq16.5$ ($MMV$). 
The grand mean, the mean of all measurements, is 6.5 ($MMM$).  We now obtain the
$T_i$:
$T_1^2=S_1^2\doteq16.5$, $T_2^2=S^2_2-S^2_1/2\doteq-5.7$, and
$T_3^2\doteq2.3$.  As we find that $T^2_2<0$, we can remove the second level of the
experiment.  Hence, we only will run one execution per binary.

\begin{table}
\tbl{Example Data (Three-level)\label{tWEData}}{

\begin{tabular}{c@{\hspace{1mm}}c@{\hspace{1mm}}c}

\begin{tabular}{|c|r@{ }r|r@{ }r|r@{ }r|}
\multicolumn{7}{c}{Raw Data} \\
\hline
$\bullet\bullet\bullet$ & \multicolumn{6}{c|}{Binaries} \\
%&& 1 & 2 & 3 \\
\hline
\multirow{2}{*}{Executions}  
 & 9 & 5  &  10 & 6  & 1 & 12\\
 & 8 & 3  &  7  & 11  & 2 & 4\\
\hline
\end{tabular}

&

\begin{tabular}{|c|r@{.}l|r@{.}l|r@{.}l|}
\multicolumn{7}{c}{Execution Means} \\
\hline
$\bullet\bullet M$ & \multicolumn{6}{c|}{Binaries} \\
%&& 1 & 2 & 3 \\
\hline
\multirow{2}{*}{Exec.}  
 & 7&0  &  8&0  & 6&5\\
 & 5&5  &  9&0  & 3&0\\
\hline
\end{tabular}

&

\begin{tabular}{|c|r@{.}l|r@{.}l|r@{.}l|}
\multicolumn{7}{c}{Execution Variances} \\
\hline
$\bullet\bullet V$ & \multicolumn{6}{c|}{Binaries} \\
%&& 1 & 2 & 3 \\
\hline
\multirow{2}{*}{Exec.}  
 & 8&0   &  8&0  & 60&5\\
 & 12&5  &  8&0  & 2&0\\
\hline
\end{tabular}

\\
&&\\
% ---------------------

\begin{tabular}{|c|c|c|c|}
\multicolumn{4}{c}{Binary Means} \\
\hline
$\bullet M M$ & \multicolumn{3}{c|}{Binaries} \\
%&& 1 & 2 & 3 \\
\hline
 & 6.3  &  8.5  & 4.8\\
\hline
\end{tabular}

&

\begin{tabular}{|c|c|c|c|}
\multicolumn{4}{c}{Variances of Exec. Means} \\
\hline
$\bullet V M$ & \multicolumn{3}{c|}{Binaries} \\
%&& 1 & 2 & 3 \\
\hline
 & 1.1  &  0.5  & 6.1\\
\hline
\end{tabular}

&

\begin{tabular}{|c|c|c|c|}
\multicolumn{4}{c}{Means of Exec. Variances} \\
\hline
$\bullet M V$ & \multicolumn{3}{c|}{Binaries} \\
%&& 1 & 2 & 3 \\
\hline
 & 10.3  &  8.0  & 31.3\\
\hline
\end{tabular}

\\
&&\\
% -------------

\begin{tabular}{|c|}
\multicolumn{1}{c}{Variance of Binary Means} \\
\hline
$V M M$ \qquad $S_3^2$\\[0.8mm]
\hline
3.6\\
\hline
\end{tabular}

&

\begin{tabular}{|c|}
\multicolumn{1}{c}{Mean Variance of Exec. Means} \\
\hline
$M V M$ \qquad $S_2^2$\\[0.8mm]
\hline
2.6\\
\hline
\end{tabular}

&

\begin{tabular}{|c|}
\multicolumn{1}{c}{Mean of Measur. Variances} \\
\hline
$M M V$ \qquad $S_1^2$\\[0.8mm]
\hline
16.5\\
\hline
\end{tabular}

\\
&&\\
% -------------

\multicolumn{3}{c}{
\begin{tabular}{|c|}
\multicolumn{1}{c}{Grand Mean} \\
\hline
$M M M$ \\
\hline
6.5\\
\hline
\end{tabular}
}

\end{tabular}
}
\end{table}

\begin{table}
\tbl{Example Data (Data from Table~\ref{tWEData} Transformed to Two-level Data)\label{tWE2Data}}{

%\begin{tabular}{c@{\hspace{1mm}}c@{\hspace{1mm}}c}

\begin{tabular}{|c|r|r|r|}
\multicolumn{4}{c}{Raw Data} \\
\hline
$\bullet\bullet$ & \multicolumn{3}{c|}{Binaries} \\
%&& 1 & 2 & 3 \\
\hline
\multirow{2}{*}{Measur.}  
 & 9 & 10 & 1 \\
 & 8  &  7 & 2 \\
 & 5  &  6 & 12 \\
 & 3  &  11 & 4 \\
\hline
\end{tabular}

\begin{tabular}{|ll|r@{.}l|r@{.}l|r@{.}l|}
\multicolumn{8}{c}{Sample Means and Variances} \\
\hline
$\bullet M$ & Binary Means & 6&3 & 8&5 & 4&8 \\
$\bullet V$ & Binary Variances & 7&6 & 5&7 & 24&9 \\[1mm]
\hline
$VM$, $S^2_2$ & Variances of Binary Means& 3&6 & \multicolumn{4}{c|}{} \\[0.8mm]
$MV$, $S^2_1$ & Mean of Measurement Variances & 12&7 & \multicolumn{4}{c|}{} \\[0.8mm]
$MM$        & Grand Mean & 6&5 & \multicolumn{4}{c|}{} \\
\hline
%\multicolumn{4}{c}{\hline} & \multicolumn{4}{c}{} \\
\end{tabular}

}
%\comment{Can we make the bottom line of the rhs table shorter?}
\end{table}

%Let us suppose that while measuring the raw data, we found that the time for
%compilation (creating one binary) is the same as that to take approximately 300
%measurements ($c_2=300$).  
Let us also suppose that we observed that the system
stabilises running a single execution reliably after 10 measurements ($c_1=10$, suppose that
Table~\ref{tWEData} already contains only the stable measurements).  We need
to find the optimum repetition counts.  As we decided to remove the second
level of the experiment, this optimum repetition count is just the number of measurements
per execution (the number of executions per binary will always be 1, and the
number of binaries we can increase any time to get more precise results). 
To find the optimum we need to recalculate the $T^2_i$ and hence the $S^2_i$ for two-level
experiment ($n=1$).  Applying the same method as before, we get
$S^2_2\doteq3.6$ and $S^2_1\doteq12.7$ (Table~\ref{tWE2Data}), hence
$T^2_1\doteq12.7$ and $T^2_2\doteq0.4$.  And thus the optimum number
of measurements per binary, $n_1$, is
$$
n_1=\left\lceil \sqrt{c_1\frac{T^2_1}{T^2_2}} \right\rceil \doteq
  \left\lceil \sqrt{10\frac{12.7}{0.4}} \right\rceil = 18
$$
We would thus run two-level experiment with $n_1=18$ measurements per
binary, executing each binary only once.

\subsection{Summarizing the Results}

\paragraph{Quantifying Performance of One System} 
We report performance of one system (in isolation) as the arithmetic mean of
all measurements $\overline{Y}$ (we omit the bullets in the notation for
simplicity).  We estimate its uncertainty using a $(1-\alpha)$ confidence
interval (i.e.  $\alpha=0.05$ gives a 95\% confidence interval):
$$ 
\overline{Y} \pm t_{1-\frac{\alpha}{2},\nu} \sqrt{\frac{S_{n+1}^2}{n_{n+1}}} =
\overline{Y} \pm
t_{1-\frac{\alpha}{2},\nu} \sqrt{ \frac{1}{n_{n+1}(n_{n+1}-1)} \sum_{j_{n+1}=1}^{n_{n+1}}
\left( \overline{Y}_{j_{n+1} \underbrace{\bullet \dots \bullet}_{n}} - \overline{Y}_{\underbrace{\bullet \dots \bullet}_{n+1}} \right)^2 }
$$
where $t_{1-\frac{\alpha}{2},\nu}$ is the $1-\frac{\alpha}{2}$-quantile
of the $t$-distribution with $\nu=n_{n+1}-1$ degrees of freedom. The
mathematical background is given in Section~\ref{sAsymIntOneSystem}.

Alternatively, we can use statistical simulation (bootstrap) to calculate
the confidence interval.  Say that we perform 1000 steps (or more if there
is time).  Within each step, we use the real data to simulate a new
experiment.  First, at the highest level, we randomly decide which
iterations of the real experiment to use (e.g.\ which binaries).  We
generate the same number of iterations as the real experiment, but some of
the real iterations can be used multiple times while some not be used at
all.  We then apply this principle to lower levels.  In the end, we get the
same number of measurements as in a real experiment, with the same structure
(repetition counts at each level and number of levels), and we calculate a
sample mean of all these measurements.  Hence, we get 1000 means, each
originating from one step of the simulation.

We form a $(1-\alpha)$ confidence interval for the mean using $\alpha/2$ and
$1-\alpha/2$ sample quantiles of these means.  If we have 1000 steps, we can
do this by ordering the simulated means and taking the 25th and 975th
values.  Pseudo-code that illustrates this procedure is shown in
Figure~\ref{fBootOneAlgo} on page~\pageref{fBootOneAlgo} and more details
are given in Section~\ref{sBootIntOneSystem}.

\paragraph{Quantifying Performance Change}
To quantify a performance change of a `new' system over an `old' system, we
use the same numbers of repetitions and levels for both.  We report the
ratio of mean execution times of the two systems.  As a measure of
uncertainty, we report a $(1-\alpha)$ confidence interval (again
$\alpha=0.05$ gives a 95\% confidence interval) as follows. For detailed notation
and derivation, see Sections~\ref{sTwoSystems} and~\ref{sFiellerRatio}.
$$
\frac{ \overline{^{O}Y} \cdot \overline{^{N}Y} \mp \sqrt{
\left(\overline{^{O}Y} \cdot \overline{^{N}Y} \right)^2 - 
  \left( \left(\overline{^{O}Y}\right)^2 - t_{\frac{\alpha}{2},\nu}^2 \cdot \frac{\oss}{n_{n+1}} \right)
  \left( \left(\overline{^{N}Y}\right)^2 - t_{\frac{\alpha}{2},\nu}^2 \cdot \frac{\nss}{n_{n+1}} \right)
}} {
 \overline{^{O}Y}^2  - 
  t_{\frac{\alpha}{2},\nu}^2 \cdot \oss \cdot {n_{n+1}}^{-1}
}
$$
The term $t_{\frac{\alpha}{2},\nu}$  denotes the $\frac{\alpha}{2}$-quantile
of the $t$-distribution with $\nu=n_{n+1}-1$ degrees of freedom.
%
%
%
%where $t_{\frac{\alpha}{2},\nu}$ is the $\frac{\alpha}{2}$-quantile of the
%$t$-distribution with $\nu$ degrees of freedom.
%
%$$
%\nu={\prod_{k=i}^{n+1} n_k} - 1.
%$$ 
%
The left-super-scripts `N' and `O' denote the new and the old system that we
compare.

Alternatively, we can use statistical simulation (bootstrap) to calculate
the confidence interval.  The algorithm is similar to the case of one
system.  We again perform a number of steps (say 1000), each producing the
metric of interest from a simulated experiment, which now is the ratio of
means.  Hence, in each step, we simulate measurements from both systems,
calculate their sample means, and then their ratio.  When we have these 1000
ratios, we take the sample quantiles for the confidence interval (i.e.  25th
and 975th for a 95\% interval).  Pseudo-code that illustrates this procedure
is shown in Figure~\ref{fBootRatio} on page~\pageref{fBootRatio}
(Section~\ref{sBootRatio}).

\paragraph{Example} 
Let us assume we have measurements from an old and a new system as shown in
Table~\ref{tWQEData}.  In practice, we would have more repetitions, but let
us use a simple example to demonstrate the method.  Let us first show how to
calculate confidence interval for the mean of one system.  We will
demonstrate this on the old system.  We need to calculate the variance at
highest level, $\ossthree$, but not the other variances, so there is less
work to do than when dimensioning the experiment.  Variances at the highest
level are simply variances of means (of binaries, in our case).  For the old
system, the mean for the first binary is 7.8, for the second 12.3 and for
the third 11.5 (all numbers rounded to one decimal place).  The grand mean
is hence 10.5 and variance of the means, $\ossthree$, is 5.8.  The 95\%
confidence interval for the mean of the old system is hence

$$
\overline{Y} \pm t_{1-\frac{\alpha}{2},\nu} \sqrt{\frac{S_{n+1}^2}{n_{n+1}}} \doteq 10.5 \pm 4.3 \sqrt{ \frac{5.8}{3} }
\doteq 10.5 \pm 6.0
$$
where 4.3 is a (rounded value of) the $0.975$ quantile of the
$t$-distribution with $2$ degrees of freedom. The confidence interval for
the mean of the new system would be obtained in the same way.

\begin{table}
\tbl{Example Raw Data (Three-level)\label{tWQEData}}{

\begin{tabular}{c@{\hspace{1mm}}c@{\hspace{1mm}}c}

\begin{tabular}{|c|r@{ }r|r@{ }r|r@{ }r|}
\multicolumn{7}{c}{Raw Data - Old System} \\
\hline
$\bullet\bullet\bullet$ & \multicolumn{6}{c|}{Binaries} \\
%&& 1 & 2 & 3 \\
\hline
\multirow{2}{*}{Executions}  
 & 9 & 11  &  16 & 13  & 15 & 7\\
 & 5 & 6  &  12  & 8  & 10 & 14\\
\hline
\end{tabular}

&

\begin{tabular}{|c|r@{ }r|r@{ }r|r@{ }r|}
\multicolumn{7}{c}{Raw Data - New System} \\
\hline
$\bullet\bullet\bullet$ & \multicolumn{6}{c|}{Binaries} \\
%&& 1 & 2 & 3 \\
\hline
\multirow{2}{*}{Executions}  
 & 10 & 12  &  9 & 1  & 8 & 5\\
 & 6 & 7  &  11  & 4  & 3 & 2\\
\hline
\end{tabular}

\end{tabular}
}

\end{table}

Alternatively, we can use the bootstrap method to calculate the confidence
interval.  Table~\ref{tWQEBoot} shows three bootstrap replicates of our
measurements.  In practice, we would need much more.  The first replicate
uses binaries 1, 2, and 1 from the original old system
(Table~\ref{tWQEData}).  For example, the third selected binary, which
corresponds to binary 1 of the old system of Table~\ref{tWQEData}, takes
executions 1 and 1 from the original binary.  For the first execution, it
takes measurements 1 and 1 from the original execution.  For the second
execution, it takes measurements 2 and 1.  The grand means of these three
replicates are, after rounding, $9.0$, $11.4$, and $10.2$.  When sorted, we
have $9.0$, $10.2$, and $11.4$.  Three replicates is certainly not enough,
but if we had, say, 1000, we would take the 25th and 975th from such sorted
sequence as the lower and upper bound of a 95\% confidence interval for the
mean of the old system.

\begin{table}
\tbl{Bootstrap Replicates of Raw Data (Old System from Table~\ref{tWQEData})\label{tWQEBoot}}{

\begin{tabular}{c@{\hspace{1mm}}c@{\hspace{1mm}}c}

\begin{tabular}{|c|r@{ }r|r@{ }r|r@{ }r|}
%\multicolumn{7}{c}{Bootstrap Replicate 1} \\
\hline
$\bullet\bullet\bullet$ & \multicolumn{6}{c|}{Binaries} \\
%&& 1 & 2 & 3 \\
\hline
\multirow{2}{*}{Executions}  
 & 9 & 11  &  12 & 8  & 9 & 9\\
 & 5 & 5  &  8  & 12  & 11 & 9\\
\hline
\end{tabular}

&

\begin{tabular}{|r@{ }r|r@{ }r|r@{ }r|}
%\multicolumn{7}{c}{Bootstrap Replicate 1} \\
\hline
\multicolumn{6}{|c|}{Binaries} \\
%&& 1 & 2 & 3 \\
\hline
%\multirow{2}{*}{Executions}  
 14 & 10  &  13 & 13  & 8 & 8\\
 15 & 7  &  8  & 12  & 16 & 13\\
\hline
\end{tabular}

&

\begin{tabular}{|r@{ }r|r@{ }r|r@{ }r|}
%\multicolumn{7}{c}{Bootstrap Replicate 1} \\
\hline
\multicolumn{6}{|c|}{Binaries} \\
%&& 1 & 2 & 3 \\
\hline
%\multirow{2}{*}{Executions}  
 7 & 7  &  11 & 11  & 9 & 9\\
 15 & 15  &  9  & 11  & 9 & 9\\
\hline
\end{tabular}

\end{tabular}
}
\end{table}

We are also interested in the 95\% confidence interval for the ratio of the mean execution times of
the new and old system from Table~\ref{tWQEData}, we proceed as follows. We
already know that $\ossthree\doteq5.8$. Using the same algorithm we
calculate $\nssthree\doteq 4.6$. The grand means are 
$\overline{^{O}Y}=10.5$ and $\overline{^{N}Y}=6.5$. From Student's
$t$-distribution, $t^2_{\frac{\alpha}{2},\nu} = t^2_{\frac{0.05}{2},2} \doteq 18.5$.
The confidence interval for the ratio of means is therefore 
$$
\frac{
  10.5\cdot6.5 \mp \sqrt{ 
   (10.5\cdot6.5)^2 -  \left(10.5^2-18.5\frac{5.8}{3}\right) \left(6.5^2-18.5\frac{4.6}{3}\right)
                        } 
} {10.5^2 - 18.5 \frac{5.8}{3}} \doteq 
\frac{68.3 \mp 60.2}{74.5}.
$$
The confidence limits are thus $0.1$ and $1.7$ (90\% performance improvement
to 70\% performance degradation).  Such a wide interval would not be useful
in practice, but this example has used only very small repetition counts.

Alternatively, we could calculate the confidence interval for the ratio of
means of the two systems using the bootstrap method.  The bootstrap
replicates would be created for both systems in the same way as we
demonstrated it for the old system.  We would then calculate the ratio of
means of these replicates (the first of the new system over the first of the
old system, the second of the new over the second of the old, etc.).  From
these ratios, we would select the respective quantiles. 

\section{For Scientists: The New Method and The Statistics Behind It}
\label{sStatistics}

In this section, we formulate the method in statistical terms, give its
assumptions and provide proofs. We also discuss alternatives. 
In the description of the statistics behind our method, we assume the
execution time of the operation of interest is a continuous random variable,
$Y$.  The range of $Y$ is a subrange of real numbers:
$\exists \mathit{BCET},\mathit{WCET} \in \mathbb{R} : P(Y<\mathit{BCET}) =
P(Y>\mathit{WCET}) = 0$,
where $\mathit{BCET}$ is the best-case execution time and $\mathit{WCET}$ is the worst-case
execution time. We assume that the expectation of $Y$ exists and that its
variance is finite, $E(Y)=\mu$ and $\var(Y)=\sigma^2$.
The `operation of interest' can be anything that is the goal of the
measurement, small or large, depending on the benchmark we use. $Y$ models
only the steady state duration of the operation, and we only focus on steady
state performance here. 

\subsection{Statistical Model for a Hierarchy of Random Effects}
\label{sModel}

\subsubsection{Two-way Classification}
\label{sTwoWay}
We first describe the model with three levels\footnote{In our text, the term
`level' always refers to levels in a benchmark experiment --- it should not
be confused with its statistical meaning in texts that address fixed effects
models.  We only have random effects.} of hierarchy.  For clarity of
exposition, we shall use our running example of random effects in
compilation, execution, and measurement.  Later we show a general
description for an arbitrary number of levels.  The intuition behind the
model is simple: the times measured in a single \emph{execution} are
randomly distributed, with a mean that is also a random variable (this
notion is later formalised using conditional expectation).  In turn, the
mean of these execution means in a \emph{binary}\footnote{We use the term
`binary' to denote a single binary executable, that is, a product of
compilation.  For the statistical model, it is just a factor.} is a random
variable.  And, finally, the mean of these binary means is an unknown
constant, the grand mean $\mu$ for the whole system we are interested in. 
We start with a formalisation of the model and its properties.

We assume that measurements from a single execution are independent
identically distributed.  Their mean differs for different executions, as
the measurements are influenced by random effects related to a particular
execution.  This can be expressed using conditional expectation as
$E(Y|[\mu_E=m]) = m$, where $\mu_E$ is a random variable which gives a mean
$m$ (a number), a mean of measurements of one particular execution.  The
expression says that the random variable $E(Y|\mu_E)$ has the value of $m$
when random variable $\mu_E$ has the value of $m$.  We use a shorter
notation for this, $E(Y|\mu_E)=\mu_E$.  We assume that the variance of
measurements within execution is a constant $\sigma_{E}^2 = \var(Y|\mu_E)$
(homoscedastic).  This says that the random variable $\var(Y|\mu_E)$ has the
value of $\sigma_{E}^2$ no matter what is the value of $\mu_E$.  We then
assume that the $\mu_{E}$ are independent identically distributed within a
given binary, but the mean of this distribution is again a random variable
within a system: $E(\mu_E|\mu_B)=\mu_B$.  We assume that the variances of
execution means within a binary are constant:
$\var(\mu_E|\mu_B)=\sigma_{B}^2$.  The binary mean $\mu_B$ is a random
variable for a given system, $E(\mu_B) = \mu_S$, where $\mu_S$ is a
constant.  We denote the variance of $\mu_B$ as $\var(\mu_B)=\sigma_{S}^2$,
where $\sigma_{S}^2$ is a constant.

It can be shown that $E(Y)=\mu_S \; (=\mu)$ and that  
$\var(Y)=\sigma_{E}^2+\sigma_{B}^2+\sigma_{S}^2 \; (=\sigma^2)$. 
We will do so later for the general case of $n$-way classification. In
summary, we have the following random effects model in two-way
classification (the distribution of the observed execution time is randomly
influenced in two ways, through execution and compilation):
$$
\begin{array}{rclrcl}
E(Y) &=& \mu, & \var(Y) &=& \sigma^2 = \sigma_{E}^2+\sigma_{B}^2+\sigma_{S}^2\\[2mm]
E(Y|\mu_E) &=& \mu_E, & \var(Y|\mu_E) &=& \sigma_{E}^2 \\[1mm]
E(\mu_E|\mu_B) &=& \mu_B, \qquad & \var(\mu_E|\mu_B) &=& \sigma_{B}^2 \\[1mm]
E(\mu_B) &=& \mu, & \var(\mu_B) &=& \sigma_{S}^2  \\
\end{array}
$$

\subsubsection{N-way Classification}
\label{sNWay}

In $n$-way classification, the measurements within an execution ($Y|\mu_1$) are
independent identically distributed with mean $\mu_1$ and variance
$\sigma_1^2$.  $\mu_1$ is a random variable.  $\sigma_1^2$ is a constant:
\begin{equation}
\label{eM1}
E\left(Y|\mu_1\right) = \mu_1, \quad \var\left(Y|\mu_1\right) = \sigma^2_1.
\end{equation}
When $n$, the number of ways of the classification, is two or more, the mean
of $\mu_1|\mu_2$ is $\mu_2$, again a random variable. In general, 
\begin{equation}
\label{eMi}
\forall i, 1 \le i \le n-1,  \quad E\left(\mu_i|\mu_{i+1}\right) = \mu_{i+1}, \quad \var\left(\mu_i|\mu_{i+1}\right) = \sigma^2_{i+1}.
\end{equation}
Finally, $\mu_n|\mu_{n+1}$ is a random variable with mean $\mu_{n+1}$, which is a
constant:
\begin{equation}
\label{eMn}
E\left(\mu_n\right) = \mu_{n+1}, \quad \var\left(\mu_n\right) = \sigma^2_{n+1}.
\end{equation}

%\comment{We need more work to get the equations nicely aligned. eqnarray is
%obsolete --- use amsmath align instead. Even so, I couldn't get nice,
%multi-column alignments.}
%
% I couldn't get better alignment either. Perhaps to be revisited for the
% final version

\begin{lem}[Rule of Iterated Expectations]
\label{lroi}
If $X$ and $Y$ are random variables and the expectations exist,
$E\left[E(Y|X)\right]=E(Y)$. \textup{\citeN{aos}, Theorem~3.24, p.~55.}
\end{lem}

We will now show that $\mu_{n+1}=E(Y)$ (note that $E(Y)=\mu$ by definition):
\begin{eqnarray*}
\mu=E(Y) &=^{(L\ref{lroi})}& E\left[E\left(Y|\mu_1\right)\right]=^{(\ref{eM1})} E\left(\mu_1\right) \\
\forall i,1 \le i \le n-1, \quad E\left(\mu_i\right) &=^{(L\ref{lroi})}& E\left[E\left(\mu_i|\mu_{i+1}\right)\right]=^{(\ref{eMi})} E\left(\mu_{i+1}\right) \\
E\left(\mu_n\right) &=^{(\ref{eMn})}& \mu_{n+1} \\
\end{eqnarray*}

\begin{lem}[Property of conditional variance]
\label{lpcv}
For random variables $X$ and $Y$, 
$\var(Y) = E\left[\var(Y|X)\right] + \var\left[E(Y|X)\right]$.
\textup{\citeN{aos}, Theorem~3.27, p.~55.}
\end{lem}

We will now show that $\sigma^2 = \var(Y) = \sum_{i=1}^{n+1} \sigma_{i}^2$
(note that $\var(Y)=\sigma^2$ by definition):
%
%\comment{REJ has corrected the first line of this proof.}
%well spotted...
\begin{eqnarray*}
\sigma^2=\var(Y) &=^{(\ref{lpcv})}& 
E\left(\var\left(Y|\mu_1\right)\right)
+ \var\left(E\left(Y|\mu_1\right)\right)=^{(\ref{eM1})}
E\left(\sigma_1^2\right) + \var\left(\mu_1\right) = \\
 &=& \sigma_1^2 + \var\left(\mu_1\right)\\
\forall i,1 \le i \le n-1,  \quad \var\left(\mu_i\right) &=^{(\ref{lpcv})}&
E\left(\var\left(\mu_i|\mu_{i+1}\right)\right) +
\var\left(E\left(\mu_i|\mu_{i+1}\right)\right)  =^{(\ref{eMi})} \\
&=& E\left(\sigma_{i+1}^2\right) + \var\left(\mu_{i+1}\right) =
\sigma_{i+1}^2 + \var\left(\mu_{i+1}\right)\\
\var\left(\mu_n\right) &=^{(\ref{eMn})}& \sigma^2_{n+1}. \\
\end{eqnarray*}

In summary, we have the following model in $n$-way classification:
$$
\begin{array}{rclrcl}
E(Y) &=& \mu, & \var(Y) &=& \sigma^2 = \sum_{i=1}^{n+1} \sigma_{i}^2 \\[2mm]
E(Y|\mu_1) &=& \mu_1, & \var(Y|\mu_1) &=& \sigma_{1}^2 \\[1mm]
\forall i,1 \le i \le n-1 \quad E(\mu_{i}|\mu_{i+1}) &=& \mu_{i+1},\qquad & \var(\mu_i|\mu_{i+1}) &=& \sigma_{i+1}^2 \\[1mm]
E\left(\mu_n\right) &=& \mu, & \var\left(\mu_n\right) &=&
\sigma^2_{n+1}.
\end{array}
$$

\subsubsection{Properties of a Sample Mean with N-way Classification}
\label{sModelProperties}
%
%\comment{we need to explain 2-way, n-way classification} 
%
We would like to estimate the unknown parameter of interest $\mu=E(Y)$ (for
example, mean execution time) based on (balanced\footnote{In `balanced'
experiments, the numbers of repetitions at each level are constant, i.e.\
every binary is executed the same number of times, every execution makes the
same number of measurements, etc.}) measurements of $Y$.  In this section,
we will show that the (sample) arithmetic mean is an unbiased estimator of
$\mu$ and that it is asymptotically normal.  We will also derive the
variance of this estimate, so that we can later construct a confidence
interval for $\mu$.  Let $n_i$ be the numbers of repetitions at each level
of the experiment.  With three levels, we thus have a 2-way classification
and $n_3$ is the number of binaries, $n_2$ the number of executions per each
binary, and $n_1$ number of steady state measurements per each execution,
and hence $n_1 n_2 n_3$ is the total number of measurements.  For $n$-way
classification, we denote the sample mean $\overline{Y}$ as
\begin{equation}
\label{esgm}
\overline{Y}=\overline{Y}_{\underbrace{\bullet \dots \bullet}_{n+1}}=
\frac{1}{\prod_{i=1}^{n+1}n_i} \left(
\sum_{j_{n+1}=1}^{n_{n+1}} \sum_{j_{n}=1}^{n_{n}} \dots
\sum_{j_{1}=1}^{n_{1}} 
Y_{j_{n+1} j_{n} \dots j_1}
\right)
\end{equation}

\begin{lem}[Lindeberg--Levy Central Limit Theorem]
\label{lclt}
Let $X_1 \dots X_n$ be independent identically distributed with mean $\mu$ and finite positive variance
$\sigma^2$. Then, $\overline{X}_{\bullet} = \frac{1}{n} \sum_{i=1}^n X_i$ has an \emph{asymptotically} normal
distribution with mean $\mu$ and variance $\sigma^2/n$,
which we denote as $\overline{X}_{\bullet} \approx N\left(\mu,\sigma^2/n\right).$

\end{lem}

\begin{lem}
\label{lpnd}
Let $X_1 \dots X_n$ be independent identically distributed normal variables, $X_i \sim N\left(\mu_i,\sigma^2_i\right)$.
From the properties of normal distribution~\cite{aos}, it follows that
$\overline{X}_{\bullet}$ has normal distribution with mean $\overline{\mu}_{\bullet}$
and variance $\overline{\sigma^2}_{\bullet}/n$, which we denote as 
$ \overline{X}_{\bullet} \sim
        N\left(\overline{\mu}_{\bullet}, {\overline{\sigma^2}_{\bullet}}/n
\right)
$
\end{lem}
%\comment{Explain $\approx$ and $\sim$.}
%done

\begin{lem}
\label{lGauss}
Let $f(t; \mu_1,\sigma_1)$,$f(t; \mu_2,\sigma_2)$ be density functions
% 
%\comment{do we need to explain?} TK: I would rather not, otherwise we would
%have to explain too many more elementary things, to be consistent.
%
of normal
variables with means $\mu_1$, $\mu_2$ and variances $\sigma_1^2, \sigma_2^2$
$$
\int f(\tau; \mu_1,\sigma_1) f(t-\tau; \mu_2,\sigma_2)) d\tau =
f\left(t; \mu_1+\mu_2,\sqrt{\sigma^2_1+\sigma^2_2}\right).
$$
\end{lem}
In other words, a convolution of density functions of normal variables
(the left-hand side of the equation) is also a density function of a normal
variable.  Moreover, the normal variable has mean $\mu_1+\mu_2$ and
variance $\sigma_1^2 + \sigma_2^2$ (by a known property of the normal
distribution).

\begin{lem}
\label{leind}
Let $X$, $Y$ be random variables with expectations and finite variance, $X \sim N\left(\mu_X, \sigma^2_X\right)$ and 
$Y|\left[X=x\right] \sim N\left( x, \sigma^2 \right)$. Then,
$Y \sim N\left( \mu_X, \sigma^2_X + \sigma^2 \right)$.
\end{lem}

\emph{Proof.} Let $f$ be the probability density function of normal
distribution with mean $\mu$ and variance $\sigma^2$:
\begin{displaymath}
f(x;\mu,\sigma) = \frac{1}{\sigma\sqrt{2\pi}}
\exp\left(-\frac{(x-\mu)^2}{2\sigma^2}\right),
\quad \mathrm{where} \; \exp(z)=e^z
\end{displaymath}
The density functions of $X$  and $Y|X$ from Lemma~\ref{leind} are:
\begin{displaymath}
f_{X}(x)=f(x;\mu_X,\sigma_X), \quad f_{Y|X}(y|x) = f_{Y|x}(y) =
        f(y;x,\sigma)
\end{displaymath}
By the definition of conditional density:
\begin{displaymath}
f_{Y,X}(y,x) = f_{Y|X}(y|x) \cdot f_X(x)
\end{displaymath}
It follows, that:
\begin{eqnarray*}
\label{eToConv}  
f_Y(y) &=& \int f_{Y,X}(y,x)\, dx = \int f_{Y|X}(y|x)f_X(x) \,dx \\
% &=& \int f(y;x,\sigma)f(x; \mu_X,\sigma_X)\, dx  \\
&=& \int
\frac{1}{\sigma\sqrt{2\pi}} \exp\left(-\frac{(y-x)^2}{2\sigma^2}\right) \cdot
\frac{1}{\sigma_X\sqrt{2\pi}}
\exp\left(-\frac{(x-\mu_X)^2}{2\sigma_X^2}\right) dx  \\
\left[ \stackrel{\mathrm{substituting}}{u=x-\mu_X} \right]
&=&
\int \frac{1}{\sigma\sqrt{2\pi}}
\exp\left(-\frac{(y-\mu_X-u)^2}{2\sigma^2}\right) \cdot
\frac{1}{\sigma_X\sqrt{2\pi}} \exp\left(-\frac{u^2}{2\sigma_X^2}\right) du 
\\
&=&
\int f(y-u; \mu_X,\sigma)f(u; 0,\sigma_X) \,du  \quad\quad\quad
\\
&=^{(L\ref{lGauss})}&
f\left(y; \mu_X, \sqrt{\sigma^2+\sigma^2_X}\right)
\end{eqnarray*}
Hence,  $Y \sim N( \mu_X, \sigma^2_X + \sigma^2). \quad \qed$

\smallskip
With these lemmas, we can infer asymptotic distributions of estimators in our
$n$-way model.  Informally, the basic idea is simple.  We get asymptotic
normal distributions for sample means of executions by the Central Limit Theorem
(Lemma~\ref{lclt}), which propagates to the grand mean $\overline{Y}$ by the
properties of the normal distribution (Lemma ~\ref{lpnd}).  Similarly, the sample
mean of means of executions within a binary ($\mu_1|\mu_2$) is
asymptotically normal by the Central Limit Theorem, which propagates to the
`mean of the grand mean' by properties of the normal distribution.  The two
are then joined by convolution, with the help of Lemma~\ref{lGauss}.

Formally, the application of Lemma~\ref{lclt} on sample means for individual
executions is described as:
$$
\forall j_{n+1} \dots j_{2}, \qquad \overline{Y}_{j_{n+1} \dots j_2 \bullet} \big|
\left[ {(\mu_{1})}_{j_{n+1} \dots j_2} = m \right] \approx N\left(
m, \frac{\sigma_1^2}{n_1} \right)
$$
The meaning is that for a particular representation $m$ of the respective
execution mean, we get a normally distributed sample mean with this
expectation $m$. The ranges for $j_{n+1} \dots j_{1}$ are as
in~(\ref{esgm}). However, for simplicity, we just will write
$$
\forall j_{n+1} \dots j_{2}, \qquad \overline{Y}_{j_{n+1} \dots j_2 \bullet} \big|
{(\mu_{1})}_{j_{n+1} \dots j_2} \approx  N\left(
{(\mu_{1})}_{j_{n+1} \dots j_2}, \frac{\sigma_1^2}{n_1} \right)
$$
Further summarisation to a higher level (i.e.\ binaries) keeps the asymptotic
normal distribution by Lemma~\ref{lpnd}:
$$
\forall j_{n+1} \dots j_{3}, \qquad \overline{Y}_{j_{n+1} \dots j_3 \bullet\bullet} \big|
{\overline{(\mu_{1})}}_{j_{n+1} \dots j_3 \bullet} \approx  N\left(
{\overline{(\mu_{1})}}_{j_{n+1} \dots j_3 \bullet}, \frac{\sigma_1^2}{n_1 n_2} \right)
$$
We now keep applying Lemma~\ref{lpnd} for all $i$, $1 < i \le n$:
$$
\forall j_{n+1} \dots j_{i+1}, \qquad \overline{Y}_{j_{n+1} \dots j_{i+1} \underbrace{\bullet \dots \bullet}_{i}} \big|
{\overline{(\mu_{1})}}_{j_{n+1} \dots j_{i+1} \underbrace{\bullet \dots \bullet}_{i-1}} \approx  N\left(
{\overline{(\mu_{1})}}_{j_{n+1} \dots j_{i+1} \underbrace{\bullet \dots \bullet}_{i-1}}, \frac{\sigma_1^2}{\prod_{k=1}^{i} n_i} \right)
$$
finally getting for $i=n+1$ (again by Lemma~\ref{lpnd})
$$
\overline{Y}_{\underbrace{\bullet \dots \bullet}_{n+1}} \big|
{\overline{(\mu_{1})}}_{\underbrace{\bullet \dots \bullet}_{n}} \approx  N\left(
{\overline{(\mu_{1})}}_{\underbrace{\bullet \dots \bullet}_{n}}, \frac{\sigma_1^2}{\prod_{k=1}^{n+1} n_k} \right)
$$
which could be written simply as 
\begin{equation}
\label{ecsm}
\overline{Y}\big|\overline{\mu_{1}} \approx  N\left(  \overline{\mu_{1}}, \frac{\sigma_1^2}{\prod_{k=1}^{n+1} n_k} \right)
\end{equation}

Now we need to derive the distribution of $\overline{\mu_1}$. This can be
done quite similarly to the distribution of
$\overline{Y}|\overline{\mu_{1}}$.  By applying the Central Limit Theorem
(Lemma~\ref{lclt}) on (unknown) means of executions, we get an asymptotically
normal distribution of the sample mean of these means:
$$
\forall j_{n+1} \dots j_{3}, \qquad {\overline{(\mu_1)}}_{j_{n+1} \dots j_3 \bullet} \big|
{(\mu_2)}_{j_{n+1} \dots j_3} \approx  N\left(
{(\mu_2)}_{j_{n+1} \dots j_3}, \frac{\sigma_2^2}{n_2} \right)
$$
The ranges for $j_{n+1} \dots j_{1}$ are as in~(\ref{esgm}). Hence, in a
2-way classification model, sample execution means $\overline{\mu_1}$ have
two sums (over binaries and over executions) and sample binary means
$\overline{\mu_2}$ have a single sum (over binaries). By applying
Lemma~\ref{lpnd}, we get
$$
\forall j_{n+1} \dots j_4, \qquad {\overline{(\mu_1)}}_{j_{n+1} \dots j_4 \bullet\bullet} \big|
{\overline{(\mu_2)}}_{j_{n+1} \dots j_4 \bullet} \approx  N\left(
{\overline{(\mu_2)}}_{j_{n+1} \dots j_4\bullet}, \frac{\sigma_2^2}{n_2 n_3} \right)
$$
Now we keep applying Lemma~\ref{lpnd} for all $i$, $1 < i < n$:
$$
\forall j_{n+1} \dots j_{i+2}, \,\, {\overline{(\mu_1)}}_{j_{n+1} \dots j_{i+2} \underbrace{\bullet \dots \bullet}_{i}} \big|
{\overline{(\mu_2)}}_{j_{n+1} \dots j_{i+2} \underbrace{\bullet \dots \bullet}_{i-1}} \approx  N\left(
{\overline{(\mu_2)}}_{j_{n+1} \dots j_{i+2} \underbrace{\bullet \dots \bullet}_{i-1}},
\frac{\sigma_2^2}{\prod_{k=2}^{i+1} n_k} \right)
$$
Finally, for $i = n$, by Lemma~\ref{lpnd} we get
$$
{\overline{(\mu_1)}}_{\underbrace{\bullet \dots \bullet}_{n}} \big|
{\overline{(\mu_2)}}_{\underbrace{\bullet \dots \bullet}_{n-1}} \approx  N\left(
{\overline{(\mu_2)}}_{\underbrace{\bullet \dots \bullet}_{n-1}},
\frac{\sigma_2^2}{\prod_{k=2}^{n+1} n_k} \right)
$$
which can be written as
\begin{equation}
{\overline{\mu_1}} |
{\overline{\mu_2}} \approx  N\left(
{\overline{\mu_2}},
\frac{\sigma_2^2}{\prod_{k=2}^{n+1} n_k} \right)
\end{equation}
The same procedure for $\mu_2|\mu_3$ gives
% \comment{FIXME: perhaps it would
% be shorter to formulate a lemma for the procedure.  though, perhaps not
% simpler to read}
%
$$
{\overline{\mu_2}} |
{\overline{\mu_3}} \approx  N\left(
{\overline{\mu_3}},
\frac{\sigma_3^2}{\prod_{k=3}^{n+1} n_k} \right)
$$
For the general case of $\mu_i|\mu_{i+1}$, $1 \le i \le n-1$, we then get
\begin{equation}
\label{emuit}
{\overline{\mu_i}} |
{\overline{\mu_{i+1}}} \approx  N\left(
{\overline{\mu_{i+1}}},
\frac{\sigma_{i+1}^2}{\prod_{k={i+1}}^{n+1} n_k} \right)
\end{equation}
The means at the highest level of non-determinism (the means of binaries in the
case of 2-way classification), the $\mu_n$, come from a single (non-conditional)
distribution with mean $\mu$ and variance $\sigma_{n+1}^2$.  By the Central
Limit Theorem, Lemma~\ref{lclt}, their sample mean is thus normally
distributed:
\begin{equation}
\label{elp}
\overline{\mu_n} \approx N\left( \mu, \frac{\sigma_{n+1}^2}{n_{n+1}} \right)
\end{equation}
By this we have the last missing bit for Lemma~\ref{leind}. Now, by using
Lemma~\ref{leind} on~(\ref{emuit}) and~(\ref{elp}), we get
$$
\overline{\mu_{n-1}} \approx N\left( \mu, \frac{\sigma_{n+1}^2}{n_{n+1}}
+ \frac{\sigma_{n}^2}{n_n n_{n+1}} \right)
$$
and then by further applications of the lemma and~(\ref{emuit}) we get
\begin{equation}
\label{emu1}
\overline{\mu_1} \approx N\left( \mu, 
\sum_{i=2}^{n+1} \frac{ \sigma_{i}^2}{\prod_{k=i}^{n+1} n_k} \right)
\end{equation}
Now we can apply Lemma~\ref{lpnd} %\comment{was \ref{elp}}
on~(\ref{ecsm}) and~(\ref{emu1}), by which
we get
\begin{equation}
\label{eMeanNormality}
\overline{Y} \approx N\left( \mu,
\sum_{i=1}^{n+1} \frac{ \sigma_{i}^2}{\prod_{k=i}^{n+1} n_k} \right)
\end{equation}
Hence, the sample arithmetic mean $Y$ is an unbiased estimator of $\mu$ with
$n$-way classification, is normally distributed, and we have an expression for
its variance. 

\subsection{Confidence Interval for the Mean of One System}
\label{sOneSystem}

Within the model described in the Section~\ref{sModel}, we can construct a
confidence interval for the mean execution time ($\mu$) of a single system. 
In Section~\ref{sTwoSystems}, we extend this to the confidence interval for
the ratio of means of two systems.

We show how to construct the interval using two alternative methods. The
bootstrap method is intuitively simple and works for additional metrics,
such as the median, as well as the mean.  The parametric method based on
asymptotic normality of the model (shown in Section~\ref{sModel}), only
works for the mean, but has the potential to provide narrower intervals for
larger sample sizes.  We compare the two methods empirically in
Section~\ref{sEvaluation}.

\subsubsection{Bootstrap Confidence Interval for One System}
\label{sBootIntOneSystem}

Let us assume that we have $n+1$ levels of hierarchy in our experiment and
that we decided to repeat $n_{n+1}, n_{n}, \dots n_{1}$ times at each level. 
When we have completed benchmarking, we would thus have $n_{n+1} n_n \cdot
\dots \cdot n_1$ measurements, which we denote as
$$
Y_{j_{n+1} j_{n} \dots j_1} \qquad \left(\forall i, 1 \le i \le n+1, \quad 1 \le j_i \le n_i \right)
$$
We assume that the measurements from a single execution, that is $Y_{j_{n+1}
\dots j_2 \bullet}$ for any fixed $j_{n+1} \dots j_2$, are independent
identically distributed, and we assume independence of means at higher
levels as described in the previous section. 

Very informally, the core idea behind bootstrap is to simulate (many)
experiments based on the real data and then calculate what we want on the
simulated data, in our case to calculate the confidence interval. So instead
of one realization of $\overline{Y}$ which comes from our real data, we can have
many realizations, generated by simulation, which is much faster than real
experimentation. From these simulated realizations, we construct the confidence
interval by selecting appropriate quantiles. Detailed discussion and
formulation of different bootstrap methods can be found in~\citeN{davison}.

The simulation of the new realizations of $\overline{Y}$ in each iteration
randomly selects a subset of real data ($Y_{\underbrace{\bullet \dots
\bullet}_{n+1}})$ with replacement.  This means that the size of the subset
will be the same as of the original set ($n_{n+1} \cdot \dots \cdot n_1$),
but some data points may not be present, while others can be presented
multiple times.  In each iteration, the simulation calculates a sample
arithmetic mean of the subset.  After all iterations finish, we estimate the
0.025 and 0.975 quantiles for a 95\% confidence interval (if we have 1000
iterations, the quantiles can be estimated as 25th and 975th ordered values,
although different estimators for quantiles exist).  Pseudo-code for the
described method is shown in Figure~\ref{fBootOneAlgo}.  For a 95\%
interval, we select $\alpha=0.05$.

Out of the wide variety of bootstrap methods, the key decisions here were
how to do resampling, and how to construct the confidence interval given the
simulated means. For resampling, depending on the underlying distribution
and the statistic of interest, it is sometimes better to resample with
replacement only at higher levels of the hierarchy, but keep the lower
levels intact~\cite{hboot,davison}. A na\"ive alternative is then also to
ignore the structure of the model and resample at random from all
measurements. We cover these alternatives in our evaluation later. The
method of resampling shown in Figure~\ref{fBootOneAlgo}, replacement at all
levels, seems to work best
(or at least not worse than others), and is a common default. 

For the construction of the confidence interval given the simulated means,
we use the percentile method.  This method is sensitive to non-symmetrical
distributions of the statistic, but it should not be much of a problem here
as we have shown earlier that $\overline{Y}$ is asymptotically normal. 
Still, one can easily plug in alternative bootstrap methods for confidence
interval construction~\cite{davison}.  Statistical software packages, such as R,
implement plenty of different methods.

%\comment{Should we be more specific about the statistical backing(for bootstrap)? My belief
%is that rather not much, because this is not our invention, as opposed to the
%statistical model before. Agree - REJ. We should make it clear that the statistical model
%is not from the literature (indeed based on it, but everything is based on
%something). Certainly - REJ. But we could include empirical distribution function and perhaps
%references to literature, other bootstrap methods.}

\begin{figure}
\begin{lstlisting}[frame=single]
Input: $n$, $(n_1, \dots, n_n)$, $Y_{j_{n+1} j_{n} \dots j_1}$ where $\forall i, \, 1 \le j_i \le n_i$, $nIterations=1000$, $\alpha=0.05$
Output: $lower$, $upper$

Uses:	mean(x) ... arithmetic average
Uses:	quantile(probability, x) ... select a sample quantile
Uses:   resample (replacement, x) ... random resampling 

$simulatedMeans$ = new vector[ $nIterations$ ]
foreach $iteration$ in 1..$nIterations$ {
  $simulatedMeasurements$ = new vector[ $n_{n+1} \cdot \dots \cdot n_1$ ]
  foreach $j_{n+1}$ in resample( 1..$n_{n+1}$, replacement = yes ) {
    foreach $j_{n}$ in resample( 1..$n_n$, replacement = yes ) {
      $\dots$
      foreach $j_{1}$ in resample( 1..$n_1$, replacement = yes ) {
        append $Y_{j_{n+1} j_{n} \dots j_1}$ to $simulatedMeasurements$
      }
      $\dots$ 
    }
  }
  $simulatedMeans$[ $iteration$ ] = mean( $simulatedMeasurements$ )
}
$lower$ = quantile( probability = $\alpha/2$, $simulatedMeans$ )
$upper$ = quantile( probability = $1-\alpha/2$, $simulatedMeans$ )
\end{lstlisting}
\caption{Bootstrap Confidence Interval for One System.}
\label{fBootOneAlgo}
\end{figure}

\subsubsection{Asymptotic Confidence Interval for One System}
\label{sAsymIntOneSystem}

Alternatively to bootstrap, we can construct a confidence interval for the
mean with $n$-way classification using the asymptotic normality of
$\overline{Y}$ ~(equation~\ref{eMeanNormality} of
Section~\ref{sModelProperties}).  For this we need to estimate the unknown
variance of the sample mean.  It is easier to do this directly than by
estimating the individual variances $\sigma_i^2$, $1 \le i \le n+1$.

We use the following estimator:
$$
S_{n+1}^2 = \frac{1}{n_{n+1}-1} \sum_{j_{n+1}=1}^{n_{n+1}} 
\left( \overline{Y}_{j_{n+1} \underbrace{\bullet \dots \bullet}_{n}} - \overline{Y}_{\underbrace{\bullet \dots \bullet}_{n+1}} \right)^2
$$
We show in Section~\ref{sVarEst} that $S_{n+1}^2 / n_{n+1}$ is an unbiased
estimator of the variance of the sample mean, that is
$$
E \left( \frac{S_{n+1}^2}{n_{n+1}} \right) = \sum_{i=1}^{n+1} \frac{ \sigma_{i}^2}{\prod_{k=i}^{n+1} n_k}
$$
Note this also means that as long as we have a hierarchical experiment ($n
\ge 1$), $S_{n+1}^2$ is \emph{not} an unbiased estimator of $\sigma_{n+1}^2$.
Relying on asymptotic normality even after the unknown variance of the
sample mean is replaced by its estimate, we get an asymptotic ($1-\alpha$)
confidence interval for $\mu$:
\begin{equation}
\label{eOSInterval}
\overline{Y} \pm u_{1-\frac{\alpha}{2}} \sqrt{ \frac{S_{n+1}^2}{n_{n+1}}} = 
\overline{Y} \pm u_{1-\frac{\alpha}{2}} \sqrt{ \frac{1}{n_{n+1}(n_{n+1}-1)} \sum_{j_{n+1}=1}^{n_{n+1}} 
\left( \overline{Y}_{j_{n+1} \underbrace{\bullet \dots \bullet}_{n}} - \overline{Y}_{\underbrace{\bullet \dots \bullet}_{n+1}} \right)^2 }
\end{equation}
We do not claim that $\overline{Y}$ has a $t$-distribution, as we do not
assume normality of $Y|\mu_1$, $\mu_1|\mu_2$, \dots, $\mu_{n-1}|\mu_{n}$,
and $\mu_{n}$.  A one-way model that makes such normality assumptions can be
found in~\citeN{mcculloch}, including the respective confidence interval,
which uses the $t$-distribution with $n_{n+1}-1$ degrees of freedom.  For
large numbers of degrees of freedom, the $t$ distribution converges to
normal, so the choice is not important.  For small number of degrees of
freedom, say smaller than 30 and definitely smaller than 20, the confidence
intervals become wider with the $t$-distribution than with the Normal
distribution.  This means that under the normality assumptions, one should
definitely use the $t$ distribution, otherwise the interval would be too
narrow (its coverage will be smaller than the projected $1-\alpha$).  In the
practice, a larger than projected coverage is usually regarded as better
than a smaller one, so it makes sense to use the $t$ distribution anyway.

\subsection{Confidence Interval for Ratio of Means}
\label{sTwoSystems}

In this section, we show how the method for constructing the confidence interval
for the mean of one system (Section~\ref{sOneSystem}) can be extended to a
confidence interval of the ratio of means of two systems.  We will refer to
these systems as `old' and `new', and make the same set of assumptions for the two systems
as we did for the single system so far (independence and identical distributions at
multiple levels).  We denote the corresponding random variables for
execution time as $^{O}Y$ and $^{N}Y$, and the means as $^{O}\mu = E(^{O}Y)$ and
$^{N}\mu = E(^{N}Y)$.  Thus, we now have a bivariate distribution ($F$) of
$^{ON}Y=(^{O}Y,^{N}Y) \sim F(x_O,x_N)$.  We assume that $^{O}Y$ and $^{N}Y$ are
independent.  The variances $^{N}\sigma_i^2$, $^{O}\sigma_i^2$ and
expectations in the two systems can differ.  Even the distributions
may differ.  The parameter of interest is now
$$
\theta = t(F) = \frac{^{N}\mu}{^{O}\mu} = \frac{\int x_N \, dF(x_O,x_N)}{\int x_O \, dF(x_O,x_N)}
$$ 
We estimate $\theta$ using measurements of the two systems.  Again, we use
balanced measurements and the same repetition counts ($n_1,\dots,n_{n+1}$)
for both systems. It can be shown that $T$,
$$
T = t(\widehat{F}) = \frac{ \overline{^{N}Y} }{ \overline{^{O}Y} }
$$
is an unbiased (plugin) estimator for $\theta$ (\citeN{davison},
Example~2.2).  Hence, we can estimate $\theta$ using the ratio of arithmetic
averages of the two systems, and we will do so both in the bootstrap and the
asymptotic parametric method.

We estimate the unknown variances $\var(^{O}Y)$ and $\var(^{N}Y)$ by
$\oss$ and $\nss$ from Section~\ref{sModel}.

\subsubsection{Bootstrap Confidence Interval for Ratio of Means}
\label{sBootRatio}

To construct a bootstrap interval for the ratio of means $\theta$, we need
to simulate many realisations of its estimator $T$.  $\alpha/2$- and
$1-\alpha/2$- quantiles of these realisations form a $(1-\alpha)$ percentile
bootstrap confidence interval for $\theta$. The pseudo-code is shown in
Figure~\ref{fBootRatio}. Discussion of more elaborate bootstrap methods for
construction of the ratio of means can be found in~\citeN{luxfra}.

\begin{figure}
\begin{lstlisting}
Input: $n$, $(n_1, \dots, n_n)$, $^{O,N}Y_{j_{n+1} j_{n} \dots j_1}$ where $\forall i \, 1 \le j_i \le n_i$, $nIterations=1000$, $\alpha=0.05$
Output: $lower$, $upper$

Uses:	mean(x) ... arithmetic average
Uses:	quantile(probability, x) ... select a sample quantile
Uses:   resample (replacement, x) ... random resampling 

function simulateMean( $oldnew$ ) {
  $simulatedMeasurements$ = new vector[ $n_{n+1} \cdot \dots \cdot n_1$ ]
  foreach $j_{n+1}$ in resample( 1..$n_{n+1}$, replacement = yes ) {
    foreach $j_{n}$ in resample( 1..$n_n$, replacement = yes ) {
      $\dots$
      foreach $j_{1}$ in resample( 1..$n_1$, replacement = yes ) {
        append $^{oldnew}Y_{j_{n+1} j_{n} \dots j_1}$ to $simulatedMeasurements$
      }
      $\dots$ 
    }
  }
  return mean($simulatedMeasurements$)
}

$simulatedRatios$ = new vector[ $nIterations$ ]
foreach $iteration$ in 1..$nIterations$ {
  $simulatedRatios$[ $iteration$ ] = $\frac{\texttt{simulateMean( oldnew = 'N' )}}{\texttt{simulateMean( oldnew = 'O' )}}$
}
$lower$ = quantile( probability = $\alpha/2$, $simulatedRatios$ )
$upper$ = quantile( probability = $1-\alpha/2$, $simulatedRatios$ )
\end{lstlisting}
\caption{Bootstrap Confidence Interval for Ratio of Means of Two Systems.}
\label{fBootRatio}
\end{figure}

\subsubsection{Asymptotic Confidence Interval for Ratio of Means}
\label{sFiellerRatio}

We can construct an asymptotic interval for $\theta$ using a theorem
by~\citeN{fieller}, which gives confidence limits for the ratio of means of
two normally distributed variables.

\begin{lem}[Fieller's Theorem~\cite{fieller}]
 \label{lfieller} Let $X,Y$ be
normally distributed random variables, not necessarily independent.  Let
$x,y$ be unbiased estimates of the means $E(X),E(Y)$.  Let $v_{xx},v_{yy}$
be variances of $x$ and $y$ (note, not of $X$ and $Y$).  Let $v_{xy}$ be
sample covariance of $x$ and $y$.  Then, the confidence limits for
$E(X)/E(Y)$ are:
$$
\alpha_1,\alpha_2 = \frac{\left(xy - t^2v_{xy}\right) \mp 
\sqrt{\left(xy - t^2v_{xy}\right)^2 - \left(x^2 - t^2v_{xx}\right)\left(y^2 - t^2v_{yy}\right)}}{x^2 - t^2v_{xx}}
$$
where $\alpha_1$ is the lower limit ($-$ sign) and $\alpha_2$ is the upper
limit ($+$ sign).  $t$ is the critical value for the two-tail
$t$-distribution, that is, for a 95\% interval it is the 0.025$^{th}$
quantile of the respective $t$-distribution.  The interval will be bounded and
non-trivial, that is $-\infty < \alpha_1 < \alpha_2 < \infty$ for $0 < t^2 <
x^2/v_{xx}$. Details on other cases can be found in~\citeN{fieller} and \citeN{luxfra}.
%$t^2$, $<t^2<t^2_{max}$, where
%
%$$
%\frac{x^2}{v_{xx}} t^2_{max} = \frac{x^2}{v_{xx}} + 
%\frac{\left(yv_{xx}-xv_{xy}\right)^2}{v_{xx}\left(v_{xx}v_{yy}-v^2_{xy}\right)}.
%$$
%
% Note there is a typo in the Fiellers 1954 paper in the first version of
% the formula for t^2_{max}, as found in LuxFra Confidence Sets for Ratios:
% A Purely Geometric Approach To Fieller’s Theorem. The second version we
% show here seems correct. Perhaps we don't need to discuss this here.
%  [I have verified they are correct - it is not that hard ]
%
\end{lem}
Note that the requirement of $t^2 < x^2/v_{xx}$ from the theorem is
intuitive.  It states that the estimator of the denominator (the sample mean
of the old system) should be statistically significantly different from
zero.  In our case with sample means of execution times, this condition is
hardly ever violated. Nevertheless, it can still happen due to the
statistical nature of the condition and hence the condition needs to be
checked.  As we detail later in our evaluation, we have seen the condition
violated for a sample size of two in our experiments.  The requirement that
$t^2>0$ is also easily met in practice, as we are not interested in
intervals for confidence close to zero.  In our case, we also assume
independence of the two variables, which implies $v_{xy}=0$.  The
$(1-\alpha)$ confidence interval for $\theta$ is hence:
\begin{equation}
\label{efi}
\frac{ \overline{^{O}Y} \cdot \overline{^{N}Y} \mp \sqrt{
\left(\overline{^{O}Y} \cdot \overline{^{N}Y} \right)^2 - 
  \left( \left(\overline{^{O}Y}\right)^2 - t_{\frac{\alpha}{2},\nu}^2 \cdot \frac{\oss}{n_{n+1}} \right)
  \left( \left(\overline{^{N}Y}\right)^2 - t_{\frac{\alpha}{2},\nu}^2 \cdot \frac{\nss}{n_{n+1}} \right)
}} {
 \overline{^{O}Y}^2  - 
  t_{\frac{\alpha}{2},\nu}^2 \cdot \oss / {n_{n+1}}
}
\end{equation}
where $t_{\frac{\alpha}{2},\nu}$ is the $\frac{\alpha}{2}$- quantile of the
$t$-distribution with $\nu=n_{n+1}-1$ degrees of freedom.
The Fieller's result can be extended for unbalanced experiments, where the
calculation of the numbers of degrees of freedom becomes more complicated. 
Details can be found in~\citeN{mratios} and \citeN{reportfieller}.  In our case of a
hierarchical experiment, this number could be modified to match the number
of degrees of freedom in our variance estimates.  
%
%We have not attempted this
%and we do not believe it is a problem in practice, as the total number of
%samples in the hierarchical experiment is large, and hence the $t$-distribution
%becomes close to normal, anyway.  
%
% no - the df is probably only n_{n+1}-1, which is small ... should have a closer look
%   this may improve some results for small numbers of binaries
%
Note that Fieller's method assumes normality of the two variables, that is
of $^{O}Y$ and $^{N}Y$, which will be violated in practice --- execution
time is not normally distributed.  Relying only on the asymptotic normality
of the mean and looking only for an asymptotic interval, we would use the
quantiles of the standard Normal distribution instead of the
$t$-distribution in the interval.  To be conservative, we would in practice
use the $t$-distribution anyway, as in the case of the interval for a
single system.  The robustness of Fieller's method to deviations from normality
is touched in~\cite{luxfra}.

For alternative methods for construction of the confidence interval for the
ratio of means we refer the reader to~\citeN{ratiocficmp} and
\citeN{luxfra}.  A common alternative is based on the delta method.

\subsection{Experiment Planning}
\label{sPlanning}

To obtain an unbiased result and realistic error bars (confidence interval),
one has to make sure that all the randomness of the system is encapsulated
by the experiment.  For example, if the system of interest has
non-deterministic compilation and different binaries also differ in
performance, one has to run multiple binaries within that experiment.  The
cost of building a new binary can be high, e.g.\ for the Mono platform,
which we use for validation of our method, the compilation time was about 20
minutes.  If the fluctuations in performance due to non-determinism in
compilation are small ($\sigma_S$ is small) compared to, say, the one due to
non-determinism in execution ($\sigma_B$), which is the usual case in our
experiments, time allocated to experimentation would be better spent by
running the existing binaries multiple times rather than building new
binaries.  We can formalise such trade-offs and, based on initial
experiments that repeat a few times at all levels to estimate the variances,
we can calculate the optimum number of repetitions at each level that
results in the most precise result for a given experimentation time.

Let us assume the random effects model in $n$-way classification of
Section~\ref{sModel}.  We have run initial experiments to obtain estimates
of the variances $\sigma^2_{1}, \sigma^2_{2}, \dots, \sigma^2_{n+1}$.  We
will discuss later which estimators can be used.  We also assume we know the
cost of a new repetition at the (non-top) levels, $c_1, c_2, \dots, c_n$, so
that the total cost of experimentation is 
\begin{equation}
\label{ecost}
c =  ( c_{n} + \dots ( c_3 + ( c_2 + (c_1 + n_1) n_2 ) n_3 \dots ) n_n )
n_{n+1}
\end{equation}
The unit of the costs ($c$, $c_i$) is the number of (the lowest-level) measurements
that could be obtained in that time. The practical meaning of the costs and
their use have been described informally in
Section~\ref{sDesignPractitioner}. With three levels only, the total cost
would be
$$
c = (c_2 + (c_1 + n_1)n_2)n_3
$$
where $n_1$ is the number of steady state measurements, $c_1$ is the number
of warm-up measurements of an execution that are not included into
summarization (cost of a new execution), $n_2$ is the number of executions,
$c_2$ is the number of measurements that could be done in the time needed to
build one binary, and $n_3$ is the number of binaries.

The costs $c_i$ can be calculated during the initial experiments that have
to be executed to estimate the variances $\sigma^2_i$.  For the purpose of
experiment planning, precise estimates of the costs are not needed
--- the planning is a back-of-the-envelope calculation. For instance, we can use
averages from the initial experiments. The optimisation problem is
to find the $n_1, \dots, n_n$ that minimize $f$,
$$
f(n_1, n_2, \dots, n_{n+1}) = \sum_{i=1}^{n+1} \frac{\sigma_{i}^2}{\prod_{k=i}^{n+1}
n_k}
$$
Function $f$ is a measure of the precision of the result, as it is a measure
of the width of the confidence interval for the mean ---
(equation~\ref{eOSInterval} in Section~\ref{sOneSystem}).  In this section,
we will show that the optimal numbers of repetitions $n_1, \dots, n_n$ are
\begin{equation}
\label{eopt}
n_1 = \sqrt{c_1 \frac{\sigma_{1}^2}{\sigma_{2}^2}}, \qquad
\forall i, 1 < i \le n, \,\, n_i = \sqrt{ \frac{c_i}{c_{i-1}} \frac{\sigma_{i}^2}{\sigma_{i+1}^2}}
\end{equation}
The optimal number of repetitions thus only depends on two adjacent levels
of the experiment, no matter what the total number of levels is. For
example, in a 2-way model, we have
$$
n_1=\sqrt{c_1\frac{\sigma^2_1}{\sigma^2_2}} \quad \textrm{and} \quad
n_2=\sqrt{\frac{c_2}{c_1}\frac{\sigma^2_2}{\sigma^2_3}} \\
$$
The number of samples $n_1$ should be large when the cost for warm-up $c_1$
is high and/or when the variance in samples $\sigma^2_1$ is higher than the
variance in executions $\sigma^2_2$.  The number of executions $n_2$ should
be large when the cost of a build $c_2$ is larger than the cost of a run
$c_1$ and/or when the variance in executions $\sigma^2_2$ is larger than the
variance in binaries $\sigma^2_3$.

\subsection{Derivation of the Optimalisation Formula}
\label{sPlanningDerivation}

For the proof, we will use the following notation
\begin{eqnarray*}
\forall i, 1 < i \le n+1, \,\, s_i &=& \sigma_{i}^2 \\
\forall i, 1 < i \le n+1, \,\, p_i &=& \prod_{k=i}^{n+1} n_k \\
k_1 &=& n_1, \qquad \forall i, 1 < i \le n, \,\, k_i = (c_{i-1} + k_{i-1})n_i
\end{eqnarray*}
Note that $k_i$ is a recurrent definition of the cost of the experiment
$c$.  In the new notation, we need to find $n_1,\dots,n_{n+1}$ that
minimize $f$ under the condition that $g=0$ for a given budget for the
experiment $c$ (in fact we only care about  $n_1,\dots,n_{n}$):
\begin{eqnarray}
\label{eFG}
f &=& f(n_1, \dots, n_{n+1}) = \sum_{i=1}^{n+1} \frac{s_i}{p_i(n_i, \dots, n_{n+1})} = \sum_{i=1}^{n+1} \frac{s_i}{p_i} \nonumber \\
g &=& g(n_1, \dots, n_{n+1}) = k_{n+1} (n_1, \dots, n_{n+1}) - c = k_{n+1} - c
\end{eqnarray}
Note that we sometimes omit the formal arguments of functions for
readability.  By the Lagrange Multiplier Theorem, optimum values can only be
among solutions of the system of equations:
\begin{eqnarray}
\label{eESystem}
\forall i,1 \le i \le n+1, \,\, \frac{\partial f}{\partial n_i} (n_1, \dots, n_{n+1}) + \lambda \frac{\partial g}{\partial n_i} (n_1, \dots, n_{n+1}) &=& 0 \nonumber \\
g (n_1, \dots, n_{n+1}) &=& 0
\end{eqnarray}
The partial derivatives are expressed in our notation as follows:
$$
\frac{\partial g}{\partial n_1} = n_2 n_3 \cdot\dots\cdot n_{n+1} = p_2 = \frac{k_1}{n_1} p_2, \quad \frac{\partial g}{\partial n_i} =^{(*)} \frac{k_i}{n_i} p_{i+1}
$$
$$
\frac{\partial f}{\partial n_1} =  -\frac{s_1}{n_1p_1}, \qquad 
\frac{\partial f}{\partial n_2} =  -\frac{s_1}{n_2p_1} -\frac{s_2}{n_2p_2}, \qquad
\frac{\partial f}{\partial n_i} = - \sum_{k=1}^{i} \frac{s_k}{n_ip_k}
$$
The derivation of $\frac{\partial f}{\partial n_i}$ is straightforward, but the derivation of the
marked $\frac{\partial g}{\partial n_i}$ deserves some explanation. From the
definitions of $k_i$ and $g$ we have that
\begin{eqnarray*}
\frac{\partial g}{\partial n_i} &=& \frac{\partial}{\partial n_i} ( k_{n+1} - c ) = \frac{\partial}{\partial n_i} k_{n+1} = 
\frac{\partial}{\partial n_i} (c_n + k_n)n_{n+1} =
\frac{\partial}{\partial n_i} (c_n + (c_{n-1} + k_{n-1})n_n \,)n_{n+1} \\
 &=& \!\!\dots = 
\frac{\partial}{\partial n_i} (c_n + (c_{n-1} + \dots (c_{i-1} + k_{i-1})n_i \dots \,)n_n \,)n_{n+1} \\
&=& \left( \frac{\partial}{\partial n_i} (c_{i-1} + k_{i-1})p_i \right) + 0 = (k_{i-1} + c_{i-1}) p_{i+1} =
p_{i+1} \frac{k_i}{n_i}.
\end{eqnarray*}

We will solve the system~(\ref{eESystem}) by substitutions and induction,
starting from partial derivatives for $\frac{\partial}{\partial n_1}$.  We
first express $n_i^2$ as follows:
\begin{eqnarray}
\label{eNI2}
\frac{\partial f}{\partial n_1} + \lambda \frac{\partial g}{\partial n_1} &=& 0 \nonumber \\
\lambda\frac{k_1}{n_1}p_2 &=& \frac{s_1}{n_1p_1} \nonumber \\ %\qquad | \cdot n_1p_1 \nonumber \\
\lambda k_1p_1p_2 &=& s_1 \qquad | \textrm{expand }k_1, p_1 \nonumber \\
\lambda p^2_2n^2_1 &=& s_1 \nonumber \\
n^2_1 &=& \frac{s_1}{\lambda p^2_2} = \frac{s_1}{\lambda n^2_2p^2_3}, \,\,
n^2_2 = \frac{s_1}{\lambda n^2_1p^2_3}, \,\, n^2_3 = \frac{s_1}{\lambda n^2_1n^2_2p^2_4}, 
\ \nonumber \\
n^2_i &=& \frac{s_1}{\lambda n^2_1n^2_2\dots n^2_{i-1}p^2_{i+1}}.
\end{eqnarray}
Now we can get a solution for $n_1$ from the equation for $\frac{\partial}{\partial n_2}$
\begin{eqnarray} 
\label{eN1}
\frac{\partial f}{\partial n_2} + \lambda \frac{\partial g}{\partial n_2} &=& 0 \nonumber \\
\lambda \frac{k_2}{n_2}p_3 &=& \frac{s_1}{n_2p_1} + \frac{s_2}{n_2p_2} \\%\qquad | \cdot n_2p_1 \nonumber \\
\lambda k_2p_1p_3 &=& s_1 + s_2n_1 \qquad | \textrm{expand }k_2,p_1 \nonumber \\
\lambda (c_1+n_1)n_1n^2_2p^2_3 &=& s_1 + s_2n_1 \qquad | \textrm{substitute }n^2_2\textrm{ using (\ref{eNI2})} \qquad \\%| \cdot n_1 \nonumber \\
(c_1+n_1) s_1 &=& s_1 n_1+ s_2n^2_1 \\
n_1 &=& \sqrt{\frac{s_1c_1}{s_2}}
\end{eqnarray}
which is the initial case of~(\ref{eopt}) which we need to prove. We follow
by induction over $i$.  Assuming~(\ref{eopt}) holds for $n_1, \dots, n_i$,
we prove it for $n_{i+1}$:
\begin{eqnarray*}
\frac{\partial f}{\partial n_{i+2}} + \lambda \frac{\partial g}{\partial n_{i+2}} &=& 0 \nonumber \\
\lambda \frac{k_{i+2}}{n_{i+2}} p_{i+3} &=& \frac{1}{n_{i+2}} \sum_{k=1}^{i+2} \frac{s_k}{p_k} =
\frac{1}{n_{i+2}} \left( \sum_{k=1}^{i} \frac{s_k}{p_k} + \frac{s_{i+1}}{p_{i+1}} + \frac{s_{i+2}}{p_{i+2}} \right) \nonumber \\%\qquad | \cdot n_{i+2}\\
\lambda k_{i+2}\cdot p_{i+3} &=^{(*)}& \lambda k_ip_{i+1} + \frac{s_{i+1}}{p_{i+1}} + \frac{s_{i+2}}{p_{i+2}} \qquad | \textrm{expand }k_{i+2} \\
\lambda (c_{i+1} + (c_{i} + k_i)n_{i+1})n_{i+2} &\cdot& p^2_{i+3}n_{i+2}n_{i+1} = \lambda k_ip^2_{i+1} + s_{i+1} + s_{i+2} n_{i+1}
\end{eqnarray*}
The transformation marked by $(*)$ is a substitution for the same equation,
but for $\frac{\partial}{\partial n_i}$.  By expanding the left hand side
and substituting $n^2_{i+2}$ using~(\ref{eNI2}), we get
\begin{eqnarray*}
c_{i+1} \frac{s_1}{n^2_1\cdot\dots\cdot n^2_in_{i+1}} + c_i \frac{s_1}{n^2_1\cdot\dots\cdot n^2_i} &=& s_{i+1} + s_{i+2}n_{i+1} \nonumber \\
c_{i+1}s_1 + c_is_1n_{i+1} &=& s_{i+1} \, n^2_1\cdot\dots\cdot n^2_i \, n_{i+1} + s_{i+2} \, n^2_1\cdot\dots\cdot n^2_i \, n^2_{i+1}
\end{eqnarray*}
It is easy to simplify $n^2_1\cdot\dots\cdot n^2_i$ by the induction assumption
for~(\ref{eopt}),
$$
n^2_1\cdot\dots\cdot n^2_i = \frac{c_1}{1}\frac{s_1}{s_2}\cdot \frac{c_2}{c_1}\frac{s_2}{s_3}\cdot \frac{c_3}{c_2}\frac{s_3}{s_4}\cdot \dots 
  \cdot \frac{c_{i}}{c_{i-1}}\frac{s_i}{s_{i+1}} = \frac{c_is_1}{s_{i+1}}
$$
and thus
\begin{eqnarray*}
c_{i+1}s_1 &=& s_{i+2} s_1 \frac{c_i}{s_{i+1}}n^2_{i+1} \\
n_{i+1} &=& \sqrt{\frac{c_{i+1}s_1s_{i+1}}{c_is_1s_{i+2}}} = \sqrt{\frac{c_{i+1}}{c_i}\frac{s_{i+1}}{s_{i+2}}}
\end{eqnarray*}
which finishes the induction step. We have solutions of~(\ref{eESystem}) for 
$$
n_1=\sqrt{\frac{s_1c_1}{s_2}}, \quad \forall i, 2 \le i \le n, \,\, n_i=\sqrt{\frac{c_is_i}{c_{i-1}s_{i+1}}}
$$
Technically, we could find a solution for $n_{n+1}$ based on the budget $c$
and solutions for $n_1,\dots,n_n$, but we are not really interested in
$n_{n+1}$.  Still, our solution of~(\ref{eESystem}) need not be the minimum
of ~(\ref{eFG}).  To verify that it actually is, we proceed as follows. First, we
reduce the optimisation problem to a single function $h$, which we get by
eliminating $n_{n+1}$ from $f$ using $g$. From~(\ref{eFG}), since $g=0$, we get
$$
(c_n+k_n)n_{n+1} -c = 0 \quad \text{hence} \quad
n_{n+1} = \frac{c}{c_n+k_n}
$$
and
\begin{eqnarray*}
f &=& f(n_1,\dots,n_{n+1}) = \frac{1}{n_{n+1}} \left( s_{n+1} + \sum_{i=1}^n \frac{s_i}{n_i\cdot\dots\cdot n_n} \right) \\
\end{eqnarray*}
For $h$ we have that 
\begin{eqnarray}
\label{eH}
h &=& f(n_1,\dots,n_{n}) = \frac{c_n+k_n}{c} \left( s_{n+1} + \sum_{i=1}^n \frac{s_i}{n_i\cdot\dots\cdot n_n} \right) = \\
  &=& \frac{1}{c} \left( c_n + n_1\cdot\dots\cdot n_n + \sum_{i=1}^{n-1} c_i \cdot n_{i+1}\cdot\dots\cdot n_n \right) \cdot
\left( s_{n+1} + \sum_{i=1}^{n} \frac{s_i}{n_i\cdot\dots\cdot n_n} \right)
\end{eqnarray}
Our optimisation problem of~(\ref{eFG}) is equivalent to minimizing $h$.
From~(\ref{eH}) we know that $h$ has the form of 
\begin{equation}
\label{ehcomb}
h(n_1, \dots,n_n) = \alpha + \sum \beta n^{*} + \sum \gamma \frac{1}{n^{*}}
\end{equation}
where $\alpha,\beta,\gamma$ are non-negative real constants and $n^{*}$ are
products of combinations of variables from $n_1,\dots,n_n$. For 1-way
classification, we have 
$$
h(n_1)=\alpha + \beta n_1 + \gamma \frac{1}{n_1}=\frac{(c_1+n_1)(s_1+s_2n_1)}{n_1c}
$$
for 2-way classification we have 
\begin{eqnarray*}
h(n_1,n_2) &=& \alpha + \beta_1 n_1 + \beta_2 n_2 + \beta_3 n_1n_2 + \gamma_1
\frac{1}{n_1} + \gamma_2 \frac{1}{n_2} + \gamma_3 \frac{1}{n_1n_2} = \\
&=& \frac{(c_2+n_2c_1+n_1n_2)(s_1+s_2n_1+s_3n_1n_2)}{n_1n_2c}
\end{eqnarray*}
From this general form it is easy to see that $h$ is continuous and positive
for $n_i > 0$. 
Every element of the sum in~(\ref{ehcomb}) is always positive. 
$h$ does not have a maximum, because it is unbounded for very small values of any and all
of its arguments, as well as for very large values. 
Each term of~(\ref{ehcomb}) is convex, hence the sum of terms, $h$, is convex,
so $h$ does not have an inflection point, but has a global minimum.\footnote{We could 
have found the solutions directly through partial differentiation of $h$.
We have done this for several values of $n$ using a symbolic algebra system, but
we chose to use the Lagrange multiplier method for the general case,
as the computation is simpler.}

\subsection{Estimating Unknown Variances}
\label{sVarEst}

Estimating the unknown variances $\sigma_1^2, \dots, \sigma_{n+1}^2$ is
harder than it may seem. Let us first define these `na\"ive' estimators

\begin{eqnarray*}
S_1^2 &=& \frac{1}{\prod_{k=2}^{n+1} n_k} \, \frac{1}{n_1-1} \sum_{j_{n+1}=1}^{n_{n+1}} \dots \sum_{j_{1}=1}^{n_1} 
\left( Y_{j_{n+1} \dots j_1} - \overline{Y}_{j_{n+1} \dots j_2 \bullet} \right)^2 \\
S_2^2 &=& \frac{1}{\prod_{k=3}^{n+1} n_k} \, \frac{1}{n_2-1} \sum_{j_{n+1}=1}^{n_{n+1}} \dots \sum_{j_{2}=1}^{n_2} 
\left( \overline{Y}_{j_{n+1} \dots j_2 \bullet} - \overline{Y}_{j_{n+1} \dots j_3 \bullet \bullet} \right)^2 \\
\end{eqnarray*}
for the general case of $i, 2 \le i \le n$ 
$$
S_i^2 = \frac{1}{\prod_{k=i+1}^{n+1} n_k} \, \frac{1}{n_i-1} \sum_{j_{n+1}=1}^{n_{n+1}} \dots \sum_{j_{i}=1}^{n_i} 
\left( \overline{Y}_{j_{n+1} \dots j_i \underbrace{\bullet \dots \bullet}_{i-1}} - \overline{Y}_{j_{n+1} \dots j_{i+1} \underbrace{\bullet \dots \bullet}_{i}} \right)^2
$$
and finally the already defined
$$
S_{n+1}^2 = \frac{1}{n_{n+1}-1} \sum_{j_{n+1}=1}^{n_{n+1}} 
\left( \overline{Y}_{j_{n+1} \underbrace{\bullet \dots \bullet}_{n}} - \overline{Y}_{\underbrace{\bullet \dots \bullet}_{n+1}} \right)^2
$$

These estimators are, apart from $S_1^2$, not unbiased estimators for
$\sigma_1^2, \dots, \sigma_{n+1}^2$. We will show later in this section that
\begin{eqnarray*}
E \left( S_1^2 \right) &=& \sigma_1^2  \\
E \left( S_2^2 \right) &=& \sigma_2^2 + \frac{\sigma_1^2}{n_1} \\
E \left( S_3^2 \right) &=& \sigma_3^2 + \frac{\sigma_2^2}{n_2} + \frac{\sigma_1^2}{n_1n_2} \\
\end{eqnarray*}
and for the general case of $i, 2 \le i \le n+1$
$$
E \left( S_i^2 \right) = \sigma_i^2 + \sum_{k=1}^{i-1} \frac{\sigma_k^2}{\prod_{l=k}^{i-1} n_{l}} = \sigma_i^2 + E\left( \frac{S_{i-1}^2}{n_{i-1}} \right)
$$
Let us define estimators $T_1, \dots, T_{n+1}$ as
\begin{eqnarray*}
T_1^2 &=& S_1^2 \\
%T_2^2 &=& S_2^2 - \frac{T_1^2}{n_1}, \\
\forall i, 1 < i \le n+1, \,\, T_i^2 &=& S_i^2 - \frac{S_{i-1}^2}{n_{i-1}}
\end{eqnarray*}
While $T_i^2$ are unbiased estimators of the unknown variances, they have
the issue that they may become negative for $i>1$.  In our case we need the
estimators only to plan the experiment.  Hence, we can simply iteratively
remove levels of the experiment where $T_i^2$ would be negative or zero. 
Maximum likelihood (ML) and restricted maximum likelihood (REML) estimators
for one-way model with normality assumption can be found
in~\citeN{mcculloch}.  Both types have alternatives for $T_i^2>0$ and
``$T_i^2<0$'', and both types are biased. Estimators for a broad range of
random and fixed effects models can be found in~\cite{varcomp}.

In the rest of this section, we will calculate the expectations of $S^2_i$.
We start with the lemmas we will use.
\begin{lem}
\label{ll1}
Let $\overline{z}=\frac{1}{n} \sum_{i=1}^{n} z_i$. Then
$
\sum_{i=1}^{n} \left( z_i - \overline{z} \right)^2 = 
  \left( \sum_{i=1}^{n} z^2_i \right) - n\left( \overline{z} \right)^2.
$
\end{lem}
%
%\emph{Proof.} This is shown easily with algebra operations only
%%
%\begin{eqnarray*}
%%
%\sum_{i=1}^{n} \left( z_i - \overline{z} \right)^2 &=&
%  \sum_{i=1}^{n} \left( z^2_i - 2z_i\overline{z} + \left( \overline{z} \right)^2 \right ) =
%  \left( \sum_{i=1}^{n} z^2_i \right) - \left( \left( \sum_{i=1}^n 2z_i \right) - n\overline{z} \right)\overline{z}\\
%%
%&=& \left( \sum_{i=1}^{n} z^2_i \right) - \left( 2n \left( \frac{1}{n} \sum_{i=1}^n z_i \right) - n\overline{z} \right)\overline{z} 
%  =  \left( \sum_{i=1}^{n} z^2_i \right) - n\left( \overline{z} \right)^2 \qed
%%
%\end{eqnarray*}
%
\begin{lem}
\label{ll2}
Let $z_i$, $i=1,..,n$ be random variables with constant expectation,
$Ez_i=Ez_1 <\infty$.  The following holds given the variances and
expectations exist:
$$
E\left( \sum_{i=1}^n \left( z_i - \overline{z} \right)^2 \right) = 
  \left( \sum_{i=1}^n \var(z_i) \right) - n \var(\overline{z}).
$$
\end{lem}
\emph{Proof.} From Lemma~\ref{ll1} and the linearity of expectation we have
that 
$$
E\left( \sum_{i=1}^n \left( z_i - \overline{z} \right)^2 \right) =^{(\ref{ll1})} 
  E\left(  \left( \sum_{i=1}^{n} z^2_i \right) - n\left( \overline{z} \right)^2 \right) =
  \left( \sum_{i=1}^n E\left( z^2_i \right) \right) -n E\left( \left(\overline{z}\right)^2 \right).
$$
From the well known property of variance, $\var(X) = E(X^2) - (EX)^2$, we
have $E(X^2) = \var(X) + (EX)^2$, and hence
\begin{eqnarray*}
\left( \sum_{i=1}^n E\left( z^2_i \right) \right) -n E\left( \left(\overline{z}\right)^2 \right) &=&
  \left( \sum_{i=1}^n \var(z_i) + (Ez_i)^2 \right) - n\left( \var\left(\overline{z}\right) + \left(E\overline{z}\right)^2\right) \\
%
%&=& \left(\sum_{i=1}^n \var(z_i)\right) + n(Ez_1)^2 - n \var(\overline{z}) - n(Ez_1)^2 \\
&=& \left( \sum_{i=1}^n \var(z_i) \right) - n \var(\overline{z}) \qed
\end{eqnarray*}
\begin{lem}
\label{ll3}
Let $z_i$\,, $i=1,..,n$ be random variables with constant variance,
$\var(z_i)=\var(z_1) <\infty$, and constant covariance,
$\cov(z_i,z_j)=\cov(z_1,z_2) <\infty$ for $i \neq j$. Then
$$
\var \left(\sum_{i=1}^n z_i \right) = n \var(z_1) + (n^2-n)\cov(z_1,z_2)
$$
and
$$
E\left( \frac{1}{n-1} \sum_{i=1}^n \left( z_i - \overline{z} \right)^2 \right) = \var(z_1) - \cov(z_1,z_2)
$$
Note that the first part of the lemma implies that
$$
\var(\overline{z}) = \frac{\var(z_1)}{n} + \frac{n-1}{n} \cov(z_1,z_2)
$$
\end{lem}
\emph{Proof.} The first part follows immediately from the property of the
variance of a sum of random variables:
$$
\var \left(\sum_{i=1}^n z_i \right) = \left( \sum_{i=1}^n \var(z_i) \right) + 2 \sum_{i<j}^n \cov(z_i,z_j) =
  n \var(z_1) + (n^2-n)\cov(z_1,z_2).
$$
The second part follows from Lemma~\ref{ll2} and from the first part:
\begin{eqnarray*}
E\left( \frac{1}{n-1} \sum_{i=1}^n \left( z_i - \overline{z} \right)^2 \right) &=&^{(L\ref{ll2})} 
  \frac{1}{n-1} \left( \sum_{i=1}^n \var(z_i) \right) - \frac{n}{n-1} \var(\overline{z}) \\
&=& \frac{n}{n-1} \var(z_1) - \frac{1}{n(n-1)}\var\left(\sum_{i=1}^n z_i \right) \\
&=& \frac{n}{n-1} \var(z_1) - \frac{1}{n-1} \var(z_1) - \frac{n(n-1)}{n(n-1)} \cov(z_1,z_2) \\
&=& \var(z_1) - \cov(z_1,z_2) \qed
\end{eqnarray*}
The plan is to use Lemma~\ref{ll3} to calculate, in turn, the expectation of
each variance estimator $S^2_i$.  We will use it for all $1 \le i \le n+1$,
although $i=1$ and $i=n+1$ are special cases which could be solved in an
easier way.  In each step, we will calculate as a side effect the variance
of $\sum_{i=1}^n z_i$, which will be handy for calculation of the term
``$\var(z_1)$'' in the next step.  For this plan, we also need the
covariances of ``$z_1,z_2$''.  We will first calculate these covariances
using the following three lemmas.
\begin{lem}
\label{llcsum}
For random variables $X_i$, $i=1,..,m$ and $Y_j$, $j=1.,,.n$ it follows from
a known property of covariance that
$$
\cov\left( \overline{X}, \overline{Y} \right) = \frac{1}{mn} \sum_{i=1}^{m}\sum_{j=1}^{n} \cov(X_i,Y_j).
$$
Note that if in addition $\cov(X_i,Y_j)$ is a constant, then $\cov(\overline{X},\overline{Y}) = \cov(X_1,Y_1)$.
\end{lem}
\begin{lem}[Law of Total Covariance]
\label{lltc}
For random variables $X$,$Y$,$Z$:  $\cov(X,Y) = E( \cov(X,Y|Z) ) + \cov(
E(X|Z), E(Y|Z))$.  
\end{lem}
This is a known property which can be shown using the law of total
expectation and the definition of covariance.

\begin{lem}[Tower Property of Conditional Expectation]
\label{ltp}
For random variables $X$,$G$,$H$: $E(X|G) = E( E(X|G,H) | G ).$
\end{lem}
The proof is
simple, but depends on ``low-level'' formal definition of conditional
expectation using $\sigma-$ algebras that we avoid here.  Details can be
found in~\cite{tprop}.

The covariance at the lowest level (measurements in one execution) can be
calculated as follows. For $k \neq l$
\begin{eqnarray*}
\cov\left( Y_{j_{n+1}\dots j_2k}, Y_{j_{n+1}\dots j_2l} \right) &=&^{(L\ref{lltc})}
  E\left( \cov\left( Y_{j_{n+1}\dots j_2k}, Y_{j_{n+1}\dots j_2l} | \left(\mu_1\right)_{j_{n+1}\dots j_2} \right) \right) + \\
&+& \cov\left( E\left( Y_{j_{n+1}\dots j_2k} | \left(\mu_1\right)_{j_{n+1}\dots j_2} \right),
              E\left( Y_{j_{n+1}\dots j_2l} | \left(\mu_1\right)_{j_{n+1}\dots j_2} \right)  \right) \\
&=&^{(*)} \, 0 + \cov\left( \left(\mu_1\right)_{j_{n+1}\dots j_2}, \left(\mu_1\right)_{j_{n+1}\dots j_2} \right) =
  \var\left( \left(\mu_1\right)_{j_{n+1}\dots j_2} \right) \\
&=&^{(\dagger)} \, \sigma^2 - \sigma^2_1.
\end{eqnarray*}
The equality marked by $(*)$ follows from independence of measurements in an
execution (an assumption of our model) and from the nesting of expectations in
our model, $E(Y|\mu_1)=\mu_1$.  The equality marked by $(\dagger)$ follows
from Lemma~\ref{lpcv} on page~\pageref{lpcv} (see the derivation of
$\sigma^2=\var(Y)$ below the Lemma). 
Covariances at higher levels require an additional step. For the second level, we have for $k \neq l$
$$
\begin{array}{l}
\cov\left( Y_{j_{n+1}\dots j_3kj_1}, Y_{j_{n+1}\dots j_3lj_1} \right) =^{(L\ref{lltc})}
  E\left( \cov\left( Y_{j_{n+1}\dots j_3kj_1}, Y_{j_{n+1}\dots j_3lj_1} | \left(\mu_2\right)_{j_{n+1}\dots j_3} \right) \right) + \\
\qquad + \, \cov\left( E\left( Y_{j_{n+1}\dots j_3kj_1} | \left(\mu_2\right)_{j_{n+1}\dots j_3} \right),
              E\left( Y_{j_{n+1}\dots j_3lj_1} | \left(\mu_2\right)_{j_{n+1}\dots j_3} \right)  \right). \\
\end{array}
$$

Note that in our model it is not important what the last index ($j_1$) is in
the above, and particularly if the two variables that are arguments of the
covariance have it the same or not.  The first term above is again zero from
the assumptions of our model, because the two executions only have in common
the mean of their means, which is the conditional.  The second term can
be expanded using Lemma~\ref{ltp} (we omit the largest index $j_{n+1}$ for
brevity)
\begin{eqnarray*}
E\left( Y_{\dots j_3kj_1} | \left(\mu_2\right)_{\dots j_3} \right) &=&^{(\ref{ltp})}
  E\left( E\left( Y_{\dots j_3kj_1} | \left(\mu_1\right)_{\dots j_3k},\left(\mu_2\right)_{\dots j_3} \right) | \left(\mu_2\right)_{\dots j_3} \right) \\
&=&^{(*)} E\left(  \left(\mu_1\right)_{\dots j_3k} | \left(\mu_2\right)_{\dots j_3} \right)  =
  \left(\mu_2\right)_{\dots j_3} 
\end{eqnarray*}
The marked equality follows from our model, where once we know the mean of
measurements in the execution $\mu_1$, the additional knowledge of the mean
of execution means in a binary $\mu_2$ makes no difference.  Also, we use
the nesting of expectations from the model, $E(Y|\mu_1)=\mu_1$.  By applying
the same approach to both arguments of the covariance we get
\begin{eqnarray*}
\cov\left( Y_{j_{n+1}\dots j_3kj_1}, Y_{j_{n+1}\dots j_3lj_1} \right) &=& 
  \cov\left( \left(\mu_2\right)_{j_{n+1} \dots j_3},  \left(\mu_2\right)_{j_{n+1} \dots j_3} \right) = 
  \var\left(  \left(\mu_2\right)_{j_{n+1} \dots j_3} \right) \\  
  &=& \sigma^2 - \sigma^2_1 - \sigma^2_2
\end{eqnarray*}
We use the same approach to calculate the covariance for all levels $1 \le i
\le n+1$.  We always use the same conditioning trick on the conditional
expectation of higher level, which reduces the conditional expectation using
the one from the previous step.  We demonstrate this at
level $i+1$.  The induction assumption is the result from the $i$-th level
\begin{eqnarray*}
\cov\left( Y_{j_{n+1}\dots j_{i+1}kj_{i-1}\dots j_1}, Y_{j_{n+1}\dots j_{i+1}lj_{i-1}\dots j_1} \right) &=&
  \sigma^2 - \sigma^2_1 - \dots - \sigma^2_i = \sum_{k=i+1}^{n+1} \sigma^2_k \\
E\left( Y_{j_{n+1}\dots j_1} | \left(\mu_i\right)_{j_{n+1}\dots j_{i+1}} \right) &=& \left(\mu_i\right)_{j_{n+1}\dots j_{i+1}}
\end{eqnarray*}
Hence at $(i+1)$-th level,
\begin{eqnarray*}
%
%cov\left( Y_{j_{n+1}\dots j_{i+2}kj_i\dots j_1}, Y_{j_{n+1}\dots j_{i+2}lj_i\dots j_1} \right) =
\cov\left( Y_{\dots kj_i\dots}, Y_{\dots lj_i\dots} \right) &=&^{(L\ref{lltc})}
  E \left( \cov\left( Y_{\dots kj_i \dots}, Y_{\dots lj_i \dots}\right) | \left(\mu_{i+1}\right)_{\dots j_{i+2}}\right) + \\
  &+& \cov \left( E\left( Y_{\dots kj_i \dots} | \left(\mu_{i+1}\right)_{\dots j_{i+2}} \right), 
             E\left( Y_{\dots lj_i \dots} | \left(\mu_{i+1}\right)_{\dots j_{i+2}} \right) \right) \\
  E \left( Y_{\dots} | \left(\mu_{i+1}\right)_{\dots j_{i+2}} \right) &=& 
    E\left( E\left(  Y_{\dots} | \left(\mu_i\right)_{\dots j_{i+1}}, \left(\mu_{i+1}\right)_{\dots j_{i+2}} \right) | \left(\mu_{i+1}\right)_{\dots j_{i+2}} \right) \\
  &=&^{(i\text{-th level})} E\left( \left(\mu_i\right)_{\dots j_{i+1}} | \left(\mu_{i+1}\right)_{\dots j_{i+2}} \right) =
    \left(\mu_{i+1}\right)_{\dots j_{i+2}}
\end{eqnarray*}
And thus 
$$
\cov\left( Y_{\dots kj_i\dots}, Y_{\dots lj_i\dots} \right) = \var\left( \left(\mu_{i+1}\right)_{\dots j_{i+2}} \right) = \sum_{k=i+2}^{n+1} \sigma^2_k
$$

Now we can finally calculate the expectations of $S^2_i$. We start at the
lowest level, $i=1$.  We define $\zone_k = Y_{j_{n+1}\dots j_2k}$.  From the
properties of our model, we have a constant variance of 
$\zone_k$, $\var(\zone_k) = \var(Y) = \sigma^2=\var(\zone_1)$. 
Also we have from the previous that $\cov(\zone_k,\zone_l)=\cov(\zone_1,\zone_2)=\sigma^2-\sigma^2_1$
for all $k \neq l$, $1 \le k,l \le n_1$. By Lemma~\ref{ll3} we get 
$$
E\left( \frac{1}{n_1-1} \sum_{k=1}^{n_1} \left( \zone_k - \overline{\zone} \right)^2 \right) =
  \sigma^2 - \sigma^2 + \sigma^2_1 = \sigma^2_1
$$
Thus, for $S^2_1$
\begin{eqnarray*}
E \left( S_1^2 \right) &=& E\left( \frac{1}{\prod_{k=2}^{n+1} n_k} \, \frac{1}{n_1-1} \sum_{j_{n+1}=1}^{n_{n+1}} \dots \sum_{j_{1}=1}^{n_1}
  \left( Y_{j_{n+1} \dots j_1} - \overline{Y}_{j_{n+1} \dots j_2 \bullet} \right)^2 \right) \\ 
 &=&  \frac{1}{\prod_{k=2}^{n+1} n_k} \prod_{k=2}^{n+1} n_k \,\, E\left( \frac{1}{n_1-1} \sum_{k=1}^{n_1} \left( \zone_k - \overline{\zone} \right)^2 \right) 
 = \sigma^2_1 \qquad \qed
\end{eqnarray*}
By Lemma~\ref{ll3} we also get the variance of the sum of $\zone_k$s, which we
will need in the following steps:
\begin{eqnarray}
\label{evar1}
\var \left( \sum_{k=1}^{n_1} \zone_k \right) &=& n_1 \sigma^2 + (n_1^2-n_1) (\sigma^2-\sigma^2_1) \\
     %= n_1^2\sigma^2 - (n_1^2 - n_1)\sigma^2_1 \nonumber \\
%
 &=&   n_1^2\sigma^2 - n_1^2\sigma^2_1 + n_1\sigma^2_1 = n_1\sigma^2_1 + n_1^2  \sum_{k=2}^{n+1} \sigma^2_k
\end{eqnarray}
Now let us calculate $S^2_2$. We now have $\ztwo_k = \overline{Y}_{j_{n+1}\dots j_3k\bullet}$.
By Lemma~\ref{llcsum} and from the previous calculation of covariances, we
have for $k \neq l$
$$
\cov(\ztwo_1,\ztwo_2) = \cov(\ztwo_k, \ztwo_l) = \cov\left( Y_{j_{n+1}\dots j_3kj_1}, Y_{j_{n+1}\dots j_3lj_1} \right) =
  \sum_{k=3}^{n+1} \sigma^2_k  
%= \sigma^2-\sigma^2_1-\sigma^2_2
$$
We get the variance of $\ztwo_k$ ($\ztwo_1$) using the previous step (Equation~\ref{evar1}):
$$
\var(\ztwo_1) = \frac{1}{n_1^2} \var\left( \sum_{j_1=1}^{n_1} Y_{j_{n+1}\dots j_3kj_1} \right) = 
  \frac{1}{n_1^2} \var\left( \sum_{j_1=1}^{n_1} \zone_{j_1} \right) =^{(\ref{evar1})} 
 \frac{\sigma^2_1}{n_1} + \sum_{k=2}^{n+1} \sigma^2_k
%\sigma^2 - \left(1-\frac{1}{n_1}\right)\sigma^2_1.
$$
Using Lemma~\ref{ll3}, for $S^2_2$ we get
\begin{eqnarray*}
E \left( S_2^2 \right) &=& E\left( \frac{1}{\prod_{k=3}^{n+1} n_k} \, \frac{1}{n_2-1} \sum_{j_{n+1}=1}^{n_{n+1}} \dots \sum_{j_{2}=1}^{n_2} 
\left( \overline{Y}_{j_{n+1} \dots j_2 \bullet} - \overline{Y}_{j_{n+1} \dots j_3 \bullet \bullet} \right)^2 \right) \\
&=& E\left( \frac{1}{n_2-1} \sum_{k=1}^{n_2} \left(\ztwo_k - \overline{\ztwo}\right)^2 \right)
  = \frac{\sigma^2_1}{n_1} + \sum_{k=2}^{n+1} \sigma^2_k - \sum_{k=3}^{n+1} \sigma^2_k
=^{(L\ref{ll3})} \sigma^2_2 + \frac{\sigma^2_1}{n_1} \quad \qed 
\end{eqnarray*}
Using the same lemma we also get the variance needed for the next step
\begin{eqnarray*}
\var \left( \sum_{k=1}^{n_2} \ztwo_k \right) &=&^{(L\ref{ll3})} n_2 \left(
  \frac{\sigma^2_1}{n_1} + \sum_{k=2}^{n+1} \sigma^2_k \right)  + 
  (n^2_2 - n_2) \left( \sum_{k=3}^{n+1} \sigma^2_k \right) =
  \frac{n_2}{n_1}\sigma^2_1 + n_2\sigma^2_2 + n_2^2 \sum_{k=3}^{n+1} \sigma_k^2 \\
&=& n_2^2 \left( \frac{\sigma^2_1}{n_1n_2} + \frac{\sigma_2^2}{n_2} + \sum_{k=3}^{n+1} \sigma_k^2 \right)
\end{eqnarray*}
We can now abstract what we should get at level $i$. 
\begin{eqnarray*}
\zi_k &=& \overline{Y}_{j_{n+1}\dots j_{i+1}k \underbrace{\bullet \dots \bullet}_{i-1}} \\
\cov\left( \zi_1,\zi_2 \right) &=& \sum_{k=i+1}^{n+1} \sigma^2_k \\
\var\left( \zi_1 \right) &=& \sum_{k=1}^{i-1} \frac{\sigma^2_k}{\prod_{l=k}^{i-1}n_l} + \sum_{k=i}^{n+1} \sigma_k^2 \\
\var\left( \sum_{k=1}^{n_i} \zi_k \right) &=& n_i^2 \left( \sum_{k=1}^{i} \frac{\sigma^2_k}{\prod_{l=k}^i n_l} + \sum_{k=i+1}^{n+1}\sigma_k^2\right)
\end{eqnarray*}
Leaving the previous as the induction assumption, we proceed at level $i+1$.
From the previous calculations of covariances we get
\begin{eqnarray*}
\zii_k &=& \overline{Y}_{j_{n+1}\dots j_{i+2}k \underbrace{\bullet \dots \bullet}_{i}} \\
\cov\left( \zii_1,\zii_2 \right) &=& \sum_{k=i+2}^{n+1} \sigma^2_k
\end{eqnarray*}
For the variance of $\zii_1$ we have, by properties of variance and from the
induction assumption, that
$$
\var(\zii_1) = \frac{1}{n_i^2} \var\left(\sum_{j_i=1}^{n_i} \zi_i \right) = \sum_{k=1}^{i} \frac{\sigma^2_k}{\prod_{l=k}^i n_l} + \sum_{k=i+1}^{n+1}\sigma_k^2
$$
Hence, by Lemma~\ref{ll3},
\begin{eqnarray*}
E\left( S_{i+1}^2 \right) &=&^{(L\ref{ll3})} \var(\zii_1) - \cov\left( \zii_1,\zii_2 \right) =
  \sum_{k=1}^{i} \frac{\sigma^2_k}{\prod_{l=k}^i n_l} + \sum_{k=i+1}^{n+1}\sigma_k^2 - \sum_{k=i+2}^{n+1} \sigma^2_k  \\
&=& \sigma_{i+1}^2 + \sum_{k=1}^{i} \frac{\sigma^2_k}{\prod_{l=k}^i n_l}
\end{eqnarray*}
To complete the induction we also need to derive the variance of the sum. We
do so again by Lemma~\ref{ll3}:
\begin{eqnarray*}
\var\left( \sum_{k=1}^{n_{i+1}} \zii_k \right) &=&^{(L\ref{ll3})} n_{i+1}\var(\zii_1) + (n_{i+1}^2-n_{i+1})\cov(\zii_1,\zii_2) \\
  &=& 
  n_{i+1} \left( \sum_{k=1}^{i} \frac{\sigma^2_k}{\prod_{l=k}^i n_l} + \sum_{k=i+1}^{n+1}\sigma_k^2 \right) +
  (n_{i+1}^2-n_{i+1}) \left( \sum_{k=i+2}^{n+1} \sigma^2_k \right)  \\
 &=&
 n_{i+1} \left( \sum_{k=1}^{i} \frac{\sigma^2_k}{\prod_{l=k}^i n_l} \right) + n_{i+1}\sigma^2_{i+1} + n_{i+1}^2 \sum_{k=i+2}^{n+1} \sigma^2_k \\
 &=&
 n_{i+1}^2 \left( \sum_{k=1}^{i+1} \frac{\sigma^2_k}{\prod_{l=k}^{i+1} n_l} + \sum_{k=i+2}^{n+1} \sigma^2_k \right) \qed
\end{eqnarray*}
For $i=n+1$, the derivation is exactly the same. We just formally define
$\sum_{k=m}^{m} \dots = 0$.  Note that as a side effect of the proof, we have
also derived the variance of the sample mean $\overline{Y}$ without the normality
assumptions,
$$
\var\left(\overline{Y}\right) = \frac{1}{n_{n+1}^2}\var\left( \sum_{k=1}^{n_{n+1}} \znn_k \right) =
  \frac{1}{n_{n+1}^2} n_{n+1}^2 \sum_{k=1}^{n+1} \frac{\sigma^2_k}{\prod_{l=k}^{n+1} n_l} =
  \sum_{k=1}^{n+1} \frac{\sigma^2_k}{\prod_{l=k}^{n+1} n_l}
$$

\section{Evaluation}
\label{sEvaluation}
%
%\comment{Should we rename this section to something like Properties of the
%method??? evaluation may be too strong...}

%\comment{Should we also attempt to validate assumptions of the model?}
% NO: the text is already too long

Both of our methods for constructing the confidence interval for the ratio
of mean execution times depend on assumptions we make about the statistical
properties of the data and on the actual sample size for which the
asymptotic properties begin to hold.  For asymptotic properties to hold, a
larger sample size is better with our method, just as it is with currently
recommended visual tests of overlapping confidence intervals and even
null-hypothesis statistical tests.  However, in practice, too large a sample
size is also undesirable.  With the currently recommended method, there is
the danger of focusing on small changes of no practical interest.  With our
method, this problem does not affect the comparison directly, because we
quantify the effect size.  However, it still has an indirect effect --- if
we choose to ignore a random factor in the experiment, however small its
impact on performance might be, an unduly large sample size will lead to
unrealistically narrow confidence intervals with too small coverage.  Too
small coverage means that the interval when constructed many times in many
experiments would cover the true unknown ratio in less than the projected
percentage (e.g. 95\%) of cases.

Note that violations of assumptions of statistical properties of the data do
not necessarily make parametric methods unusable in practice --- for
example, the $t$-test and ANOVA are known to be more robust than was initially
assumed~\cite{cfirobust,robustsim}.  References to works on the robustness of
ANOVA can be found in~\citeN{maxwell}.
%
%Still, any such additional evidence
%on robustness only applies to certain deviations from normality, which may
%not be the ones we are facing in computer performance measurements, so
%validation remains important.

The goal of our evaluation is to sanity check our method on real data and
demonstrate the trade-offs between sample size (experimentation time),
number of levels in the experiment, actual coverage, false alarm rate, and
the threshold for change that we still care about.  We tested our method on
a set of real benchmarks, for which we run orders of magnitude more
repetitions that could normally be afforded in a real comparison study.

%Based on the data, we perform a number of statistical simulations.
%\comment{We need a little more explanation here - how's this?}
%For example, we might set up two simulations based on data drawn from the same
%real experiment. Here we know that there should be no difference in performance
%between the two, but we can evaluate the results that different methods give for
%these metrics.
%
% I agree the mention of statistical simulations may be confusing at this
% point. I think it is better to remove it from here completely - the
% content is elsewhere, and the text is already very long.

% to
%evaluate our quantification method.  We focus on the actual coverage of the
%confidence interval and on the rate of false alarm when used for comparison
%of two systems.  By false alarm we mean a situation when the experimenter
%would, based on results quantified by our method and a given threshold of
%`how large difference we care about', conclude that there is a performance
%difference between the two systems, while in fact there is no difference.

\subsection{Benchmarks}
\label{sBenchmarks}

We chose the Mono platform~\cite{mono}, an open-source implementation of
.NET, for our experiments. We expect that our conclusions with regard to the
statistical method should apply to other managed platforms, such as Java,
and to some extent to any runtime system.  We run four benchmarks, each in
multiple variants, so that in total we have 10 tests.  FFT is a Fast Fourier
Transformation benchmark adapted from the SciMark2
suite~\cite{scimark2,scimarkcs}.  TCP Ping and HTTP Ping are simple remote
procedure call benchmarks, which include two processes that communicate via
TCP and HTTP channels.  Rijndael is an encryption benchmark\footnote{Source
code for the benchmarks is available from
\url{http://www.cs.kent.ac.uk/~tk243/esize.tgz}}.  For each benchmark we ran
a variant with the default optimisations enabled and then an `OPT' variant
with all optimisations enabled.  The motivation was that different
optimisation levels should lead to different performance, and may also lead
to difficulties when quantifying performance.  
%Our findings on the false
%alarm rates and the coverage, however, were in the end similar for the
%default and the OPT variants.

%In the end, however, the statistical properties of the OPT and
%the default variant did not differ in our experiments.
%
%\comment{This is not very clear and I worry that a reviewer might misread it to
%say that we didn't find anything. I think you are trying to say that, no matter
%whether you used the OPT or the default variant for the simulations, your
%findings on false alarm rates, coverage, etc, was similar?}
%
%
%~\comment{We could
%actually use the data to quantify the performance improvement of the
%optimisations.  But could not validate this in any way.}.  
%
For the FFT benchmark, we also run one variant that allocates a new FFT
buffer for every measured iteration, and another variant `NA' that re-uses
the same buffer for all of its execution.
We introduce the `NA' variant because we found that the default version
violates the independence assumption of our measurements within executions. 
The reason is subtle: memory placement influences conflict misses, which in
turn significantly influence the measured performance of FFT.  It might seem
that re-allocating the buffer before each measurement would nicely randomize
out this effect.  But it appears that memory locations get re-used by the
allocator almost regularly, hence creating a dependence in the measurements. 
With the `NA' version, the location of the buffer does not change within a
single execution, but instead changes between executions, so the assumption
is not violated.  The lesson learned from this is that it is better to have
non-determinism where we can control it with the experiment, and that
randomization to avoid measurement bias is harder than it may seem.  Note that the
solution of `not reallocating' within an execution is not a general one
--- if the collector was a moving one, there would again be non-determinism
in location out of our control.

The benchmarks were run on an Intel Pentium 4 under Fedora 4 (Linux 2.6.11)
with Mono 1.1.13.  Our experiments had three levels: we repeated
compilations, executions, and measurements.  The re-compilation involved a
complete re-build of the Mono platform, which took about 20 minutes.
Benchmark executions took roughly 15s with FFT, 6s with Rijndael, 4s with
HTTP, and 102ms with TCP). 
We used 150 builds of the platform for the experimentation and 100 executions
of each benchmark.  For the FFT and Rijndael benchmarks, we took 64 raw
measurements per execution, for TCP we took 256 and for HTTP we took 512. 
We chose these numbers to get more measurements for benchmarks where one
measurement was fast and where the variation in measurements seemed high. 
However, for our evaluation it should matter only that we have a sufficient
number of measurements to estimate the variance.
%
%\comment{We must justify this.} \comment{I tried, but cannot do much better.
%There is always experimenter's decision involved, which is based on
%experience and guess.}.
%
%
We target our evaluation at steady state performance. To ensure reaching the
steady state, we dropped the initial 30\% of measurements from each
execution.  By manual inspection on selected graphs, we verified that this
is a safe choice.  In a real performance study, we would drop much less with
these benchmarks. 
We observed that the build of the Mono platform itself is not deterministic
and impacts performance at least in some benchmarks, so re-building was
necessary.  But, the choice of (only) three levels was an arbitrary one. 
Partial re-building (the runtime, the compiler, the class libraries) could
introduce more levels, perhaps saving overall experimentation time, but we
did not attempt that.

\subsection{Variation in the Data}

%\comment{Do we need to say more, explicitly, about bias?}
% added a footnote to that effect ; and clarified reference to bias earlier 
% in the paper (consistent use of ``measurement'' bias)

The variation in the (real) measurements for the study is shown in
Table~\ref{tVariations}.  The table shows sample relative (percentage)
variation at each level of the experiment, which is the square root of the
variance estimate ($S^2_3$, $S^2_2$, $S^2_1$), normalized against the sample
mean of all measurements $\overline{Y}$.  With our data, the $S^2$ estimates
were almost identical to the unbiased~\footnote{The bias of statistical
estimators should not be confused with measurement bias. Both lead to
systematically wrong results, but the causes are different. Measurement bias
is due to poor experiment design. Bias in estimators is due to limited 
statistical methods.} $T^2$ estimates (shown in
parentheses when they differ by more than 0.1\%).  All the FFT benchmarks
have noticeable performance variation due to non-deterministic compilation
(3.4\% to 4.4\%).  In case of the `NA' versions it is even more than the
variation due to non-deterministic measurement.  The non-FFT benchmarks have
variation due to compilation below 1\%.  The FFT benchmarks also have high
variation due to execution (6.7\% to 8.2\%), which is far more than the
variation due to measurement (1.4\% to 4.6\%).  The non-FFT benchmarks have
higher variation due to measurements than execution. The HTTP and TCP
benchmarks have no variation due to execution (the $S^2_2$ estimate in this
case is only positive because of its bias).  In a real experiment, we would
thus remove repetition of execution for these benchmarks.  The new variation
due to binaries would then be 0.2\% for both HTTP benchmarks and 0.5\% for
both TCP benchmarks (not shown in the table).  The level of optimisation
(OPT or default) does not seem to have much impact on the relative
variation.  And (not shown in the table), it did not have too much impact on
execution time, either.

%                    sd3 sd2  sd1
%fft_scimark_na_oall 3.4 8.2  1.4
%fft_scimark_na      3.4 7.8  1.4
%fft_scimark_oall    4.4 8.2  4.3
%fft_scimark         4.1 6.7  4.6
%http_ping_oall      0.2 0.8 22.5
%http_ping           0.2 0.7 21.5
%rijndael_oall       0.4 3.8  9.3
%rijndael            0.4 3.5  9.1
%tcp_ping_oall       0.6 1.7 41.8
%tcp_ping            0.6 1.8 38.6

\begin{table}
\tbl{Relative Percentage Variation in the Real Data\label{tVariations}}{
\begin{tabular}{|r|r@{.}l|r@{.}l|r@{.}l|}
\hline
& \multicolumn{2}{c|}{ Compilation [\%] } & \multicolumn{2}{c|}{ Execution [\%] } & \multicolumn{2}{c|}{ Measurement [\%] } \\
\hline 
FFT NA OPT   & 3&4 & 8&2 & 1&4 \\
FFT NA       & 3&4 & 7&8 & 1&4 \\
FFT OPT      & 4&4 & 8&2 & 4&3 \\
FFT          & 4&1 & 6&7 & 4&6 \\\hline
HTTP OPT     & 0&2 & 0&8 (0) & 22&5 \\
HTTP         & 0&2 & 0&7 (0) & 21&5 \\\hline
Rijndael OPT & 0&4 & 3&8 (3.5) & 9&3 \\
Rijndael     & 0&4 & 3&5 (3.2) & 9&1 \\\hline
TCP OPT      & 0&6 & 1&7 (0) & 41&8 \\
TCP          & 0&6 & 1&8 (0) & 38&6 \\
\hline
\end{tabular}
}
\begin{tabnote}
\Note{Source:}{Estimated relative variation for compilation,
execution, and measurement, $S$ ($T$).  The $T$ estimates are only shown
when they differ from $S$ by more than 0.1\%.  All estimates are normalized
against grand mean $\overline{Y}$.}
\end{tabnote}

\end{table}

\subsection{False Alarm Rate}

When applied to two identical systems, our method should ideally always
report 1
%
%\comment{NB.  I consistently read `one' as `an' rather than '1'!}
%\comment{I used ``one'' because I was advised by a reviewer last time that
%numbers smaller than six (6) should always be spelled as letters.  I have
%seen also a rule that says any number than 10 should be spelled...
%REJ: I agree with that rule but not for data, which is best with numerals.}
%
as the ratio of the means and a (narrow) confidence interval around it.  Due
to statistical nature of performance, though, we may get a not so narrow
interval and it may not include 1.  If it does not include 1 and we care
about changes of any size, we conclude incorrectly that the two compared
systems have different performance (a false alarm).  If, say, we cared only
about differences above 2\% (our threshold was 2\%), we only would get a
false alarm if the lower bound of the interval was above 1.02, or the upper
bound below 0.98.  This would be less likely than with a zero threshold, but
could still happen.  In addition to the threshold, a significant factor
influencing the false alarm rate is the number of binaries we use to
estimate the interval.  More binaries means a larger sample size, but also
much more expensive experiments (adding a binary not only adds compilation
time, but also time for executions and measurements).

We ran a statistical simulation to quantify these trade-offs. In each
iteration, the simulation takes at random two sets of binaries with
replacement,\footnote{`Replacement' means that after choosing an element
from the set at random, it is replaced in the set and available to be chosen
again.  This contrasts with `no replacement' methods where, once an element
is chosen, it cannot be chosen again.} producing `two' systems to compare. 
It then applies either of our two methods to compute 95\% confidence
interval for the ratio of means.  Finally, the simulator makes a decision
whether the two systems differ, based on a pre-defined threshold: we used
0\%(the `significance' approach), 1\%, 2\%, \dots, 5\%.  We report the
percentages of decisions that the systems are different.  Such a decision is
always a false alarm, because we fed the method with data from the same system.

\subsubsection{Bootstrap vs. Asymptotic}
%\comment{Do we want to say more about bootstrap?}
% I think if we wanted to say more, we should do earlier in the text where
% we describe the method

\begin{figure}[!ht]
\centering
\subfigure[FFT NA OPT, RRR-Bootstrap.\label{fFAFFTb}]{
  \includegraphics[width=0.45\linewidth,trim=0cm 0.5cm 1cm 2cm,clip=true]{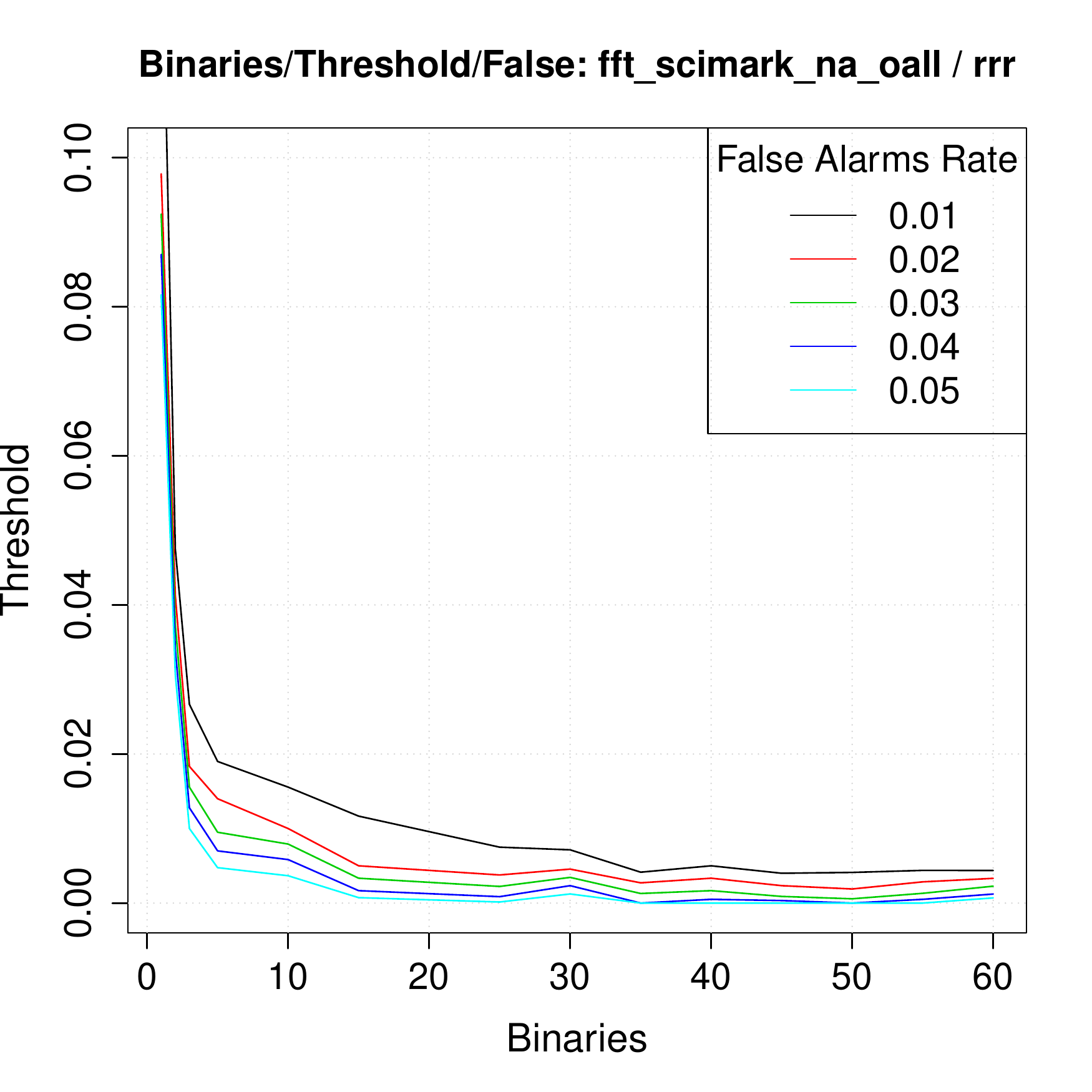}
}
\subfigure[FFT NA OPT, Asymptotic.\label{fFAFFTa}]{
  \includegraphics[width=0.45\linewidth,trim=0cm 0.5cm 1cm 2cm,clip=true]{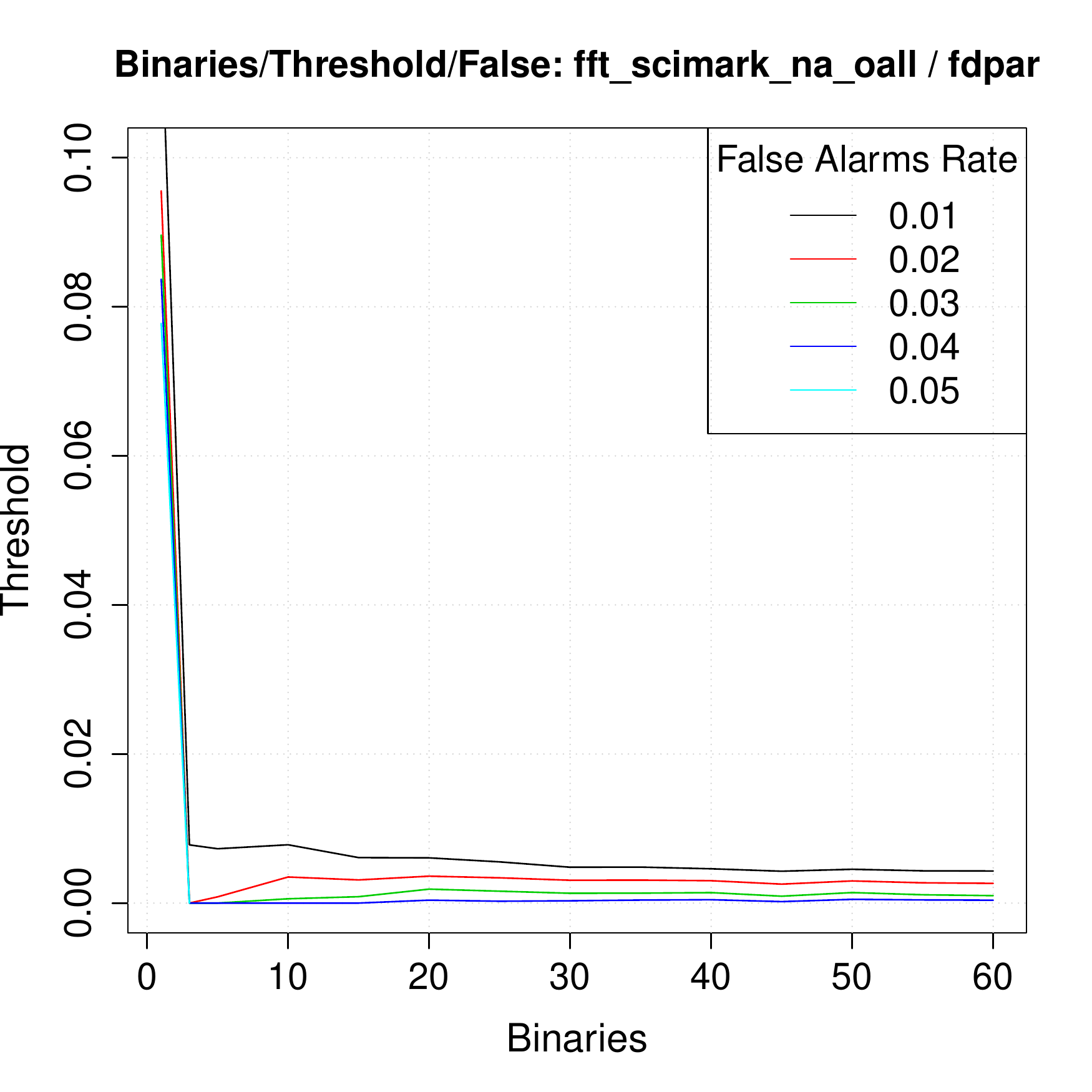}
}
\subfigure[Rijndael, RRR-Bootstrap.\label{fFARijndaelb}]{
  \includegraphics[width=0.45\linewidth,trim=0cm 0.5cm 1cm 1.5cm,clip=true]{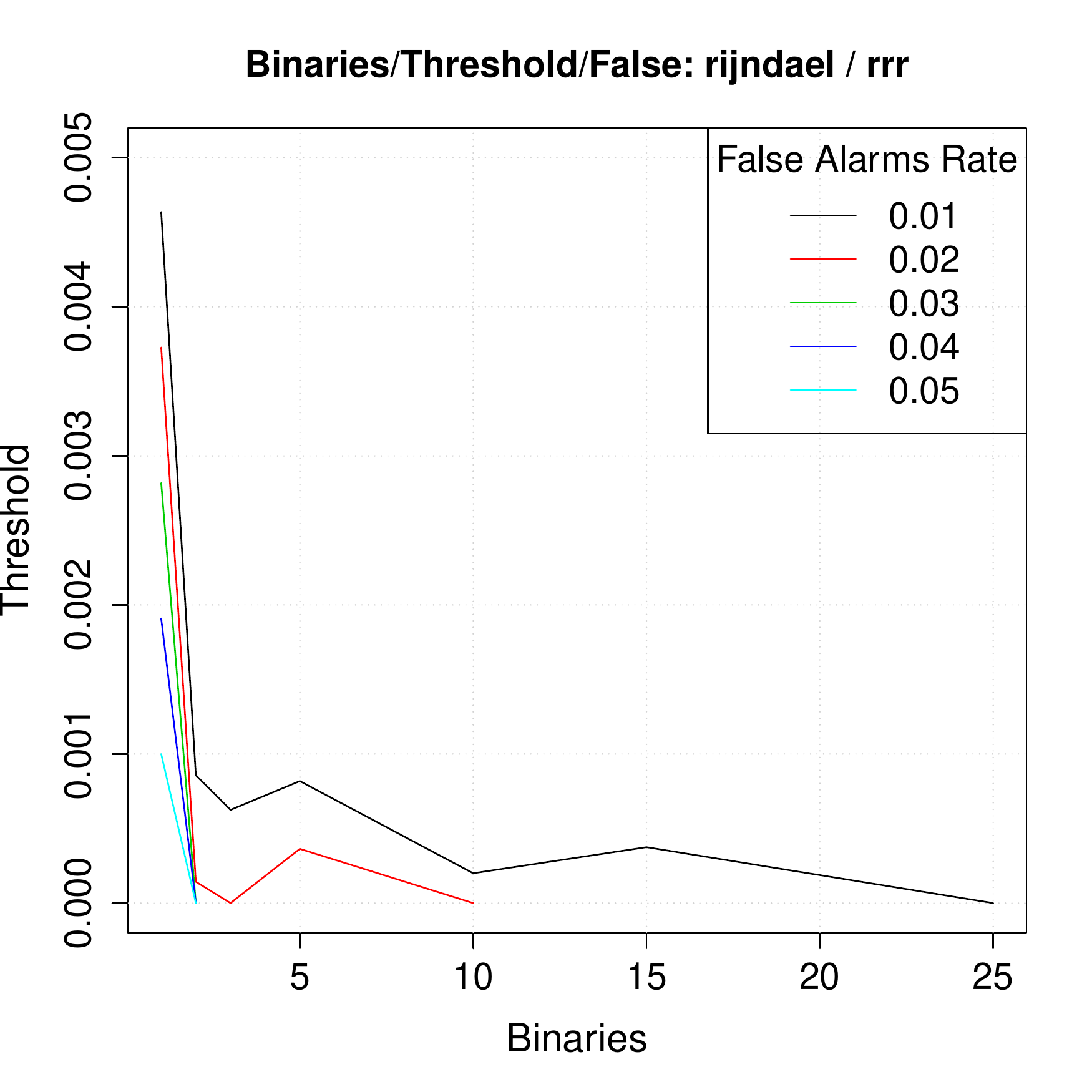}
}
\subfigure[Rijndael, Asymptotic.\label{fFARijndaela}]{
  \includegraphics[width=0.45\linewidth,trim=0cm 0.5cm 1cm 1.5cm,clip=true]{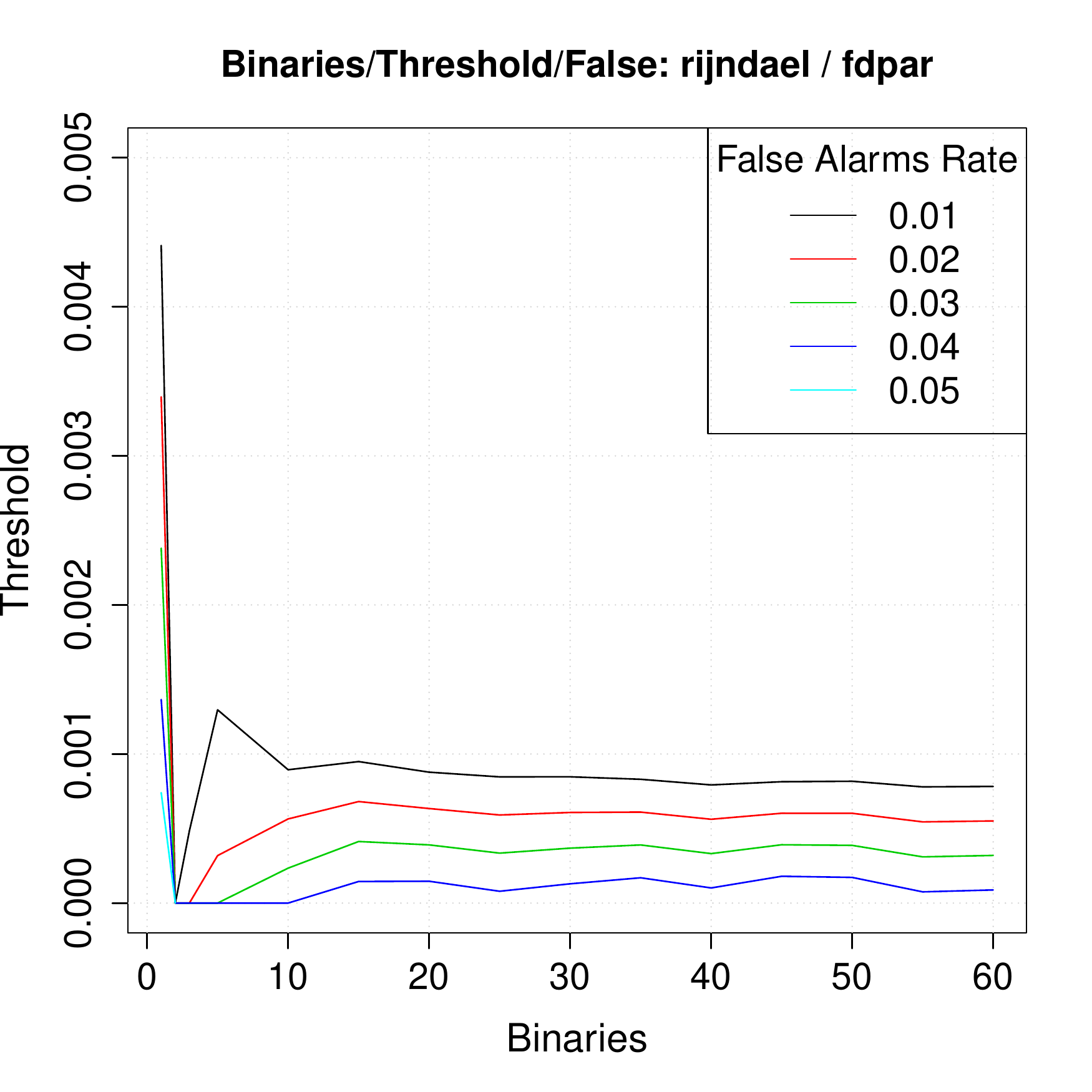}
}
\caption{False alarm rate depending on number of binaries and a threshold
for performance change.  Each curve is an isoquant contour, connecting the
minimal combinations of thresholds and binaries that lead to no more than a
given false alarm rate.  A smaller rate is better, a smaller number of
binaries is cheaper (better) and a smaller threshold is better.  Thus,
combinations to the top and right of a contour work as well, but none to the
bottom and left do.  Note that the shown thresholds for FFT are up to 10\%,
but only 0.5\% for Rijndael --- this means that for Rijndael, binaries
do not matter.}
\label{fFAFFT}
\end{figure}

Results for the FFT benchmark (NA OPT) are shown in Figures~\ref{fFAFFTb}
and~\ref{fFAFFTa}.  Figure~\ref{fFAFFTb} is for the RRR-bootstrap interval,
which is the default and best performing bootstrap, with replacement at all
three levels.  Figure~\ref{fFAFFTa} is for our asymptotic interval.  These
two-dimensional plots capture a three-dimensional function --- the lines
plotted are (isoquant) contour lines representing thresholds.  For example,
the red line connects all minimal combinations of binaries/thresholds that
give a false alarm rate of up to 2\%.
%
%\comment{added: }
%
In other words, the curves show how many binaries are needed to obtain less
than a certain false alarm rate for a given threshold of interest.  All
combinations to the top and right (higher thresholds and more binaries) of
the line then do as well, but none to the bottom and left do.  We construct
a 95\% confidence interval.  Hence, for large number of binaries, the false
alarm rate with 0\% threshold should converge to 5\%.  The light blue line
representing the 5\% false alarm rate should hence approach the $x$-axis as
the number of binaries increases.  The line can also disappear --- if the
false alarm rate gets below these 5\% for a given number of binaries and
remains there also for all larger numbers, it disappears in the graph at
that given number of binaries.

% This was with the biased estimator of variance... indeed better numbers,
% because wider intervals
%
%The plots show that the asymptotic method needs only 3 binaries to stay
%within the 5\% false alarm rate, for a 0\% threshold of interest.  Nearly 60
%binaries are, however, needed for 4\% of false alarms.  With the bootstrap
%method, we cannot reach the 0\% threshold (even with 60 binaries and 5\%
%false alarm rate).  The plots clearly show the importance of selecting a
%non-zero threshold for the comparison.  For a threshold of 2\%, we can reach
%a 1\% false alarm rate with only 10 binaries, using both methods.  Larger
%thresholds are even cheaper, and a threshold of 1\% already helps a lot.

The plots show that the false alarm rate with both methods and 0\% threshold
gets close to the 5\% false alarm rate with an increasing number of
binaries.  The asymptotic method seems better than bootstrap for small
numbers of binaries, say 2 to 20, but there is no practical difference for
larger numbers of binaries.  While being able to use less than 20 binaries
in experiments would indeed be desirable, the reason for a good false alarms
rate with the asymptotic method is a bad one, as we will show in
Section~\ref{sCoverage} --- the intervals are wider than they should be.
On the other hand, the plots show that we could use non-zero thresholds to
reduce the number of binaries needed.  For a threshold of 2\%, we can reach
a 2\% false alarm rate with only 3 binaries, using either method.  Larger
thresholds are even cheaper, and a threshold of 1\% already helps a lot.

Technical note: the plots are based on simulated measurements for selected
numbers of binaries and simulated comparisons for selected thresholds.  The
minimum threshold that works with a given number of binaries and a given allowed
false alarm rate is found using linear interpolation.  The plots also
include one binary only, which corresponds to a two-level experiment in
which compilation is not repeated at all (1-way classification model with
the asymptotic method).

Results for all the variants of the FFT benchmark are similar.  The `NA'
versions of FFT produce slightly fewer false alarms than the default, which
could be caused by less dependence in the data, but the difference is almost
negligible, and we did not attempt to verify the cause.  For all FFT
benchmarks, 3 binaries would be enough to obtain no more than a 5\% false
alarm rate with a threshold of 2\%, using either method.  The false alarms
rate with 0\% threshold converges to 5\% as it should.

Results for the Rijndael benchmark are shown in Figures~\ref{fFARijndaelb}
and~\ref{fFARijndaela}.  Using either method, 2 binaries are enough to get
about a 5\% false alarm rate even with threshold of 0\%.  With the asymptotic
method and the 0\% threshold, the false alarm rate converges to 5\% as it
should.  From 2 until about 30 binaries, the false alarm rate increases. 
With the bootstrap method, however, the false alarm rate is far too small
--- it is about 3\% for 5 binaries and then keeps decreasing until it gets
about 0.5\% for 60 binaries.  The reason for this is again the coverage, as
we will show in Section~\ref{sCoverage}.  Other non-FFT benchmarks behaved
similarly.  With 0\% threshold the false alarm rate converges to 5\% with
the asymptotic method, but to a much smaller value with the bootstrap
method.  The Ping benchmarks were more susceptible to non-deterministic
compilation than Rijndael with the bootstrap method --- more binaries were
needed for the same false alarm rate.  With 10 binaries and more, the
bootstrap method with 0\% threshold provides up to 5\% false alarms, with any
non-FFT benchmark.  A non-zero threshold here helps as well: with 1\%
threshold and 2 binaries, any non-FFT benchmark with either method provides
up to 1\% of false alarms.

With the asymptotic method, we have seen (rare) instances of the violation
of the condition in Lemma~\ref{lfieller} (Section~\ref{sFiellerRatio}), and
hence could not apply the method.  These instances were only with 2
binaries.  While this can in theory happen even for large numbers of
binaries, it becomes even less likely.  The bootstrap method does not have a
similar problem.

\paragraph{Summary} The results show that choosing a non-zero threshold for
comparison, even a small one (say 1\%), can drastically reduce the number of
binaries (and hence experimentation time) needed to get a given false alarms
rate.  The results also show that the asymptotic and the bootstrap methods
differ in the false alarm rate even of large numbers of binaries and
non-FFT benchmarks (we find the cause in Section~\ref{sCoverage}).  The
results show that increasing the number of compilations does not always
reduce the false alarm rate (again we find the cause in
Section~\ref{sCoverage}).

\subsubsection{Ignoring Non-deterministic Compilation}

\begin{figure}
\centering
\subfigure[FFT NA OPT, Asymptotic.\label{fFA2LFFT}]{
  \includegraphics[width=0.45\linewidth,trim=0cm 0.5cm 1cm 1.5cm,clip=true]{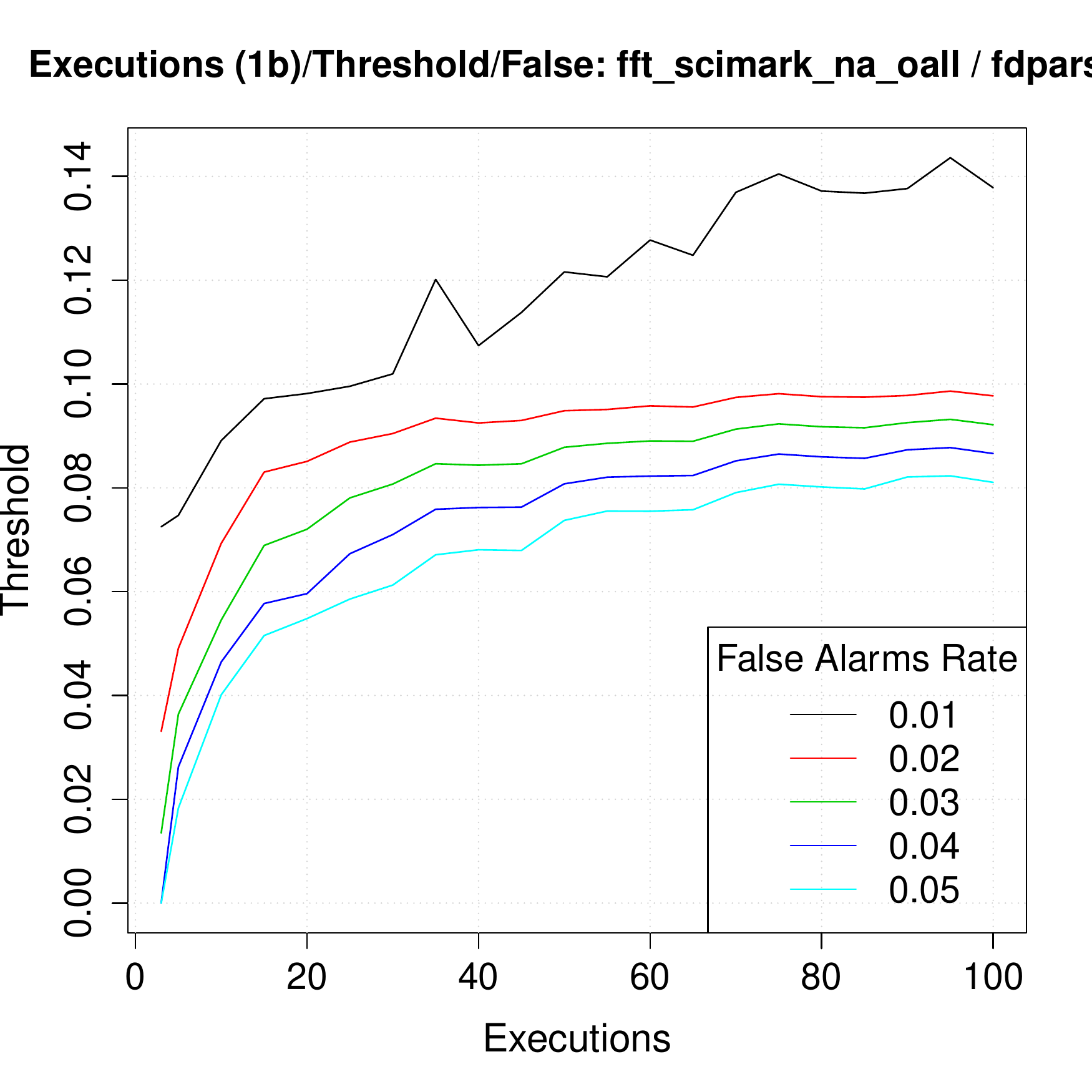}
}
\subfigure[TCP Ping OPT, Asymptotic.\label{fFA2LTCP}]{
  \includegraphics[width=0.45\linewidth,trim=0cm 0.5cm 1cm 1.5cm,clip=true]{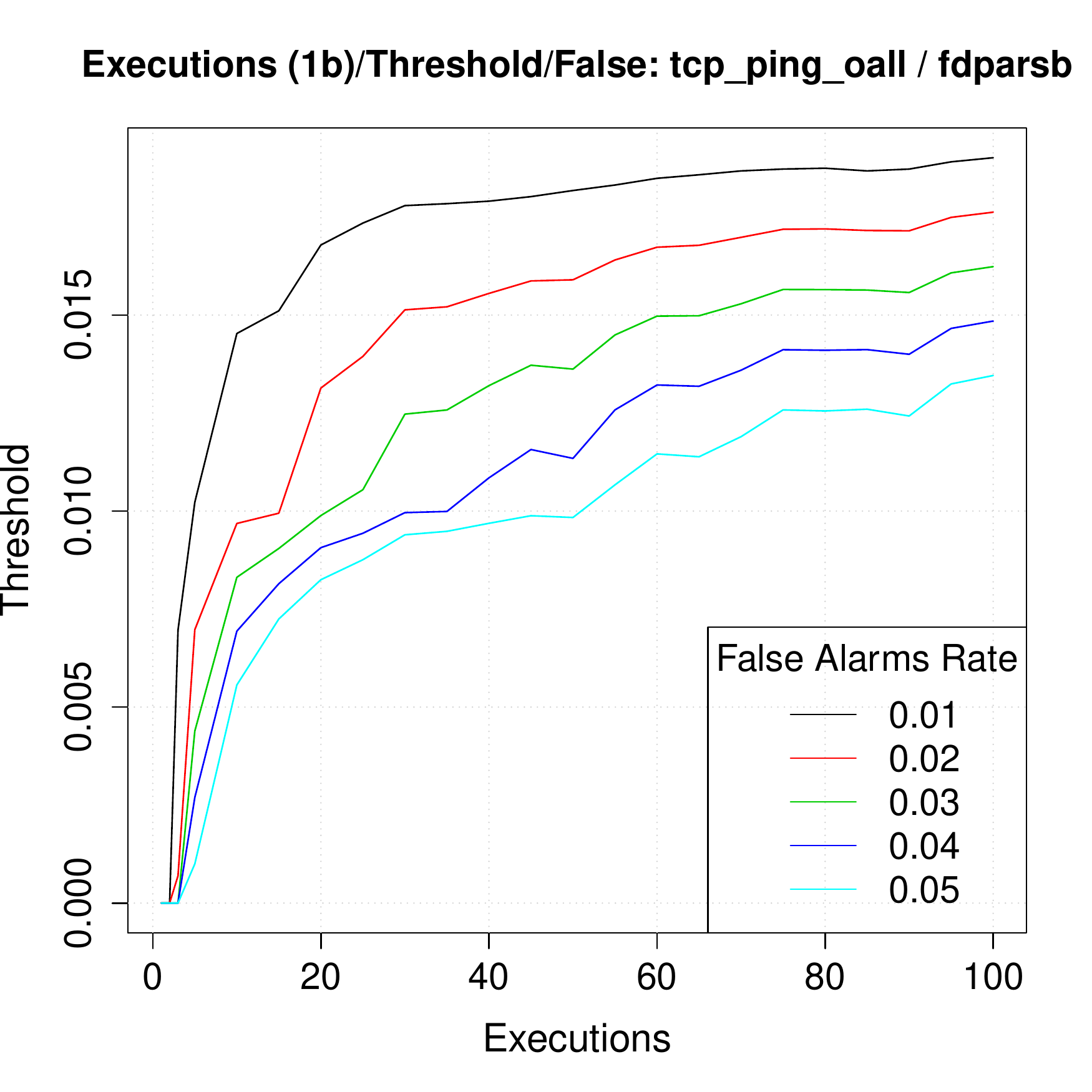}
}

\caption{False alarm rate depending on the number of executions and a
threshold for performance change.  These numbers represent a 2-level
experiment, in which non-deterministic compilation is ignored.  However, it
is not ignored in the simulation, and hence an increased sample size leads
to a high false alarm rate with some benchmarks.  Note that the thresholds
needed for TCP Ping are nearly 8$\times$ smaller than those needed for FFT. 
}

\label{fFA2L}
\end{figure}

Running experiments for multiple builds of the system is expensive,
while Table~\ref{tVariations} shows that the variation due to
non-deterministic compilation is often small.  Hence we quantify in more
detail the impact of using only 1 binary on the false alarm rate.  Results
for FFT (NA OPT) and TCP Ping (OPT), with asymptotic method, are shown in
Figure~\ref{fFA2L}.  For both benchmarks, the false alarm rate increases
with the increasing number of executions.  With a zero threshold, the false
alarm rates are very high (nearly 60\% for 100 executions).  Selecting a
non-zero threshold, however, helps.  With TCP (OPT), we get a 5\% false
alarm rate with a threshold below 2\% for up to 100 executions.  With FFT
(NA OPT), the threshold would have to be nearly 9\%.  We also calculated a
value for one execution only, that is for a flat experiment where even
non-determinism due to execution is ignored (not seen in the plots).
With FFT (NA OPT) and the flat experiment, the false alarm rate with a 0\%
threshold is about 95\%, and it is expected to be very high.  A 20\%
threshold gets the rate down to about 9\%.  With TCP (OPT), the false
alarm rate in a flat experiment is negligible, below 0.1\% already for a 0\%
threshold.  This is in line with Table~\ref{tVariations}, where the $T^2_2$
estimate (executions) is zero and $T^2_3$ (measurements) is negligible.

The results are similar for all FFT benchmarks, except that `NA' versions do
a bit better than non-`NA' (note that the variation due to non-deterministic
compilation is smaller for `NA' versions).  All the non-FFT benchmarks show
similar trends, only the false alarm rates differ corresponding to different
variations (Table~\ref{tVariations}).  We can get a 5\% false alarm rate
with up to 100 executions with a 1\% threshold and Rijndael or HTTP
benchmarks.  And also with 2\% threshold and TCP benchmarks.

Results for RRR bootstrap look similar with the FFT benchmarks, except that
the bootstrap false alarm rates are slightly higher, and much higher for
smaller numbers of executions.  With the bootstrap, the rate is smallest at
5 executions (decreasing from 1 to 5, then increasing to 100).  With the
asymptotic method, it is increasing from 3 executions.  For 2 executions, we
discovered a violation of the condition in Lemma~\ref{lfieller}, and hence did
not check the trend.  For the HTTP and particularly the TCP Ping benchmarks, the
RRR bootstrap false alarm rates are significantly smaller than the
asymptotic ones, but the trends are the same.  We will see in
Section~\ref{sCoverage} that this is because the bootstrap method gives too
wide confidence intervals for these benchmarks (their actual coverage is
above 95\%).  For the Rijndael benchmarks, the bootstrap method provides
higher rates (is worse) for small number of executions, but comparable to
the asymptotic method or better for large number of executions.

\paragraph{Summary} It is indeed possible and desirable to measure only a
single binary with benchmarks where the variability due to non-deterministic
compilation is low, but a non-zero threshold and a reasonable number of
executions is necessary.  Oversampling by too many executions leads to
increased numbers of false alarms.

\subsubsection{Resampling with Replacement vs. Without}

Our bootstrap method uses replacement at all levels of the hierarchy in the
data.  While this is a natural default, it was not self-evident that it
would perform the best.  In some cases, it has been suggested that
resampling only at higher levels of the hierarchy, leaving the rest of the
data intact, is better~\cite{hboot,davison}.  We therefore compare
replacement at all levels (RRR) with replacement at the two higher levels
(RRN) and replacement only at the highest level (RNN).  We also implement a
na\"ive solution of flat resampling with replacement (FLAT), where
measurements from all executions and binaries are joined into a single set
from which they are selected, ignoring the hierarchy and thus losing the
original structure of the data.  This also tells us what to expect in cases
where a systematic source of non-determinism in the experiment has not been
identified, or the hierarchy of random effects has simply been ignored and
measurements were treated as if from one single execution, as is sometimes
the case in current practice.  The FLAT resampling, however, only
corresponds to non-determinism that happens in the repeated parts of the
experiment.  Any source of non-determinism above (not repeated) will result
in bias, instead.

\begin{figure}
\centering
\subfigure[FFT.\label{fFARFFT}]{
  \includegraphics[width=0.45\linewidth,trim=0cm 0.5cm 1cm 2cm,clip=true]{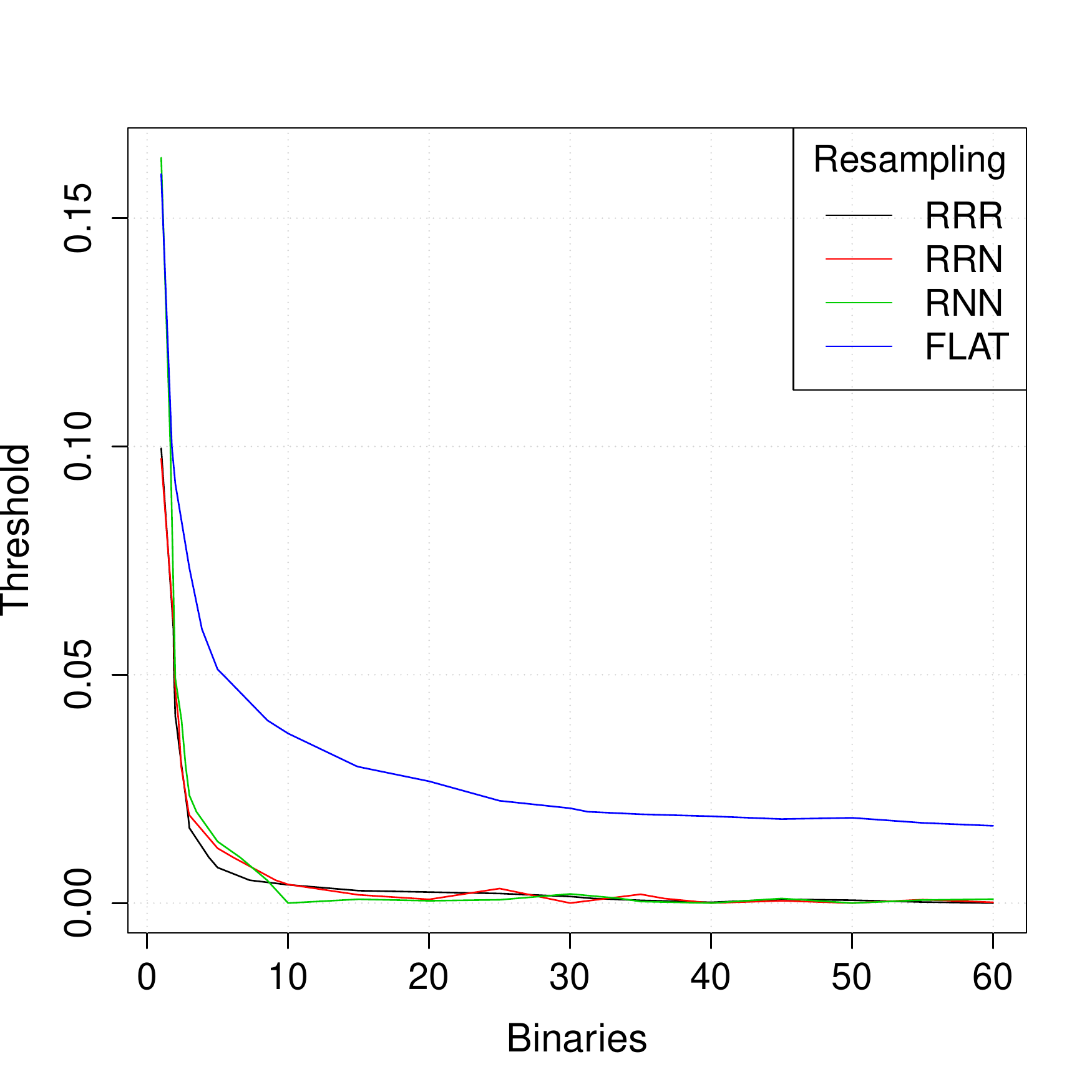}
}
\subfigure[TCP Ping.\label{fFARTCP}]{
  \includegraphics[width=0.45\linewidth,trim=0cm 0.5cm 1cm 2cm,clip=true]{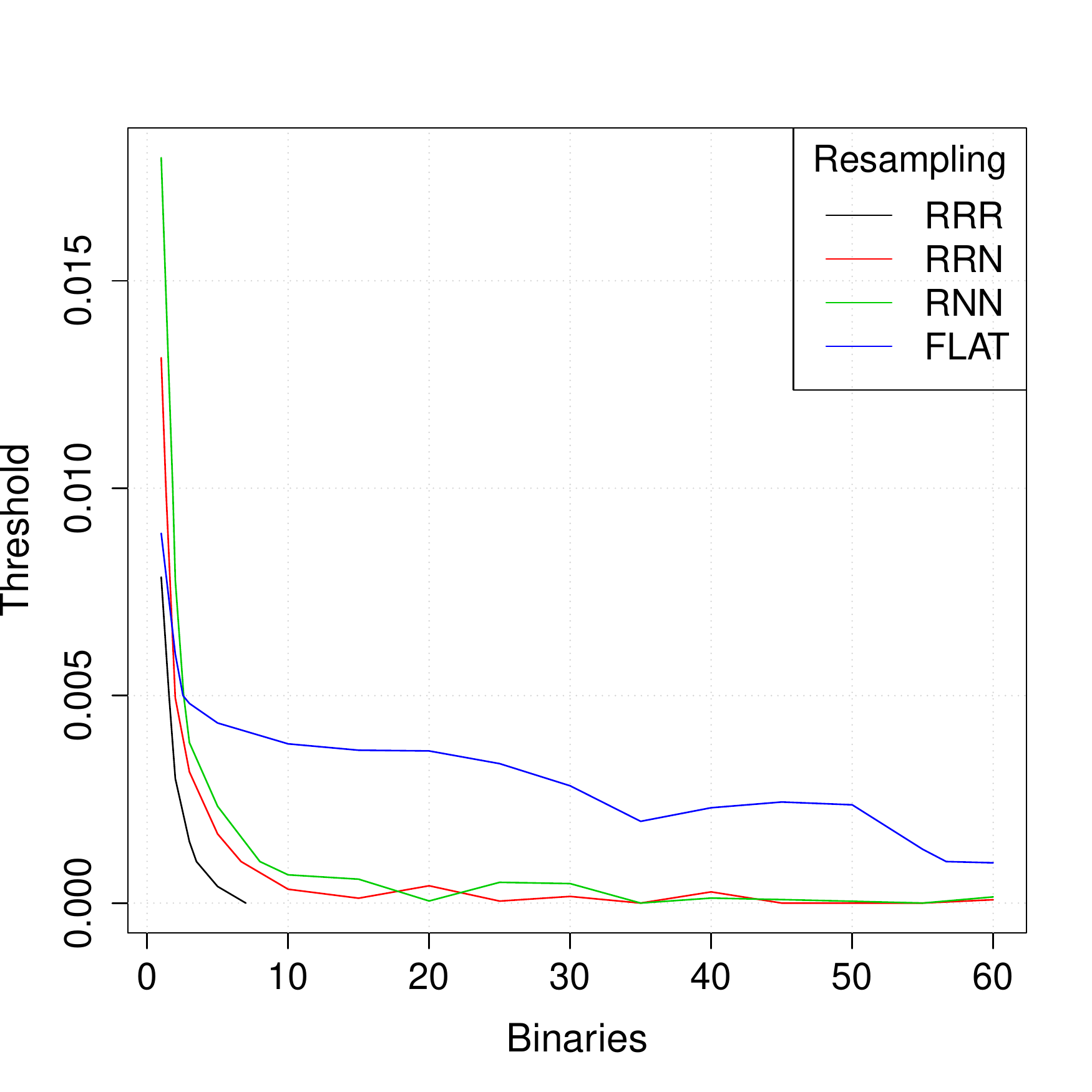}
}
\caption{Minimum numbers of binaries and minimum permissible thresholds for
a 5\% false alarm rate with different resampling techniques and the
bootstrap method.  A smaller number of binaries and a smaller threshold are
better.  Note that the thresholds needed for TCP Ping are about 10$\times$
smaller than those needed for FFT.}
\label{fFABoot}
\end{figure}

The results are shown in Figure~\ref{fFABoot} for the FFT benchmark (default
settings) and the TCP Ping benchmark.  Flat resampling is clearly the worst
for both benchmarks, though for TCP it is probably still good enough, as it
increases the comparison threshold needed by less than 0.5\%.  This is no
surprise as TCP has no variation at the second level (executions) and very
small variation at the third level (compilations).
All of RRR, RRN, and RNN perform about the same for FFT.  For TCP Ping, RRR
seems somewhat better than RRN and RNN, but the corresponding difference in
possible threshold is negligible.  RRR ought to be better, as the other
methods leave the measurements intact, while measurements are the key source
of variation in this benchmark.  All of these observations also apply to
other benchmarks of their class (FFT, non-FFT).  In the Rijndael benchmarks,
though, the RRN resampling gets close to RRR.  Note that in contrast to the
Ping benchmarks, Rijndael's variation is by and large due to
non-deterministic execution.  In summary, RRR seems as good a choice as any,
does not perform worse than other methods, and we use it in all other
experiments.

Note that flat resampling is particularly bad for the FFT benchmarks, which
are susceptible to random effects at higher levels of the hierarchy
(Table~\ref{tVariations}).  To reliably evaluate performance changes for an
FFT benchmark with a 5\% false alarm rate, one would need a threshold of 2\%
and with it as many as 35 binaries.  Note that RRR needs only 3 binaries
under the same conditions, which means more than a tenfold reduction in
experimentation time.  On the other hand, for non-FFT benchmarks, selection
of the resampling method has no practical impact, and even flat resampling
is acceptable.  

\paragraph{Summary} RRR resampling seems to be a safe choice. Replacement
should be done at all the levels of the real experiment.

\subsection{Coverage}
\label{sCoverage}

%\comment{We've mentioned coverage quite a bit before. We should probably explain
%it a little somewhere earlier, maybe with a forward reference to here.}
%
% done -to some extent 

The false alarms evaluation so far provides only part of the story. A method
that would always give a confidence interval for performance change of say
0\% $\pm$ 0.5\% no matter what the data were would score best, as it would
have no false alarms.  Ideally, we would like to validate that the intervals
our method produces are really 95\% confidence intervals for the ratio of
means --- i.e.\ that they include the ratio in 95\% of cases (their coverage
is 95\%).  The consequences of improper coverage, apart from simply lack of
precision in the reported uncertainty estimate, depends on how the quantified
performance change is to be used.  If it is to support, say, introduction of
a new optimisation, which seems fairly common, the conservative default
is to make the performance change seem smaller (or even zero), and hence
smaller coverage is a problem, but higher coverage does not matter so much. 
If, on the other hand, the goal is, say, to show that a new feature has
small overhead, the conservative default is to make the overhead look big,
and hence higher coverage is a problem, but smaller coverage does not matter
so much.

For our evaluation of coverage, we need to know what the true, but normally
unknown, ratio of means is, and hence we need to assume a particular model
of the data.  Moreover, it has to be a simple model, so that we know the
effect size based on the model parameters.  Here we use a hierarchical
normal model: measurements within an execution are independent identically
distributed with means that also come from a normal distribution.  At each
level of the experiment, the distribution of the means for the lower level
is normal.  The grand mean ($\overline{Y}$) is then the same as the mean of
the means at highest level ($\mu_B$ or $\mu_{n}$ in $n$-way classification,
Section~\ref{sModel}).

The parameters of the model are the mean $\mu=\mu_{n}$ at the highest level
and the variances at all levels, $\sigma_i$, $1 \le i \le n+1$.  For the
experiment, we choose a 5\% true performance improvement, that is $\theta =
^{N}\!\!\!\mu / ^{O}\!\mu = 0.95$.  We use the same variances for both
systems.  Hence, the parameters of the experiment are $^{O}\!\mu$ and
$\sigma_i$, $1 \le i \le n+1$.  Values of the parameters are likely to
influence the resulting coverage, so we have to explore multiple selected
combinations.  To make the selection realistic, we feed the model with
parameters estimated from the measured data, for each benchmark.  We use the
$S^2_i$ estimates for the variances and use three levels for all benchmarks.

Our evaluation is a statistical simulation of experiments. In each iteration
we generate synthetic data for the two systems to compare --- we simulate
binary means, execution means, and finally individual measurements.  On the
simulated data of the two systems we apply our quantification methods, the
(RRR) bootstrap and the asymptotic method.  We report the ratio of cases in
which the constructed confidence interval includes the real effect size of
$\theta=0.95$.  This ratio, which is our estimate of the coverage of the
interval, should be 95\%, as we construct 95\% confidence intervals.
Note that in a way, the false alarm rate experiments also provided a
coverage estimate (with a 0\% threshold, the false alarm rate should have been
5\% with 95\% intervals, because the true effect size was $\theta=1$).  Here
we complement those experiments with coverages for $\theta=0.95$ (a real
change) and using synthetic data.  Hence, in contrast to the false alarm
rate experiments, the results now cannot be affected by deviations from
normality, initialisation noise, lack of independence, or by
heteroscedasticity.

\subsubsection{Coverage in Three-level Experiment}

\begin{figure}
\centering
\subfigure[RRR Bootstrap Confidence Interval.]{
  \includegraphics[width=0.45\linewidth,trim=0cm 0.5cm 1cm 2cm,clip=true]{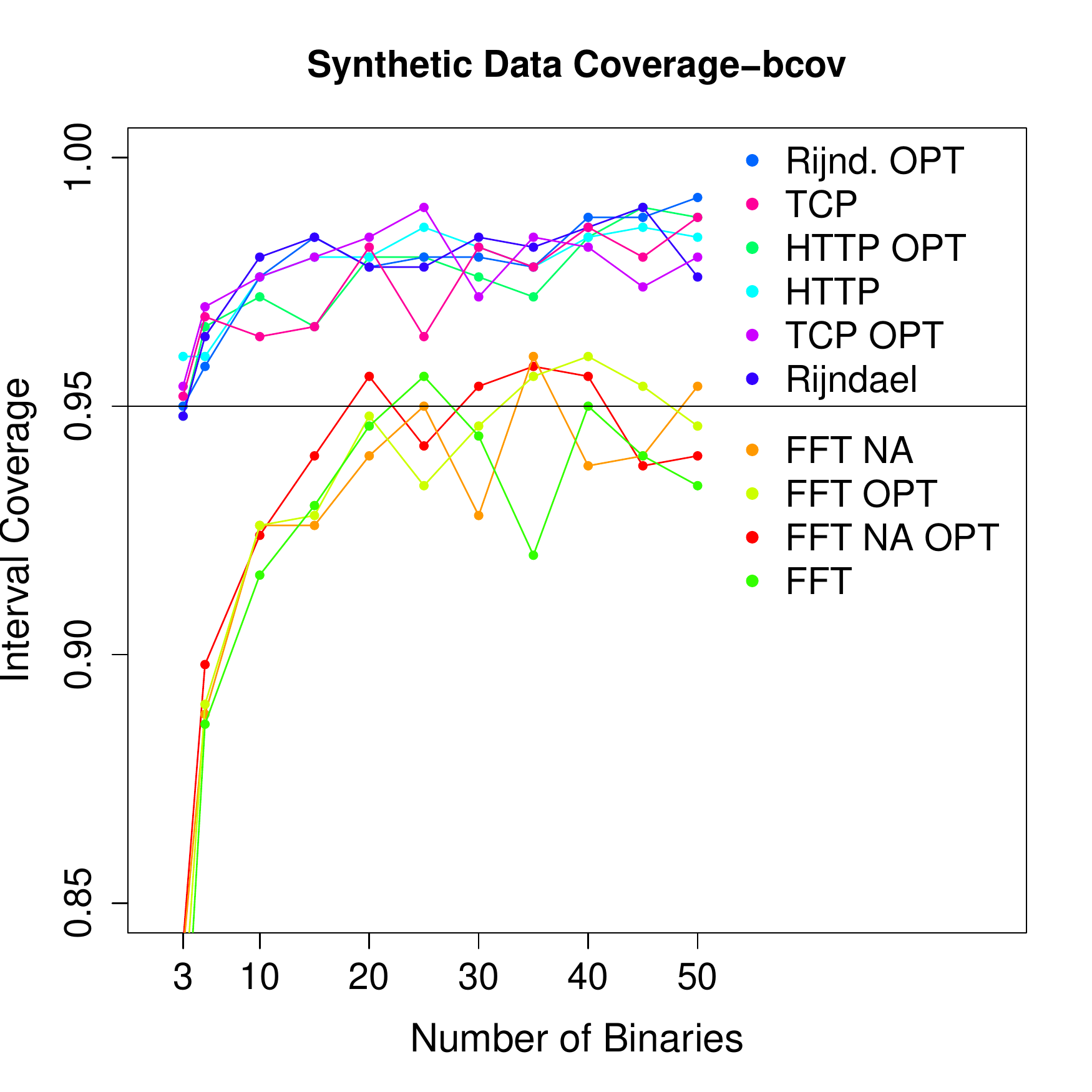}
}
\subfigure[Asymptotic Confidence Interval.]{
  \includegraphics[width=0.45\linewidth,trim=0cm 0.5cm 1cm 2cm,clip=true]{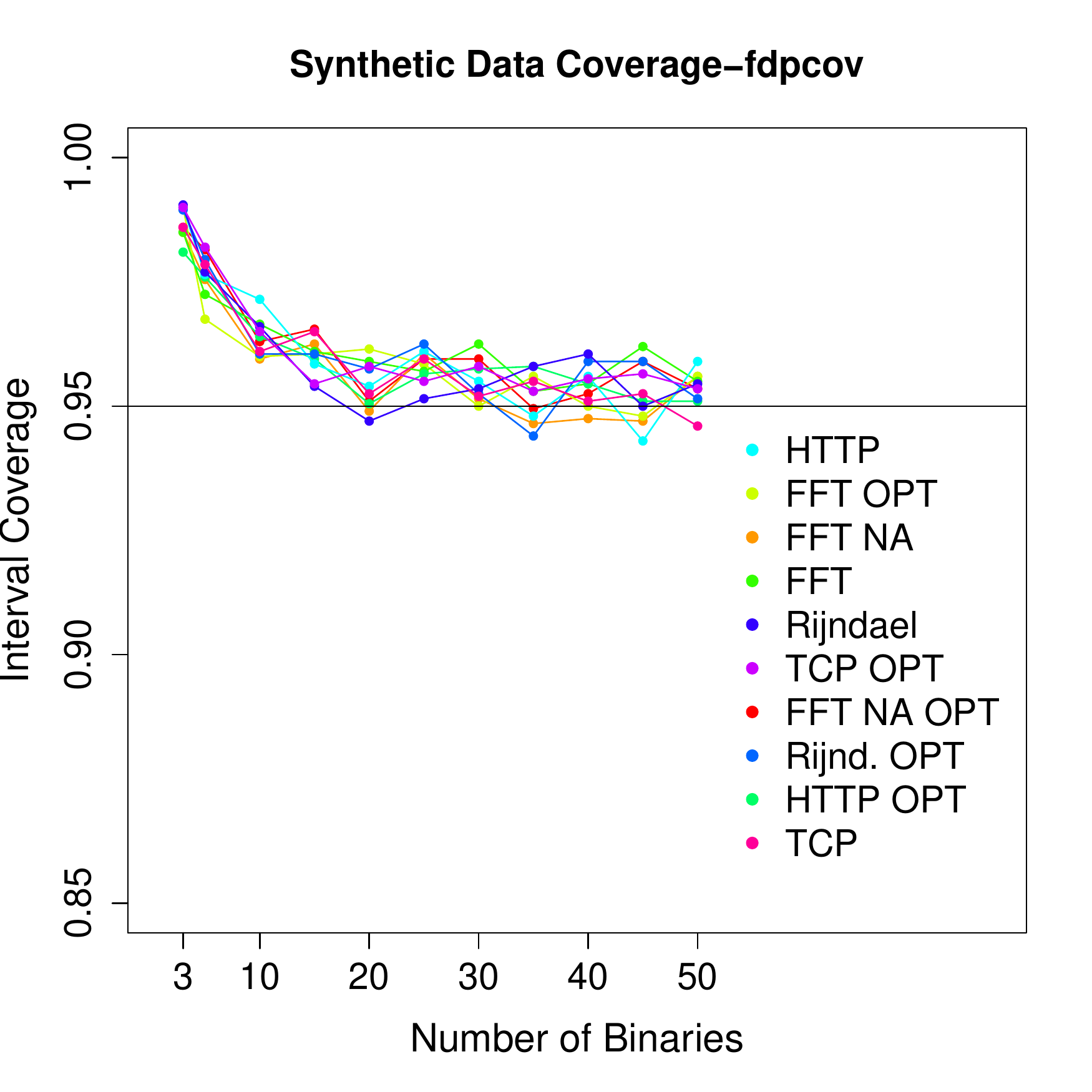}
  \label{fCOVasy}
}

\caption{Actual coverage of a 95\% confidence interval for the ratio of
means, estimated in a hierarchical normal model, fed by estimated parameters
from actual benchmarks.}
\label{fCOV}
\end{figure}

Results of the simulation for a large number of samples at all levels (100
at each level) are shown in Figure~\ref{fCOV}.  
%
% The intervals are thus wider then they need to be, which is safe
%(does not generate false alarms \comment{Is this the right --- or a natural
%--- definition of safe?  Safety seems to suggest not missing true alarms
%rather than worrying about false ones.}), but reduces the power of
%comparisons (fewer differences would seem significant, for a given
%threshold, than projected).  
%
% in statistics, the error of missing a true alarm is regarded as less of a
% problem than creating a false alarm ; significance methods are based on
% this ; and in studies when we want to show that our optimisation is better
% then it is safe ; true indeed, in studies where we want to show that our
% feature has low overhead, this is not safe
%
% added few sentences to that effect to the beginning of the section
%
Coverages with the asymptotic method, Figure~\ref{fCOVasy}, converge to the
projected 95\% as the number of binaries increases.  The coverages are
similar for different benchmarks.  They are about 99\% for 3 binaries, below
98\% for 10 binaries, and below 97\% for 20 binaries.  For 50 binaries they
are 95-96\%.  With the bootstrap method, coverages are different for FFT and
non-FFT benchmarks.  For FFT benchmarks, the coverages also converge to 95\%
as with the asymptotic method, although they are smaller for small numbers
of binaries.  The non-FFT benchmarks with the bootstrap method have high
coverage of about 95\% for just 3 binaries and their coverage further
increases up to 98-99\% for 50 binaries.

%It is not without interest that
%very similar over-conservative intervals for non-FFT benchmarks are also
%obtained with the asymptotic method in case that conservative $S^2_i$
%estimates are used for $\sigma^2_i$ to get variance estimate for the sample
%mean (not shown in the plot).

\begin{figure}
\centering
\subfigure[FFT NA OPT (from 94.8\% to 98.9\%).\label{fCOVGridFFT}]{
  \includegraphics[width=0.45\linewidth,trim=0cm 0.5cm 1cm 2cm,clip=true]{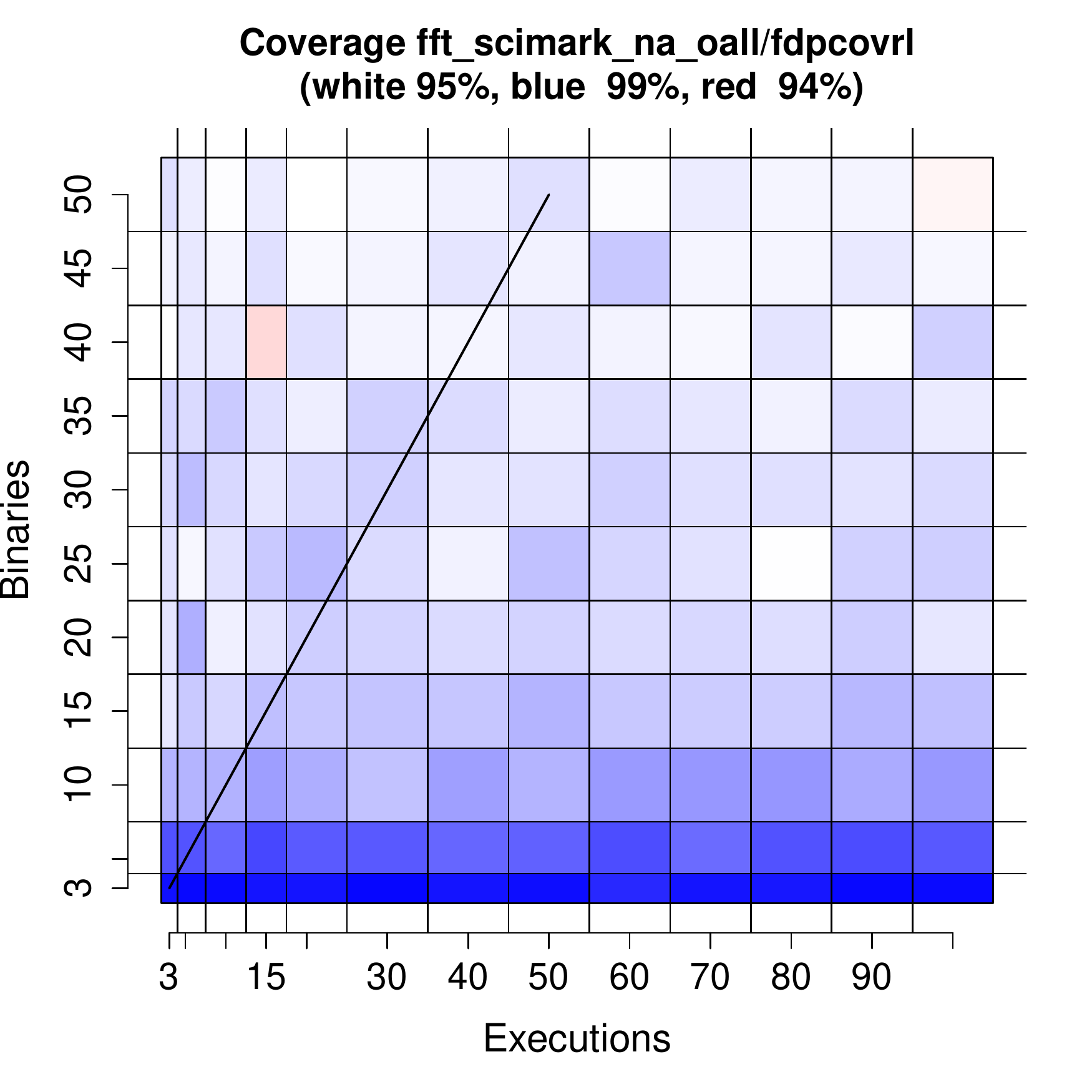}
}
\subfigure[Rijndael OPT (from 95\% to 98.9\%).\label{fCOVGridRijndael}]{
  \includegraphics[width=0.45\linewidth,trim=0cm 0.5cm 1cm 2cm,clip=true]{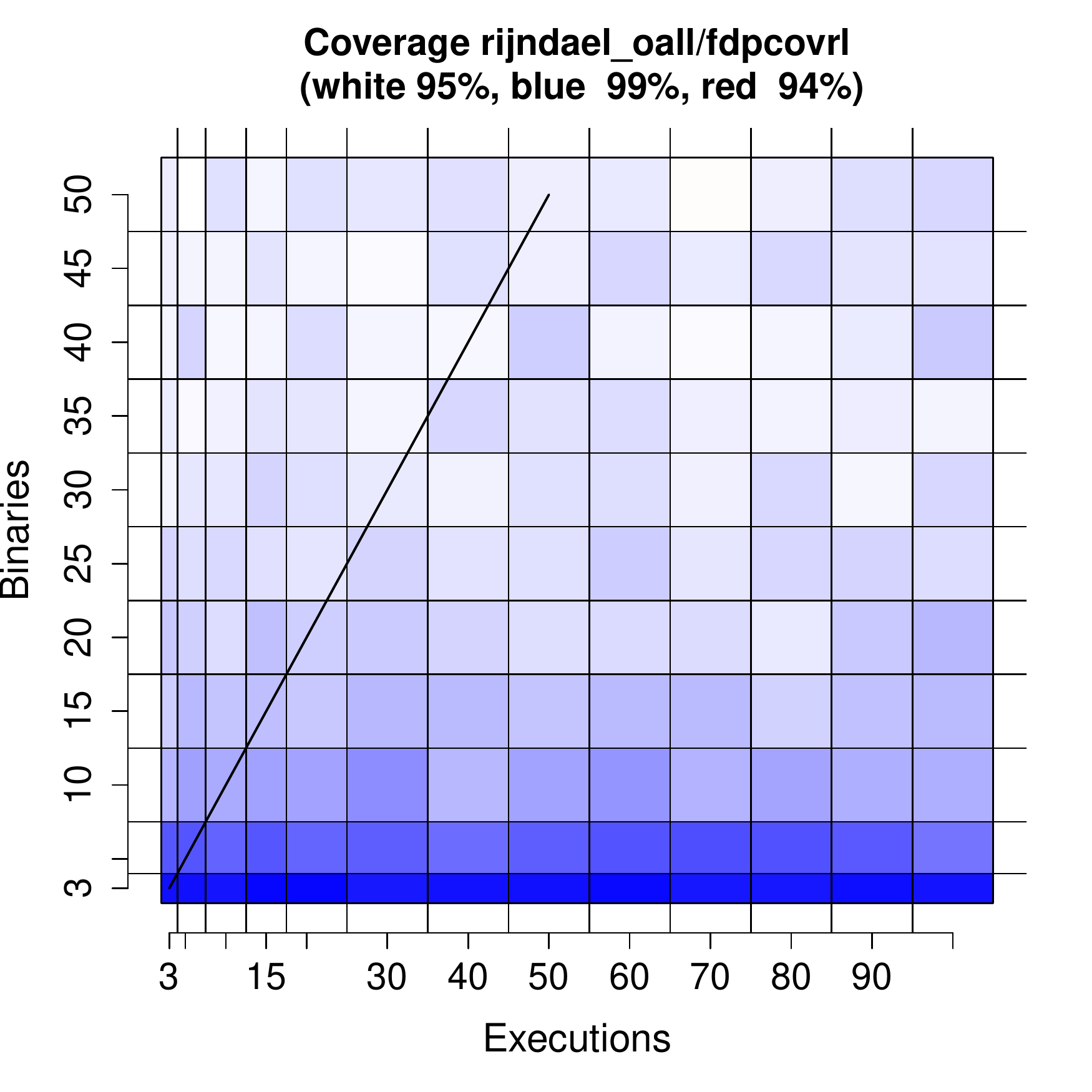}
}

\caption{Actual coverage of a 95\% asymptotic confidence interval for the
ratio of means, estimated in a hierarchical normal model, fed by estimated
parameters from actual benchmarks.  White is the 95\% coverage.  Red is
smaller coverage (darker is worse-smaller), blue is higher coverage (darker
is worse-higher).  Overall, white is ideal, darker is worse than lighter,
and red is usually worse than blue.  The diagonal line denotes the same
numbers of binaries and executions.  If printed in greyscale, note that the
plots are almost completely blue --- the coverage is only rarely below 95\%,
and if so, only slightly.}
\label{fCOVGrid}
\end{figure}

The coverages in Figure~\ref{fCOV} are all estimated for a large number of
executions per binary.  In practice, experimentation time is a precious
resource.  We hence looked if we could get similar coverages for smaller
numbers of executions, thus saving some experimentation time. 
Figure~\ref{fCOVGrid} then shows the results for for FFT (NA OPT) and
Rijndael (OPT) benchmarks.  We ran the experiment for all benchmarks, but
only with the asymptotic method as the bootstrap method would require too
high computation costs to evaluate.  In Figure~\ref{fCOVGrid}, the coverages
are colour coded.  White is the projected ideal coverage of 95\%.  Blue
denotes too high coverages, darker blue is worse (higher coverage).  Red
would denote smaller coverages, darker red would be worse (smaller
coverage).  However, in this plot, the coverage was always 95\% or only
slightly below, so no red colour is present.  The diagonal line shows cells
for the same numbers of binaries and executions.

Figure~\ref{fCOVGridFFT} shows that we can get close to the projected
coverage of 95\% with a high number of binaries, which is in line with
Figure~\ref{fCOV}.  Increasing the number of executions per binary does not
seem to impact the coverage.  Figure~\ref{fCOVGridRijndael} shows that the
Rijndael benchmark performs similarly.  The other benchmarks do as well. 
This suggests that the repetition count at the highest level is most
important for coverage.  Note that we have shown in Section~\ref{sPlanning}
that the repetition at some lower level can, however, be important for
getting most precise results (narrowest intervals) within given
experimentation time.

%With
%the conservative biased estimates $S^2_i$ of the variances, the coverage of
%non-FFT benchmarks is always too large (not in the plot).  With these
%estimates, the coverage of FFT benchmarks is then too large for a high
%number of binaries and a small number of executions, but it improves as the
%number of executions increases.
%
%This is so because the
%denominator in the ratios that form the bias becomes larger.
%
% but why not for TCP, HTTP ? would the number of executions needed be 
% much larger ?

We also explored coverages with the asymptotic method, but using only the
normal approximation instead (``normal-asymptotic'') of the $t$-distribution
(``asymptotic'').  See Section~\ref{sFiellerRatio} for more details.  With
the normal-asymptotic method, the coverages are too small for small number
of binaries (around 88\% for 3 binaries), but converge to the projected 95\%
as the number of binaries increases.  They get above 90\% for 5 binaries and
above 94\% for 15 binaries.  Similarly to the asymptotic method, it is the
number of binaries that impacts (improves) the coverage, not the number of
executions.  We also ran the false alarms experiments with the
normal-asymptotic method.  The false alarm rate with 0\% threshold converged
to 5\% with all benchmarks.  As in practice a too-high coverage is often
worse than too-low coverage, it makes sense to use the asymptotic method
($t$-distribution) even in cases when the normality assumptions cannot be
made.

\paragraph{Summary} With the asymptotic method, the coverage is too large
for small numbers of binaries, but converges to the projected number as the
number of binaries increases. With the bootstrap method, the coverage of FFT
benchmarks is too low for small numbers of binaries, but then converges to
the projected number. The coverage of non-FFT benchmarks with the bootstrap
is always too large. With the asymptotic method but only the normal
approximations, the coverages also converge to the projected number (all
benchmarks), but are too small for small number of binaries. In most cases,
the asymptotic method would be the best choice. The number of executions
does not have an impact on the coverage.

\subsubsection{Ignoring Non-deterministic Compilation}

\begin{figure}
\subfigure[RRR Bootstrap Confidence Interval.\label{fCOVRBoot}]{
  \includegraphics[width=0.45\linewidth,trim=0cm 0.5cm 1cm 2cm,clip=true]{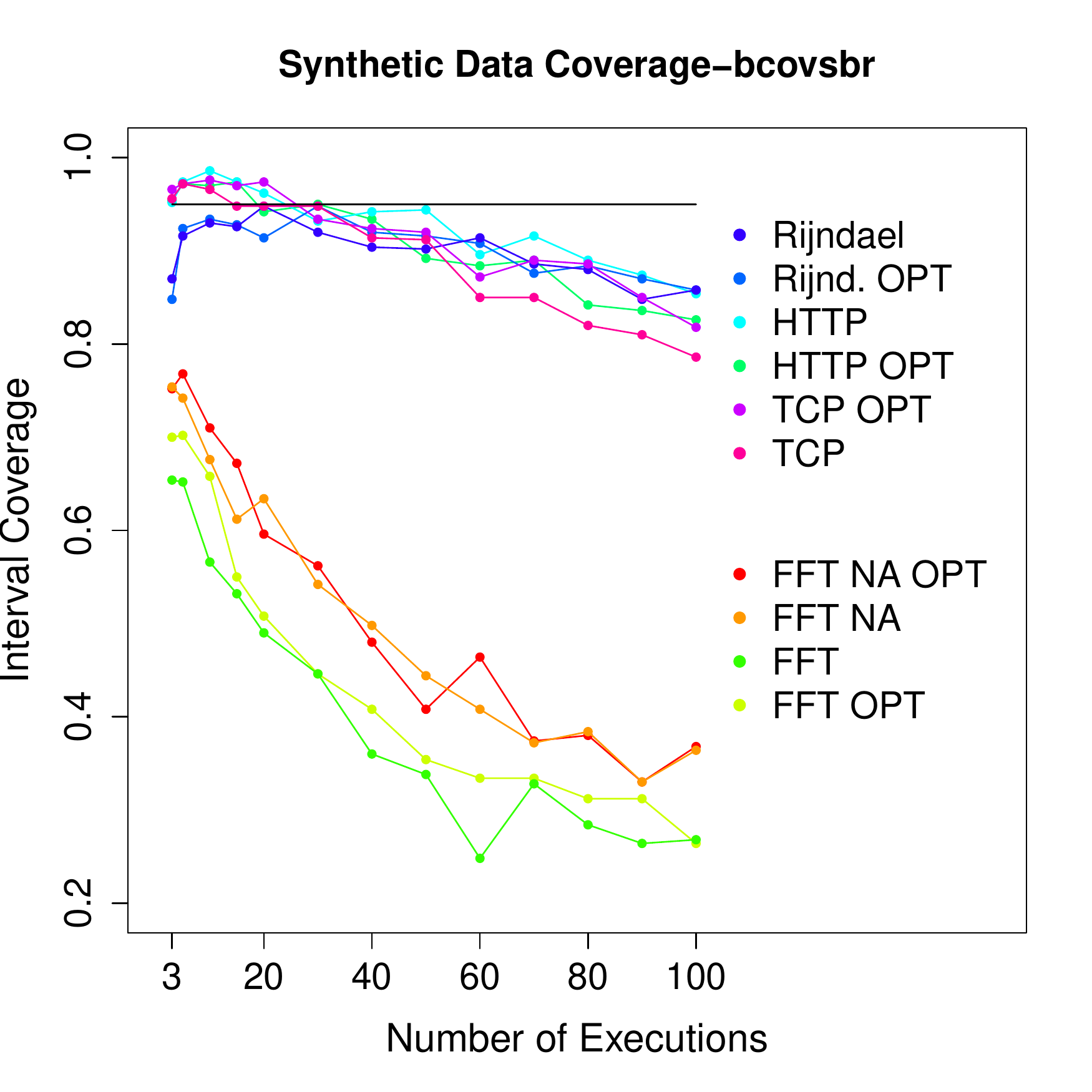}
}
\subfigure[Asymptotic Confidence Interval.\label{fCOVRPar}]{
  \includegraphics[width=0.45\linewidth,trim=0cm 0.5cm 1cm 2cm,clip=true]{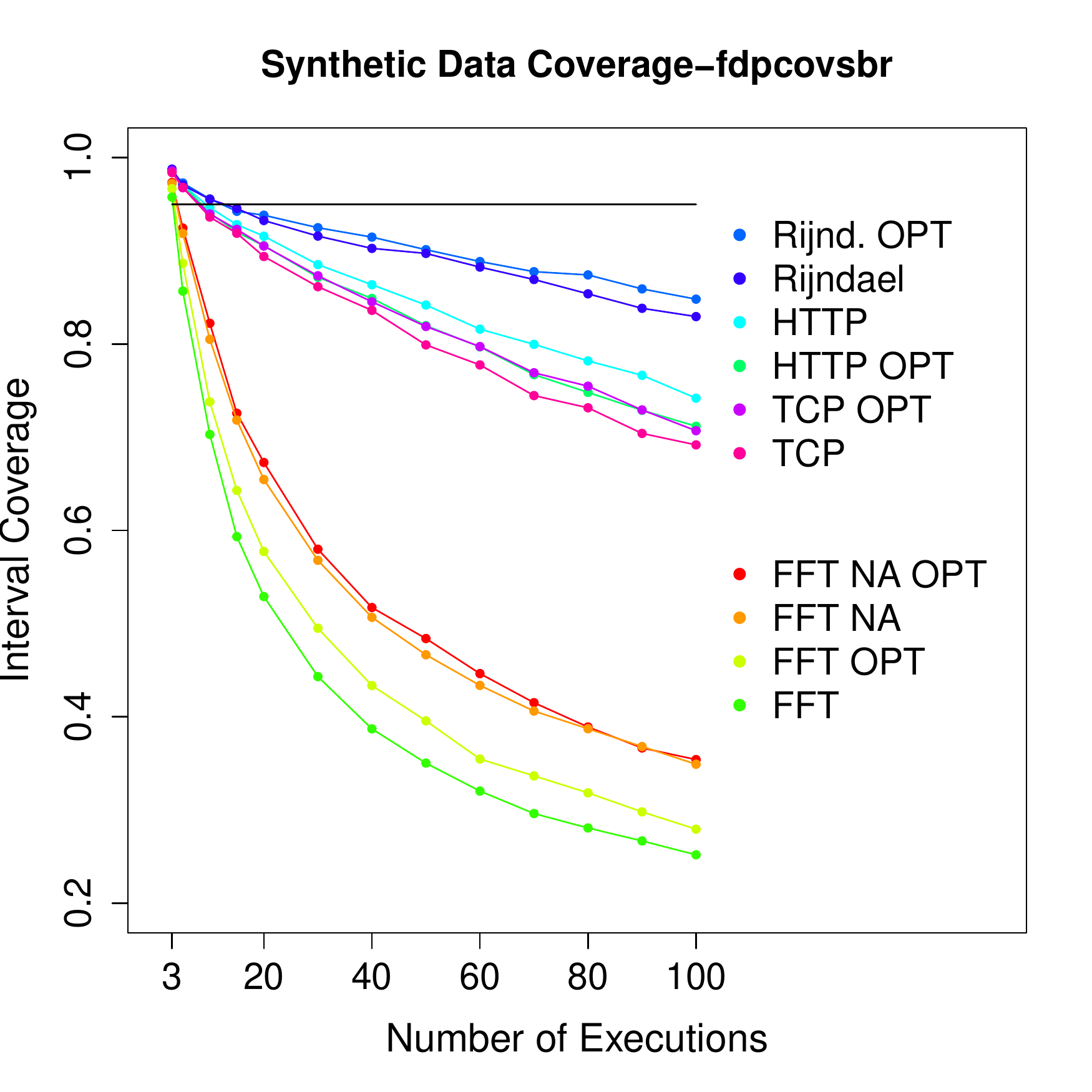}
}

\caption{Actual coverage of a 95\% confidence interval for the ratio of
means, estimated in a hierarchical normal model.  In contrast to
Figure~\ref{fCOV}, this plot simulates a 2-level only experiment (a scenario
in which the experimenter would choose to ignore uncertainty due to
compilation).  Closer to 0.95 is better.}
\label{fCOVR}
\end{figure}

With the non-FFT benchmarks, the performance variation due to
non-deterministic compilation is very small (1\%, Table~\ref{tVariations}
below). In practice, one would probably choose not to repeat compilations
with these benchmarks, but rather select an appropriate comparison
threshold.  In another simulation, we look at what the true coverage would
be if we choose to do this.  The experiment we simulate ignores variation
due to non-deterministic compilation, but the simulation does not. 
Figure~\ref{fCOVR} shows the results, for the asymptotic method and varying
numbers of executions.  The coverage decreases with increasing number of
executions.  This is expected, because as the intervals get narrower, they
are more likely to miss the true mean that takes many different binaries
into account.  This reduction is first good with the asymptotic method (all
benchmarks) and the bootstrap method (non-FFT benchmarks), because it
corrects for the over-coverage of these methods.  But, the coverage soon
gets unacceptably low and not only so with the FFT benchmarks, which are
prone to non-deterministic compilation to a high degree.  The coverage also gets
too low with an unduly large number of binaries with the non-FFT
benchmarks where this impact is very small.

\begin{figure}
\centering
\subfigure[FFT NA (from 34\% to 98\%).]{
  \includegraphics[width=0.45\linewidth,trim=0cm 0.5cm 1cm 2cm,clip=true]{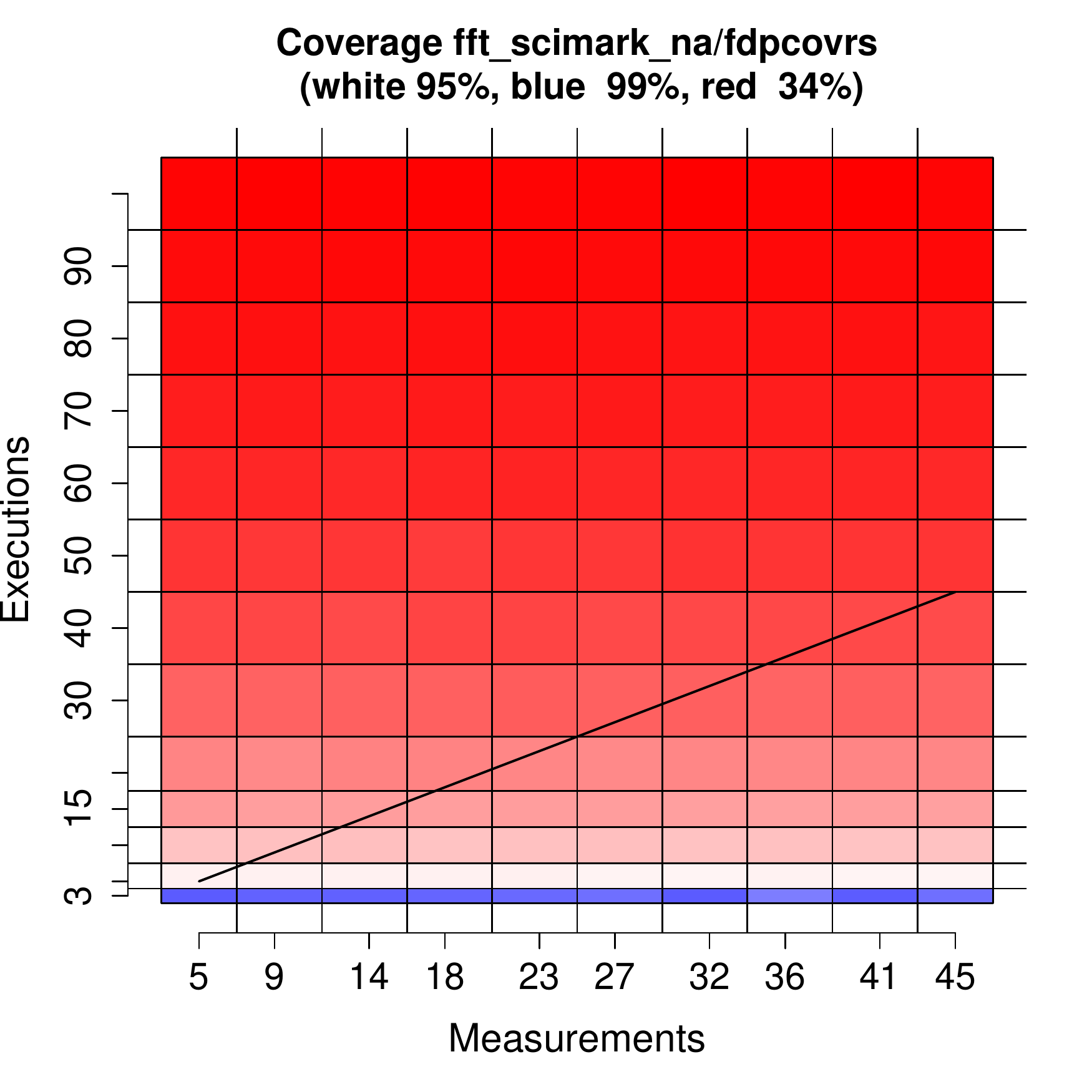}
}
\subfigure[HTTP OPT (from 72\% to 99\%)]{
  \includegraphics[width=0.45\linewidth,trim=0cm 0.5cm 1cm 2cm,clip=true]{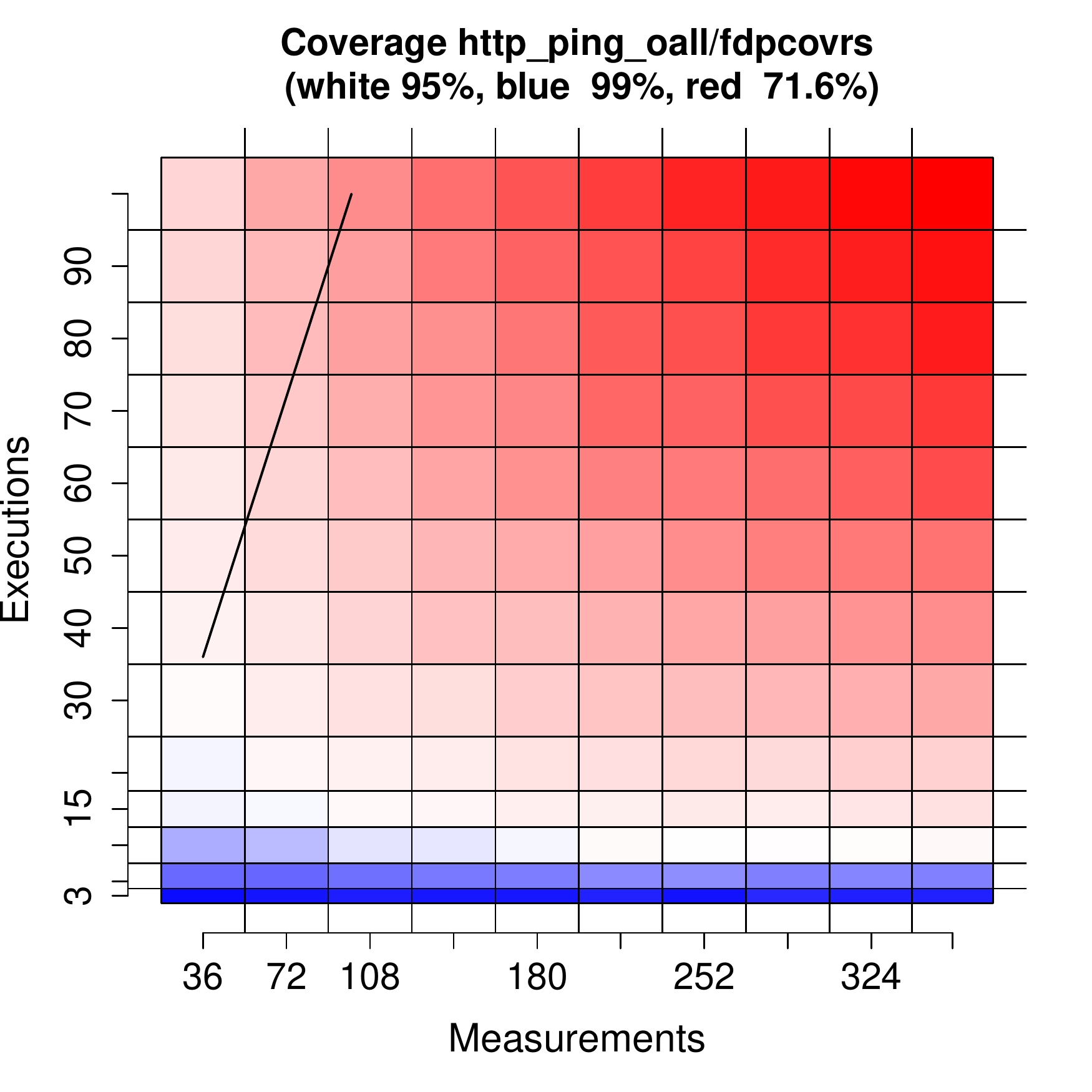}
}

\caption{Actual coverage of a 95\% asymptotic confidence interval for the
ratio of means, estimated in a hierarchical normal model.  In contrast to
Figure~\ref{fCOVGrid}, this plot simulates a 2-level only experiment (a
scenario in which the experimenter would choose to ignore uncertainty due to
compilation).  If printed in greyscale, note that the plots have a blue line
of cells at about 3 executions, while all cells above are red.  The tone of
the red colour does not match in the two plots --- the coverage of FFT is far
worse.  }
\label{fCOVRSGrid}
\end{figure}

We also looked at the trade-offs between the repetition counts for
measurements per execution and executions per binary, with respect to
coverage.  We ran the experiments with the asymptotic method only. 
Figure~\ref{fCOVRSGrid} shows the results for the FFT NA and HTTP Ping OPT
benchmarks.  With the FFT NA benchmark, the number of measurements per
execution do not seem to make a difference.  The too-high coverage gets
first better and then too small, as the number of executions increases, as
in Figure~\ref{fCOVRPar}.  Other FFT benchmarks behave similarly.  With the
HTTP Ping OPT benchmark, the number of measurements per execution matters
once the number of executions is high.  Increasing the number of
measurements makes the coverage worse, no matter from how many executions
these measurements are.  Other Ping benchmarks behave similarly, except that
the coverages are higher than with FFT.  The Rijndael benchmarks lie in
between --- increasing the number of measurements per execution makes the
coverage worse, but perhaps not as strongly as with the Ping benchmarks.
The differences between benchmarks can be explained with the variations at
different levels (Table~\ref{tVariations}).  In this experiment, anything
that makes the interval narrower makes also the coverage worse.  With FFT
benchmarks, variation due to non-deterministic execution is much larger than
that due to non-deterministic measurement, and hence the coverage decreases more
with increased executions. With the other benchmarks and particularly the
Ping benchmarks, variation due to execution is far smaller than that due to
measurement, and hence the number of executions is not important in reducing
the coverage. 

\paragraph{Summary} When non-deterministic compilation has minimal impact on
variability in performance, it makes sense to use only one binary.  However,
over-sampling (too many executions and/or measurements, depending on the
benchmark) can inflate the influence of ignored variability due to
compilation and result in a too low coverage.

\subsection{Dimensioning the Experiments}
\label{sDimensioning}

We compare the results precision that would be obtained with our method
against the default single-level method.  

Let us first focus on the FFT benchmark (descriptions of the benchmarks were
given in Section~\ref{sBenchmarks}) and let us assume we have a window of 6
hours for our experiments.  From Table~\ref{tVariations} we see that the
benchmark has high variation at all three levels, including compilation, and
hence multiple binaries have to be measured.

The usual one-level approach would execute each binary once, obtaining one
measurement (after warmup).  One execution with one measurement (and 19
warm-up measurements) takes about 5 seconds.  Compilation takes about 20
minutes (1200 seconds), so we can take about $6*3600/1205 = 17$ samples
within our 6 hours.  We would calculate the sample mean from these 17
samples and provide an asymptotic confidence interval based on t-test like
e.g.  in~\cite{lilja,jain,georges07}.  The half-width of such 95\% interval
would be about 4.7\% of the mean with the FFT benchmark ($\pm 4.7\%$), as we
can derive from Equation~\ref{eMeanNormality} on
page~\pageref{eMeanNormality} and the definition of the t-test:
$$
\left(1/\overline{Y}\right)t_{1-\frac{\alpha}{2}} \sqrt{var(\overline{Y})} =
\left(1/\overline{Y}\right)t_{1-\frac{\alpha}{2}}\sqrt{\frac{ \sigma_3^2 + \sigma_2^2 + \sigma_1^2}{17} }
$$
We substitute the sample mean and the unbiased estimates $T_i^2$ of
variances $\sigma_i^2$.

With the method we propose in this work, we dimension the experiment using
Equation~\ref{ecost} on page~\pageref{ecost}.  We would collect $n_1$
measurements from each execution and run each binary $n_2$ times, where:
$$
n_1=\left\lceil\sqrt{c_1\frac{\sigma^2_1}{\sigma^2_2}}\right\rceil \quad \textrm{and} \quad
n_2=\left\lceil\sqrt{\frac{c_2}{c_1}\frac{\sigma^2_2}{\sigma^2_3}}\right\rceil \\
$$
The costs are $c_1=19$ (19 warmup measurements) and $c_2=5343$ (number of
measurements that could roughly be done in 20 minutes, the time needed for
compilation).  Substituting the unbiased estimates $T_i^2$ of variances
$\sigma_i^2$, we get $n_1=4$ and $n_2=28$.  One execution will take within 6
seconds (4 measurements and warmup) and all executions of one binary will
take about 150 seconds, and so within 6 hours we can build and measure
$\left\lfloor 6*3600/1350\right\rfloor = 16$ binaries, obtaining 1792
samples.  Using Equation~\ref{eOSInterval} on page~\pageref{eOSInterval} we
would calculate a 95\% confidence interval for the mean, half-width of which
would about 2.3\% of the mean ($\pm 2.3\%$ instead of the $\pm 4.7\%$ with
the single-level method).

Interval half-widths for all benchmarks and for experimentation windows of
3,6, and 9 hours are shown in Table~\ref{tPrecisions}.  Our method would use
the 3-level model for all FFT and Rijndael benchmarks.  With HTTP and TCP
benchmarks, it would only execute each binary once, but collecting multiple
measurements.  It would not repeat executions because variation due to
execution in those benchmarks is small (and the $T_2^2$ estimate turns
negative).  Our method is always better than the default method (and by
design it should never be worse).  The benefits are particularly big when
the variation due to non-deterministic compilation is far below variations
at lower levels (see Table~\ref{tVariations}).  

Note that when running multiple benchmarks on different versions of the same
managed runtime, one can re-use the binaries of the runtime.  Even in this
case there would be a balance between spending time on compiling or on
running existing binaries, but the optimisation procedure would need to be
extended and will require a summarization technique over different
benchmarks.

%                    opt1 naive1 opt2 naive2 opt3 naive3
%fft_scimark_na_oall  3.3    7.5  2.0    4.6  1.5    3.6
%fft_scimark_na       3.2    7.1  1.9    4.4  1.5    3.4
%fft_scimark_oall     4.3    8.6  2.5    5.3  2.0    4.1
%fft_scimark          3.6    7.6  2.3    4.7  1.8    3.7
%http_ping_oall       0.5   18.8  0.3   11.5  0.3    9.1
%http_ping            0.4   18.0  0.3   11.1  0.2    8.7
%rijndael_oall        1.3    8.3  0.2    5.1  0.1    4.0
%rijndael             0.6    8.1  0.2    5.0  0.2    3.9
%tcp_ping_oall        1.3   34.9  0.8   21.5  0.6   16.9
%tcp_ping             1.3   32.2  0.8   19.8  0.6   15.5

\begin{table}
\tbl{Result Precision with Our Method and the Default Method\label{tPrecisions}}{
\begin{tabular}{|r|r@{.}l@{\%\,/\,}r@{.}l|r@{.}l@{\%\,/\,}r@{.}l|r@{.}l@{\%\,/\,}r@{.}l|}
\hline
& \multicolumn{12}{c|} { 95\% Confidence Interval Half-width (smaller is better) } \\
& \multicolumn{4}{c|}{ 3 hours exp. [our/default]} & \multicolumn{4}{c|}{ 6 hours exp. [our/default]} & \multicolumn{4}{c|}{ 9 hours exp. [our/default]} \\
\hline 
FFT NA OPT   & 3&3 & 7&5\% &   2&0 & 4&6\% &   1&5 & 3&6\% \\
FFT NA       & 3&2 & 7&1\% &   1&9 & 4&4\% &   1&5 & 3&4\% \\
FFT OPT      & 4&3 & 8&6\% &   2&5 & 5&3\% &   2&0 & 4&1\% \\
FFT          & 3&6 & 7&6\% &   2&3 & 4&7\% &   1&8 & 3&7\% \\
HTTP OPT     & 0&5 &18&8\% &   0&3 &11&5\% &   0&3 & 9&1\% \\
HTTP         & 0&4 &18&0\% &   0&3 &11&1\% &   0&2 & 8&7\% \\
Rijndael OPT & 1&3 & 8&3\% &   0&2 & 5&1\% &   0&1 & 4&0\% \\
Rijndael     & 0&6 & 8&1\% &   0&2 & 5&0\% &   0&2 & 3&9\% \\
TCP OPT      & 1&3 &34&9\% &   0&8 &21&5\% &   0&6 &16&9\% \\
TCP          & 1&3 &32&2\% &   0&8 &19&8\% &   0&6 &15&5\% \\
\hline
\end{tabular}
}
\begin{tabnote}
\Note{Source:}{Relative half-widths of 95\% confidence interval for the mean
(relative to the grand mean) with our method and the default single-level
method.  Smaller is better.  Shown for 3, 6, and 9 hours time windows for
experimentation.  Our method repeats executions and/or measurements where it
is beneficial to improve result precision, but the default method can only
use one execution and measurement from a binary.  With FFT and Rijndael
benchmarks, our method uses both multiple executions per binary and multiple
measurements per execution.  With TCP and HTTP, it uses a single execution
per binary but multiple measurements, as it leads to better precision.  Our
method is always better, and particularly more so when the variation due to
compilation is smaller than variations due to effects at lower levels.  }
\end{tabnote}

\end{table}

\subsection{Summary of Evaluation}

Our evaluation has demonstrated that both the bootstrap and the asymptotic
method work reasonably well.  For the bootstrap, resampling with replacement
at all levels (RRR) is a safe choice.  For the asymptotic method, using the
$t$-distribution even when normality assumptions cannot be made seems a safe
choice.  It is particularly more conservative (produces wider intervals)
than relying just on asymptotic normality.  The asymptotic method in theory
may fail to provide a result when the sample mean of the ``old'' system
appears not significantly different from zero.  We have seen it happen, but
only for 2 or 3 binaries/executions, never for larger sample sizes.  The
coverages with the asymptotic method converge to the projected value.  With
the bootstrap method, this is however not the case with all the benchmarks. 
The non-FFT benchmarks, which means those of our benchmarks with high
variation due to non-deterministic compilation, always have too high
coverages.  What has not been shown in the analysis is that the bootstrap
method is easier to implement, and perhaps to understand, than the
asymptotic method, but takes more computation time.  The computation time,
though, should not be a problem in a regular application of the method, when
only a few intervals need to be constructed.

We demonstrated that choosing a non-zero threshold for comparisons
simplifies the quantification considerably --- it reduces the necessary
experimentation time (total number of samples, expensive repetitions of
compilation and execution) and/or the false alarm rate.  It may even
completely eliminate the need for repetition at the highest level (i.e. 
compilations) when the variation caused by non-determinism at that level is
small.  Repeating compilations was necessary with the FFT benchmarks.  With
the other benchmarks, using a single binary with a well chosen threshold
would provide the same results and indeed reduce experimentation time.  The
Ping benchmarks would not need repetition of executions, either.  Hence,
even with the same system to test (in our case Mono), one can expect
different benchmarks to have very different needs for statistical modeling. 
Selection of good repetition counts is important --- too low and too high
counts lead to poor coverage and a high false alarm rate, but selecting a
sensible (non-zero) threshold reduces this problem in practice.

Note that the necessity of repeating compilation of FFT benchmarks in Mono
is by itself an important observation.  FFT benchmarks are (still) used as
we have seen in our survey of the ``2011'' papers mentioned previously (15
of all 122 papers measured also FFT).  In one case, the evaluation with FFT
even used the Mono platform.  As compilation is expensive one needs to
balance carefully whether to spend time on more executions of existing
binaries or to produce more binaries.  A similar tradeof is between running
more measurements of the same execution, or investing into starting (and
warming up) a new execution.  These tradeoffs are different for each
benchmark, and out method allows to find the optimum numbers.

\section{Conclusion}

Empirical evaluations in computing, particularly in programming languages
and systems research, are dominated by quantification of performance change
measured by the ratio of execution times.  Regrettably, we find that
evaluations reported even at premier venues commonly fail to report
uncertainty in measurements or to cater for non-determinism from various
sources.

In this work we attack both problems --- we have created a statistical model
that captures such non-determinism and we show how to construct a confidence
interval for the ratio of means within the model.  Our model is based on
general assumptions and caters for random factors that influence performance
and factors that the experimenter intentionally randomizes to reduce bias. 
We evaluate our method experimentally using statistical simulation on a set
of benchmarks.

The best method for quantification of performance change recommended in the
field to date is based on a visual test for overlap of confidence intervals
and only provides a binary answer as to whether the difference seen is
unlikely to be by pure chance.  Our method can provide the same answer, if
needed.  However, such an answer is usually not needed --- what is needed is
an uncertainty estimate for the ratio of means, and our method provides that
as well.

We have learned a number of lessons along the way.  Introductory statistical
textbooks offer insufficient guidance to advance evaluation practice in
computer systems, because these are not updated fast enough nor with
computer systems in mind.  Books specializing in computer systems
performance evaluation explain statistical methods in the context of
computer science, but still only include mostly introductory statistics,
which leads to recommendations of limited applicability.  It may be that
lack of applicability also contributes to the poor adoption of their
recommendations, though it may also be that there is simply not enough
pressure to do better evaluations.  To move forward, we need to consult
methods in the original statistical publications, and we sometimes need to
adapt them to our field.  As a part of that, we need to advance our
knowledge of the various factors in computer systems that influence
performance.  The work we present here is based on both of these activities.

While our method is better than the best recommended one, and even more so than
current practice, it does not solve all problems.  More work needs to be
done to incorporate fixed effects, that is effects of factors such as
hardware platform or operating system, which only have a small set of values
controlled by the user and experimenter.  This should be doable, and fixed
effects have been addressed even in books on computer systems performance
evaluation. There is a need for further work on rigorous summarization over
different benchmarks.  Apart from better statistics, we need better
benchmarks, benchmarks better analysed, and we could and should improve the
experimentation practice in our field by bringing more of the scientific
method used in ``real'' science, particularly physics and natural sciences
--- thorough reports, archival of experimental artifacts, repeatability and,
most importantly, reproducibility demonstrated through independent
reproduction studies.

% Acknowledgments
%\begin{acks}
%\end{acks}

% Bibliography
\bibliographystyle{acmsmall}
\bibliography{bib/paper}

% History dates
%\received{February 2007}{March 2009}{June 2009}

\end{document}